\documentclass[fleqn,usenatbib]{mnras}

\usepackage{newtxtext,newtxmath}

\usepackage[T1]{fontenc}
\usepackage{ae,aecompl}

\usepackage{graphicx}	

\usepackage{amsmath}	
\usepackage{amssymb}	
\usepackage{subfig}
\usepackage{booktabs}
\usepackage{afterpage}
\usepackage{float}
\usepackage{tablefootnote}
\usepackage{threeparttable}

\newcommand{\sgr}{$s_{gr}$}
\newcommand{\sbv}{$s_{BV}$}
\newcommand{\msgr}{\mathrm{s}_{gr}}
\newcommand{\dm}{\Delta m}
\newcommand{\dmf}{\Delta m_{15}}
\newcommand{\nickel}{$^{56}$Ni }
\newcommand{\mni}{M_\mathrm{Ni56} }
\newcommand{\msol}{M_\odot}
\newcommand{\MyIndent}{\hspace*{0.18cm}}%

\title[Type Ia luminosity function]{The ZTF-BTS Type Ia supernovae luminosity function is consistent with a single progenitor channel for the explosions}

\author[Sharon \& Kushnir]{
Amir Sharon$^{1}$\thanks{E-mail: amir.sharon@weizmann.ac.il}
and Doron Kushnir$^{1}$
\\
$^{1}$Department of Particle Physics and Astrophysics, Weizmann Institute of Science, Rehovot 76100, Israel\\
}

\date{Accepted XXX. Received YYY; in original form ZZZ}

\pubyear{2021}

\begin{document}
\label{firstpage}
\pagerange{\pageref{firstpage}--\pageref{lastpage}}
\maketitle

\begin{abstract}
We construct the Type Ia supernovae (SNe Ia) luminosity function (LF) using the Zwicky Transient Facility Bright Transient Survey (BTS) catalogue. While this magnitude-limited survey has an unprecedented number of objects, it suffers from large distance uncertainties and lacks an estimation of host extinction. We bypass these issues by calculating the intrinsic luminosities from the shape parameters of the light curve's $ g $ and $ r $ bands, with the luminosities calibrated from the well observed SNe Ia sample of the Carnegie Supernova Project, allowing us to construct, for the first time, the intrinsic LF of SNe Ia. We then use a novel tight relation between the color stretch and the synthesized $^{56}$Ni mass, $\mni$, to determine the $\mni$ distribution of SNe Ia. We find that the LFs are unimodal, with their peaks in line with previous results, but have a much lower rate of dim events and luminous events. We show that the features on top of the unimodal LF-derived distributions are all compatible with statistical noise, consistent with a single progenitor channel for the explosions. We further derive, for the first time, the SNe Ia distribution of host galaxy extinction, and find a mean selective extinction of $E(B-V)\approx0.1$ and a non-negligible fraction with large, $ >1\,\text{mag} $, extinction in the optical bands. The high extinction is typical for luminous SNe, supporting their young population origin. 
\end{abstract}





\section{Introduction}
\label{sec:intro} 

Type Ia supernovae (SNe Ia) are widely accepted to be the result of thermonuclear explosions of white dwarfs (WDs), but their progenitor systems and explosion mechanism are still under debate \citep[for a review, see, e.g.,][]{Maoz2014}. The luminosities of SNe Ia, powered by the radioactive decay chain of \nickel, span a significant range \citep{Phillips1993}. The peak bolometric luminosities vary between $ 10^{42}$ and $10^{43}\,\text{erg}\,\text{s}^{-1} $ and the synthesized \nickel masses, $\mni$, vary between $ \sim0.1$ and $1\msol $ \citep[see][for a recent compilation]{Sharon2020}.  

There is an ongoing debate in the literature on whether the observed range of SNe Ia properties can be explained by a single progenitor channel or whether multiprogenitor channels are required. The latter was favoured in \cite{Pakmor2013}, supported by the bimodal distribution of the $ B $ band magnitude decline during the first $15\,$d after the peak, $ \dmf(B) $, taken from the CfA3 sample \citep{Hicken2009}. Subsequent works that analysed the $ \dmf(B) $ distribution have repeated this claim \citep{Ashall2016,Hakobyan2020}. However, the $ \dmf(B) $ bimodality is driven by a clustering of low-luminosity SNe Ia in a small range of $ \dmf(B) $, where the peak luminosity is not a monotonic function of $ \dmf(B) $ \citep{Burns2018}, questioning the reality of the bimodal distribution. Multiple explosion channels are also favoured by \citet{Polin2019}, which divided the population of SNe Ia using the Si II velocity and the peak magnitude. Other works argue for a single progenitor channel. These include the continuous and relatively uniform properties of SNe Ia, which would require a fine tuning of the multiprogenitor channels \citep{Maoz2014}. This perspective has been recently highlighted by the tight correlation of $ \mni $ with the gamma-ray escape time, $ t_0 $ \citep{Wygoda2019,Sharon2020b}.

An accurately measured SNe Ia luminosity function (LF), which describes the intrinsic luminosity distribution of these SNe, can constrain the progenitor systems. For example, consider the case of two progenitor channels with significant rates but different intrinsic LFs. In this case, the total LF will show some structure where the two channels overlap (if the two channels do not overlap, the LF would not be unimodal). However, in order to obtain the intrinsic LF, the host galaxy's extinction must be taken into account, which usually cannot be easily done. Instead, many surveys in the last decades obtained a pseudo LF, in which the host galaxy's extinction is not removed. While the pseudo LF is useful on its own for many applications, the relation to the intrinsic LF is not clear. In fact, we show in this work that there is a significant difference between the pseudo LF and the intrinsic LF of SNe Ia. This is because a host extinction in optical wavelengths is typically a few tenths of a magnitude and can be larger than $ 1\,\text{mag} $ \citep[][see also Section~\ref{sec:extinction}]{Phillips2013}. The difference between the pseudo LF and the intrinsic LF complicates comparison of the predictions of various (extinction-free) models with the observations \citep[see e.g.,][]{Ruiter2013,Shen2017}. 

A prime example of a survey that aimed to construct the pseudo LF out of a complete volume-limited sample of SNe is the Lick Observatory Supernova Search \citep[LOSS;][]{Leaman2011,Li2011,Li2011b,Maoz2011}. The SNe Ia sample consists of 74 SNe discovered between 1998 and 2006, all with associated host galaxies within a distance of $ 80\,\text{Mpc} $. Using the known distances to the galaxies, an absolute peak magnitude pseudo LF within the range $\approx[-16.7,-19.5]\,\rm{mag}$ (of a filter that most closely matches the $ R $ band) was constructed. Subsequent works have analysed the host galaxies' properties \citep{LOSSrevisitedI,LOSSrevisitedII}, re-examined the SNe classifications \citep{Shivvers2017}, and constructed the $\mni$ distribution \citep{Piro2014}.

More recent surveys are also useful for constraining the LF. The CNIa0.02 \citep{Chen2020} is a volume-limited survey of nearby SNe with $ z<0.02 $ discovered by the All-Sky Automated Survey for Supernovae \citep[ASAS-SN;][]{Kochanek2017}, with 240 SNe Ia. The survey is complete for SNe with a $ V $ band peak magnitude of $ m_V<16.5\,\text{mag} $, which corresponds to absolute magnitudes of $ M_V\lesssim-18.2\,\text{mag} $ at $ z=0.02 $. Since the dimmest SNe Ia peak magnitudes reach $ \approx-17\,\text{mag} $ \citep{Taubenberger2017,Burns2018}, and extinction further increases the observed magnitudes, a volume correction would be required for the low-luminosity end of the LF. 

The Zwicky Transient Facility (ZTF) Bright Transient Survey \citep[BTS;][hereafter P20]{Fremling2020,Perley2020} seeks to identify and classify extragalactic transients in the Northern sky. The objects are observed in the $ g $ and $ r $ bands and are spectroscopically classified. The number of objects substantially increased compared to previous surveys, with over a thousand of identified SNe per year, most of them SNe Ia. If available, the redshift of the transient is determined from the host galaxy, or otherwise from the SN spectral features. In practice, a significant fraction of SNe lacks a spectroscopic redshift of the host. In P20, the SNe with an observed peak magnitude of $ m<18.5\,\text{mag} $ are used to construct the pseudo LF and measure the total rate of each SNe type. They found for SNe Ia an absolute peak magnitude range of $\approx[-16.5,-20.5]\,\text{mag} $ and a total rate of $\approx 2.35\times10^4\,\text{Gpc}^{-3}\,\text{yr}^{-1} $.

Another well-known SNe survey is the Carnegie Supernova Project \citep[CSP;][and others]{Contreras2010,Stritzinger2011,Krisciunas2017,Burns2018}. The CSP's Ia sample consists of 123 well-observed SNe Ia, with independent distance estimates. The high cadence of observations over a large range of wavelengths enables the estimation of the total extinction using templates \citep{Burns2018}, and also the construction of the bolometric luminosity \citep{Scalzo2019,Sharon2020}. However, the sample is not complete and cannot be used to study SNe rates.

\begin{figure}
	\includegraphics[width=\columnwidth]{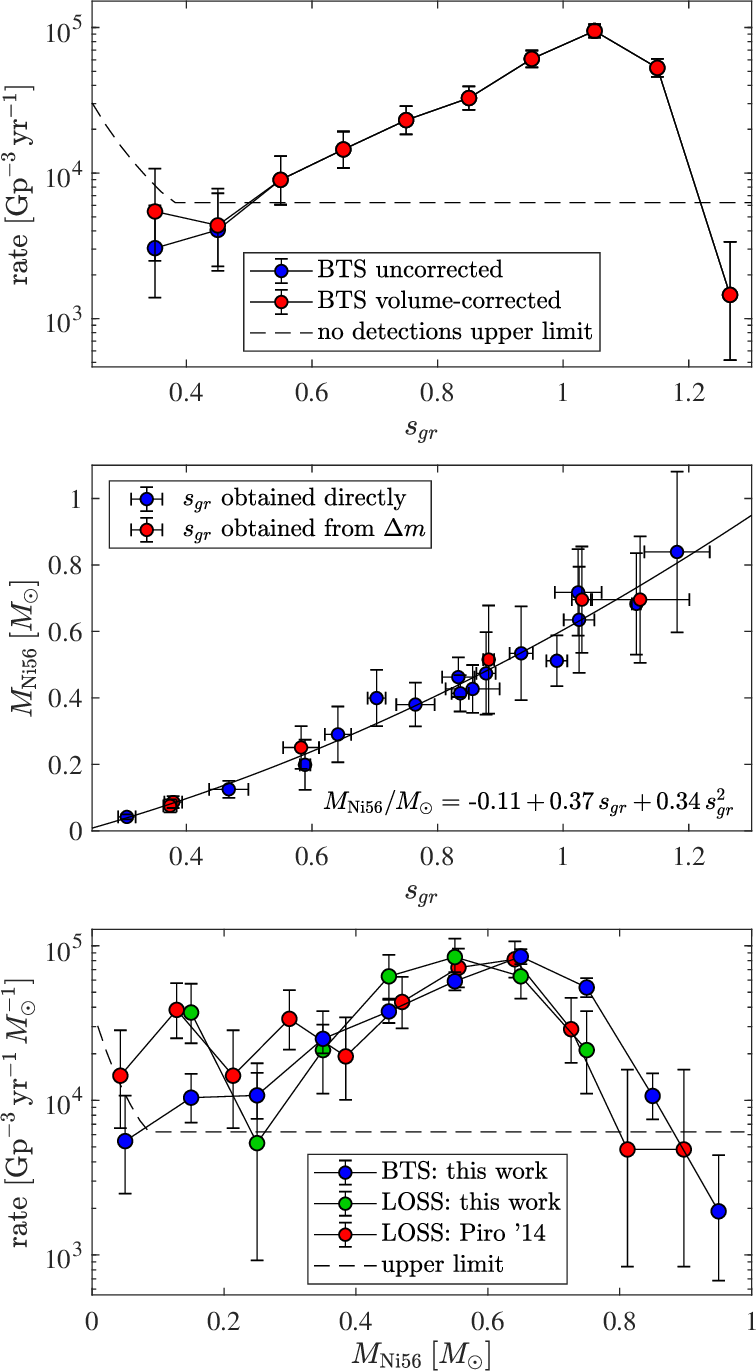}
	\caption{The main results of the paper. Top panel: The \sgr\ distribution of the $ 180\,\text{Mpc} $ volume-limited BTS sample. Volume-uncorrected (-corrected) results are shown as blue (red) symbols, with the volume-uncorrected rates scaled to match the total volume-corrected rate (only the dimmest bin, with three SNe, requires a non-negligible volume correction). The black, dashed line indicates the $95$ per cent upper limit for no detections in a bin of 0.1. Middle panel: The $\mni-$\sgr\ distribution of the CSP sample. Blue symbols indicate SNe Ia with a direct \sgr\ measurement, while red symbols indicate SNe Ia with \sgr\ obtained through a decline-rate relation. Bottom panel: The $\mni$ distribution from BTS (this work, blue), LOSS (this work, green), and of \citet{Piro2014} (red). The black, dashed line indicates the 95 per cent upper limit for no detections in a bin of $ 0.1\,M_\odot $}
	\label{fig:LF Ni}
\end{figure}

In this work, we use a combination of the BTS public catalogue\footnote{https://sites.astro.caltech.edu/ztf/bts/explorer.php, downloaded on July 2021, with 1519 SNe Ia that passed the quality cuts described in P20.} and the CSP Ia sample to accurately determine the intrinsic LF of SNe Ia. We first compare in Section~\ref{sec:II} the light curve shapes of the CSP and the BTS samples to show that the host extinction and the distance uncertainties significantly affect the luminosity distribution. In order to partly bypass these difficulties, we use the tight correlations between the light curve color stretch, \sgr\ \citep{Ashall2020} and the intrinsic luminosity, as calibrated from the CSP Ia sample. The use of \sgr\ allows the accurate determination of the intrinsic luminosity for the entire luminosity range of SNe Ia \citep[this is a variant of $ s_{BV}$, introduced by][]{Burns2014,Burns2018}. We determine \sgr\ values for the vast majority of the BTS survey in Section~\ref{sec:CSP}. In Section \ref{sec:completeness}, we estimate the SNe distances (for the ones that lacks a spectroscopic redshift of the host) and investigate the completeness of the sample. We find that most of the sample is complete up to $ \sim180\,\text{Mpc} $, and we use this distance to construct a volume-limited subsample that is used for our main results. In Section~\ref{sec:LF}, we construct, for the first time, the intrinsic \sgr\ distribution (upper panel of Figure~\ref{fig:LF Ni}) and the intrinsic LF (Figure~\ref{fig:LF peak r}). We find that the rate of dim events is lower by almost an order of magnitude than the corresponding pseudo-LF values of P20, and that the dimmest event has a peak magnitude of $ M_r\approx-17.5\,\text{mag} $, much more luminous than the dimmest pseudo-LF events of P20 with $ M_r\approx-16.7\,\text{mag} $. We find a total rate of $ \approx2.91^{+0.58}_{-0.45}\times10^4\,\text{Gpc}^{-3}\,\text{yr}^{-1}  $ (per comoving element in the redshift range $ z\approx [0.01,0.04] $), consistent with previous studies \citep{Dilday2010,Graur2011,Li2011b,Frohmaier2019,Perley2020}.

We then use in Section~\ref{sec:Ni} the CSP Ia sample to calibrate a novel tight relation between \sgr\ and $\mni$ (the middle panel of Figure~\ref{fig:LF Ni}). $\mni$ is determined using the methods of \citet{Sharon2020}, which use the bolometric light curve and require accurate distances and measurements over long periods of time and over a wide range of wavelengths.  While these requirements are fulfilled for many SNe in the CSP sample, they are difficult to achieve in a complete survey with thousands of objects per year, such as the ZTF BTS. Combining the $\mni-$\sgr\ relation with the \sgr\ distribution allows us to determine the $\mni$ distribution, which is the main result of this paper (the bottom panel of Figure~\ref{fig:LF Ni}, blue symbols). In Appendix~\ref{app:LOSS}, we also apply our methods to the LOSS sample to determine the $\mni$ distribution (green symbols). For comparison, the results of \citet{Piro2014}, based on the LOSS sample, are shown as red symbols.

We find that the LF, the $s_{gr}$, and the $\mni$ distributions are unimodal. The peaks of the distributions are at $ M_r\approx-19.2 $ and $ \mni\approx0.6\,M_\odot $ \citep[in agreement with P20 and][respectively]{Piro2014}. The rates of dim events ($ \mni\lesssim0.4\,M_\odot $) and luminous events ($ \mni\gtrsim0.8\,M_\odot  $) are lower by at least a factor of $5$ from the rate of the most common, $\mni\approx0.6\,M_{\odot}$, events. 
We perform Monte Carlo (MC) simulations to show that the features on top of the unimodal-derived distributions are all consistent with statistical noise (Section~\ref{sec:unimodal}). The derived distributions, therefore, are consistent with a single progenitor channel for the explosions.

We construct similar distributions for different host properties in Section~\ref{sec:host}. We show that the BTS LF of SNe Ia in early-type galaxies is more inclined towards low-luminosity events than the LF of SNe Ia in star-forming galaxies \citep[similar to the results of][]{Li2011,Ashall2016}. In Section \ref{sec:extinction}, we construct, for the first time, the distribution of the host galaxy extinction for a subsample of the BTS with accurate distances and for the LOSS survey. We find a mean $ E(g-r) $ value of $ \sim0.1\,\text{mag} $, and that a non-negligible fraction of SNe, especially in star-forming galaxies, exhibit large extinction values, $\gtrsim1\,\text{mag} $, in the optical bands. We find that the host galaxy extinction, $A^\text{h}$, is highly correlated with the luminosity of the SNe. For example, in the $r$ band, luminous SNe with $\msgr\gtrsim1$ suffers from $ A^\text{h} \approx1\,\rm{mag}$ for a non-negligible fraction of the SNe. We found no low \sgr\textbf($ <0.8 $) SNe with $A^\text{h}$ larger than 1 mag. These results are in line with the idea that luminous SNe Ia originate from young populations, while low-luminosity SNe Ia require old progenitors \citep{Howell2009,Lampeitl2010,Maoz2014}. We summarize our results in Section~\ref{sec:discussion}, highlighting the importance of acquiring spectroscopic redshift for all SNe hosts, which would tightly constrain the intrinsic LF. 

In Appendix~\ref{app:LOSS}, we elaborate on our analysis of the LOSS sample. In Appendix~\ref{app:k corr}, we analyse the effects of the SNe redshift distribution and the filter transmission functions on the decline rates and peak magnitudes. In Appendix~\ref{app:volume corr}, we describe a few different methods to calculate the LF and we compare between volume-limited LF and magnitude-limited LF. In Appendix~\ref{app:CSP}, we provide the properties of each SNe Ia from the BTS, CSP, and LOSS samples that are used in this work. The provided data can be used to reproduce the main results of this paper.

\section{BTS-CSP comparison}
\label{sec:II}

In this section, we analyse the pseudo-LF derived in P20 by comparing the BTS light curves to the CSP sample.
The ZTF BTS public catalogue \citep[][P20]{Fremling2020} includes, for each transient, the light curves in the $ g $ and $ r $ bands, the observed peak magnitude, the galactic extinction in the direction of the transient and the redshift. The provided absolute peak magnitude is inferred from the observed peak magnitude, Galactic extinction, and the distance (determined from the redshift through a cosmological model with $ \Omega_M=0.3,\Omega_\Lambda=0.7 $, and $ h=0.7 $). \cite{Fremling2020} reported that about $44$ per cent of the objects were associated with host galaxies that have a catalogued spectroscopic redshift, which was obtained by searching for galaxies within $ 2 $ arcsec of the SN. For the other cases, the redshift is determined from the SN spectral features, with a non-negligible uncertainty of  $ \Delta z\approx0.005 $ \citep[][P20]{Fremling2020}. For construction of the pseudo-LF the host extinction is not required \citep[note, however, that in optical wavelengths the host extinction is typically a few tenths of a magnitude and can be larger than $ 1\,\text{mag} $;][see also Sections~\ref{sec:completeness} and~\ref{sec:extinction}]{Phillips2013}. 

In this work, using the NASA Extragalactic Database (NED)\footnote{https://ned.ipac.caltech.edu/} and the Sloan Digital Sky Survey (SDSS)\footnote{https://www.sdss.org/} catalogues, we obtained the spectroscopic redshift of $ \approx35$ per cent of the SNe. This was achieved by using the methods described in \cite{Ofek2014} and \cite{Soumagnac2018}, and a search radius of $ 10\,\text{kpc} $ around the SN location. The angular size of the search radius was estimated with the provided redshift.

\begin{figure*}
	\includegraphics[width=0.8\textwidth]{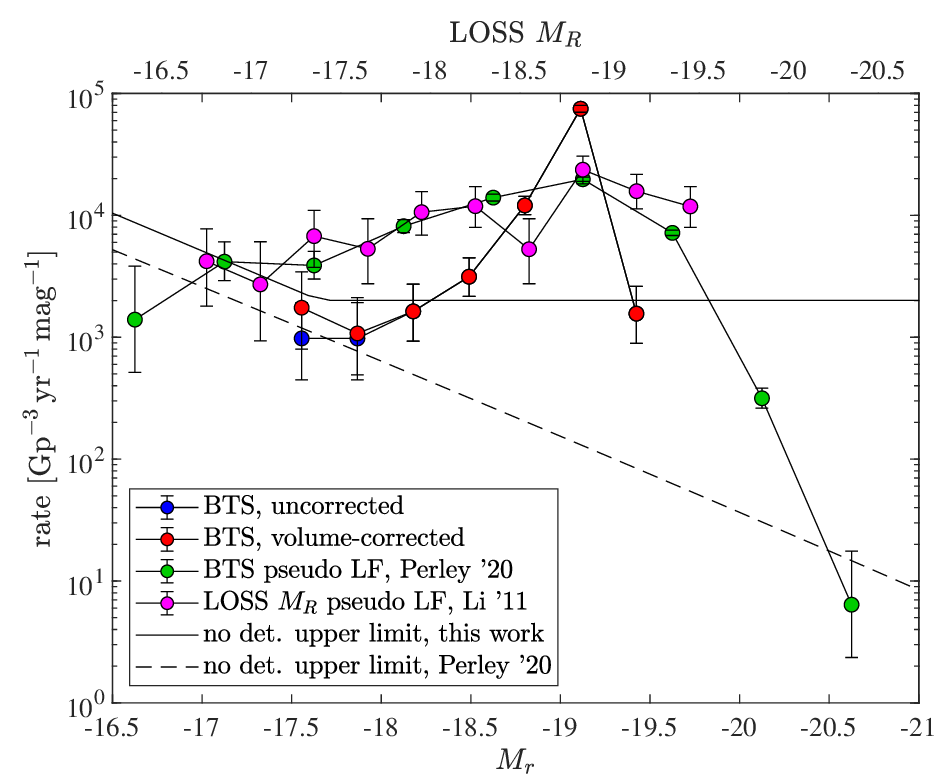}
	\caption{The $ M_r $ LF. The BTS volume-corrected (-uncorrected) intrinsic LF of this work is indicated by red (blue) symbols (only the dimmest bin requires a significant volume correction). The pseudo-LF of P20 and of LOSS \citep{Li2011} are indicated by green and magenta symbols, respectively. All the LFs are scaled so that their total rate matches the total BTS volume-corrected rate. The solid (dashed) black lines indicate the $95$ per cent upper limit for no detections in a bin of $ 0.33\,(0.5)\,\text{mag} $ in our (P20) analysis.}
	\label{fig:LF peak r}
\end{figure*}

To build the pseudo-LF and to calculate the SNe rate, the authors of P20 only considered objects with observed peak magnitudes brighter than $ m_\text{lim}=18.5\;\text{mag} $. They considered several completeness factors and calculated the volumetric rate by:
\begin{equation}\label{eq:rate}
R=\frac{1}{T}\frac{1}{f_\text{global}}\sum_{i=1}^{N}w_i,
\end{equation}
where $ T =2.12\,\text{yr}$ and $N=875$ are, respectively, the time and number of SNe Ia of the survey, $ f_\text{global} =0.172$ is the fraction of observed SNe due to the sky coverage, galactic extinction effects and recovery fraction, and $ w_i $ are the weights of each SN. The weights are given by $w_i= (f_{\text{cl},i} V_\text{max,i})^{-1}$, where $ f_{\text{cl},i} $ is the classification efficiency that depends on the observed magnitude ($ 15\,\text{mag}<m<18.5\,\text{mag} $), and $ V_\text{max,i} $ is the volume within each SN can be detected, given the limiting magnitude:
\begin{equation}
\label{eq:max vol}
V_\text{max,i} = \frac{4\pi}{3}\left(10^{\frac{m_\text{lim}-M_i}{5}-8}\right)^3\,\text{Gpc}^3,
\end{equation}
where $ M_{i} $ is the absolute peak magnitude of the SN. P20 obtained an SNe Ia magnitude range of $[-16.7,-20.66] \,$mag (in this method, the dimmest events could be the result of large host extinction). Their pseudo-LF, shown in Figure \ref{fig:LF peak r}, peaks at $ m\approx-19\,\text{mag} $, and the total rate is $ (2.35\pm0.24)\times10^4\,\text{Gpc}^{-3}\,\text{yr}^{-1} $. 

\begin{figure}
	\includegraphics[width=\columnwidth]{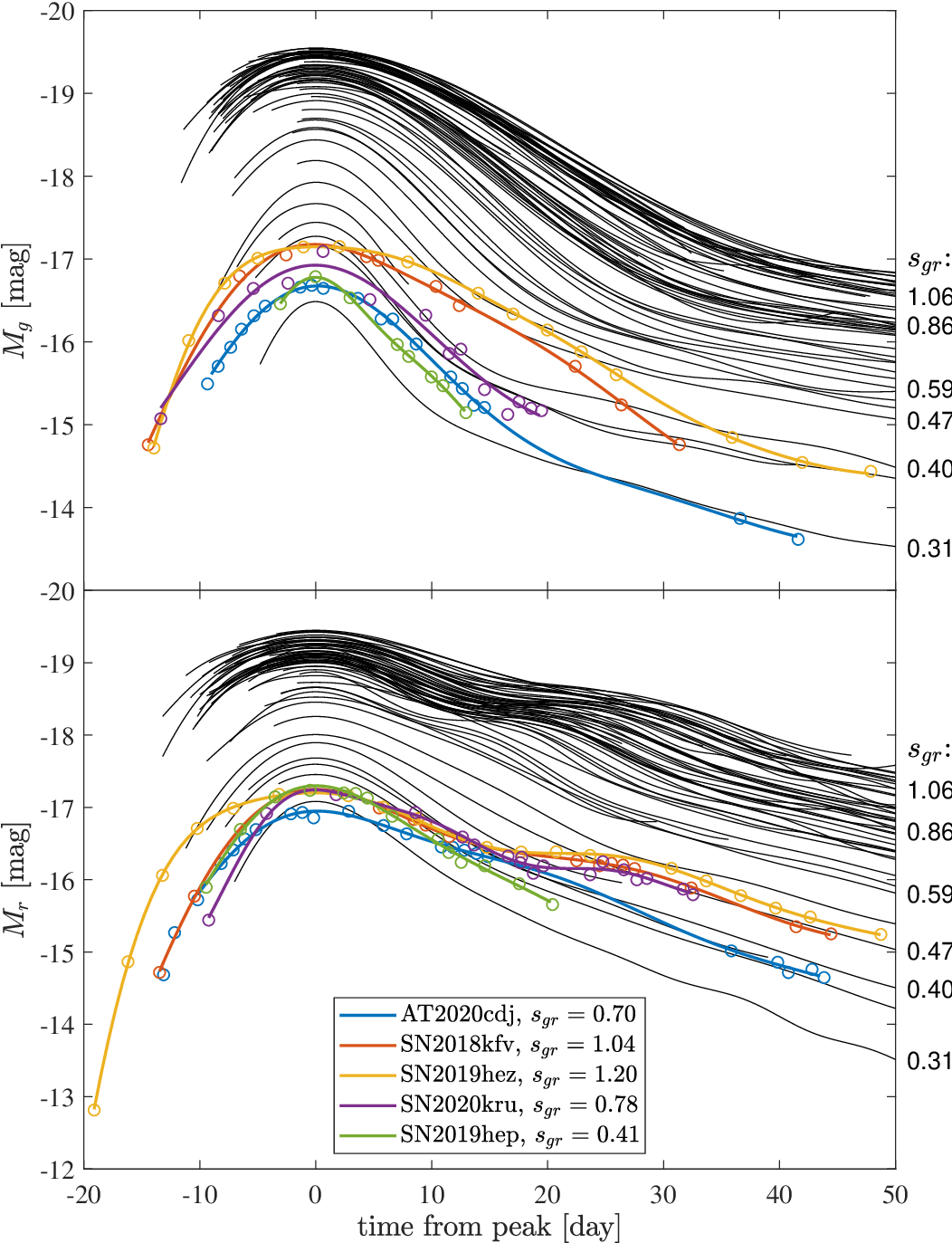}
	\caption{CSP $ g $- and $ r $-band light curves and several BTS SNe with low absolute magnitudes, as determined by P20. The \sgr\ of the BTS SNe is given in the legend, and the \sgr\ of several representing SNe from the CSP are displayed to the right of the figure. The range of the \sgr\ of the BTS sample is between $ \approx0.4$ and $ 1.2 $, while the \sgr\ of the CSP SNe for the same luminosities reaches a maximum of $ \approx0.4 $.}
	\label{fig:templates-faint}
\end{figure}

The luminous and the dim ends of the P20 pseudo-LF demonstrate the effects of host extinction and distance uncertainties. The highest reported peak magnitude of the sample is $ M=-20.66\,\text{mag} $, which is $ \sim1\,\text{mag} $ higher than the highest peak magnitude in the CSP sample (SN 2005eq). Given the estimate $ \mni\approx0.85,M_\odot $ for SN 2005eq \citep{Sharon2020}, and using the SN 2005eq bolometric correction from the $r$ band, we find $\mni \approx 2\,M_\odot $ for the most luminous P20 event. There are also $ \sim20 $ SNe with $ M\lesssim-20\,\text{mag} $ (see examples in Table~\ref{tab:BTS edges}), corresponding to  $ \mni\gtrsim1.2\,M_\odot $. We argue below that the luminosity of these objects is, in fact, much lower. We next compare in Figure~\ref{fig:templates-faint} the $ g $ and $ r $ absolute magnitude light curves of some of the dimmest SNe of P20 with the CSP sample. The absolute magnitude light curves of the BTS sample were constructed by simply adding the differences between the given absolute and observed peak magnitudes to the observed light curves. For the CSP sample, we use the provided distances and extinction values. As can be seen in the figure, some SNe from the BTS sample display features that are absent in low-luminosity SNe of the CSP sample, such as a shoulder in the $ r $ band after the peak or a much slower decline of the $ g $ band, and are therefore more likely to be more luminous events. We list in Table~\ref{tab:BTS edges} the absolute peak magnitudes of these dim SNe, as well as those of the most luminous SNe in the sample. For a reference, we provide the \sgr\ and peak magnitudes of these events from the calibration of Section~\ref{sec:CSP}, demonstrating the large differences between our and the P20 calculations. None of the objects in Table \ref{tab:BTS edges} have a spectroscopic redshift in our analysis.

\begin{table}
	\caption{Five of the dimmest and most luminous SNe Ia, as given by the BTS explorer, and are a part of the P20 LF. The absolute magnitudes of these SNe, as calculated from $\msgr$, are in tension with the absolute magnitudes reported by P20. None of these SNe have a spectroscopic redshift.}
	\label{tab:BTS edges}
	\begin{threeparttable}
	\renewcommand{\TPTminimum}{\linewidth}
	\makebox[\linewidth]{%
	\begin{tabular}{lccc}
		SN & $ M_\text{peak}^{z} $\tnote{a}  & $ M_\text{peak}^{\msgr} $\tnote{b}  & \sgr\ \\\midrule 
		2020cdj & -16.98 & -18.76 & 0.70 \\
		2018kfv & -17.01 & -19.18 & 1.04 \\
		2019hez & -17.21 & -19.27 & 1.20 \\
		2020kru & -17.24 & -18.91 & 0.78 \\
		2019hep & -17.32 & -17.81 & 0.41 \\\midrule
		2020fxm & -20.21 & -19.27 & 0.88 \\
		2019fck & -20.30 & -19.46 & 1.13 \\
		2019tmp & -20.30 & -19.59 & 1.26 \\
		2020yjf & -20.48 & -19.52 & 1.14 \\
		2019phj & -20.66 & -19.48 & 1.14 \\
	\end{tabular}}
	\begin{tablenotes}
		\item [a] Absolute magnitude of the BTS explorer, calculated from the redshift. 
		\item [b] Absolute magnitude of this work, calculated from $\msgr$.
	\end{tablenotes}
\end{threeparttable}
\end{table}

In order to quantify the tension between the luminosities and the light curve shapes, we use the well-known correlation between them \citep{Phillips1993}. While historically the magnitude decline, $ \dmf $, at a given band was used to calibrate the peak magnitude, \citet{Burns2014,Burns2018} showed that the color stretch parameter, $ s_{BV} $, provides a much better correlation, specifically for low-luminosity SNe Ia. The parameter \sbv is obtained by measuring the time difference between the $ B $ band peak time and the $ B-V $ color peak time:
\begin{equation}\label{eq:sbv def}
s_{BV} = \frac{t_{\mathrm{max},B-V}-t_{\mathrm{max},B}}{30\,\text{d}}.
\end{equation}
While originally defined in the Johnson photometric system, an equivalent useful parameter, $\msgr$, can be defined for the Sloan filters, with the $ g $ and $ r $ bands replacing the $ B $ and $ V $ bands, respectively \citep{Ashall2020}. 

\begin{figure}
	\includegraphics[width=\columnwidth]{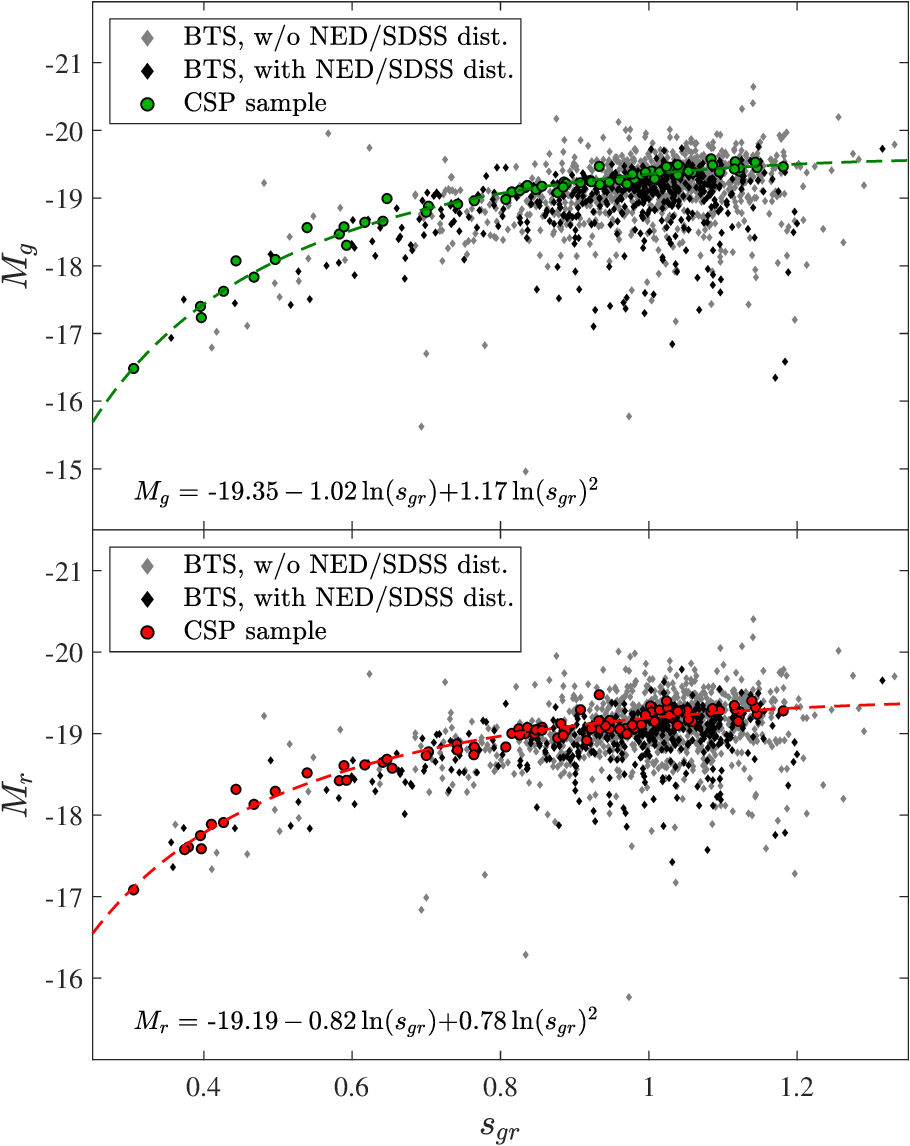}
	\caption{$\msgr -\, $peak magnitude distribution of the CSP sample and of the BTS sample (using the redshift for the BTS sample). Upper panel: $ g $-band absolute magnitude as a function of $\msgr$. The CSP values are indicated by green circles, and the BTS values by black (grey) diamonds for SNe with (without) a spectroscopic redshift. The fit to the CSP sample is indicated by a dashed line and is displayed in the bottom of the panel. Bottom panel: Same as the top panel but for the $ r $ band with red symbols.}
	\label{fig:peak mag sgr}
\end{figure}

We downloaded the sample of BTS objects after 3.13 yr of observations; it consists of 1519 SNe Ia (note that only a subsample is used to construct the LF, see Appendix \ref{app:volume corr}). We obtained the spectroscopic redshifts of 528 SNe with the use of the NED and SDSS catalogues. We measure the light curves' properties (and estimated the errors of the derived values) by using Gaussian processes interpolations as implemented by the SNooPy package \citep{Burns2011}. The results of the CSP and BTS samples are given in Tables~\ref{tab:CSP sample} and~\ref{tab:BTS sample}, respectively (see Section~\ref{sec:CSP} for details). The peak magnitude-\sgr\ distributions of the CSP and the BTS samples are shown in Figure~\ref{fig:peak mag sgr}. In this plot, the peak magnitudes of both samples are calculated from the observed magnitudes, the provided distance, and the galactic extinction. We further correct for the host extinction of the CSP sample and apply a $ K $-correction for both the CSP and the BTS samples (which somewhat differs from the $ K $-correction of P20, see Appendix~\ref{app:k corr} for details). 

\begin{figure}
	\includegraphics[width=0.95\columnwidth]{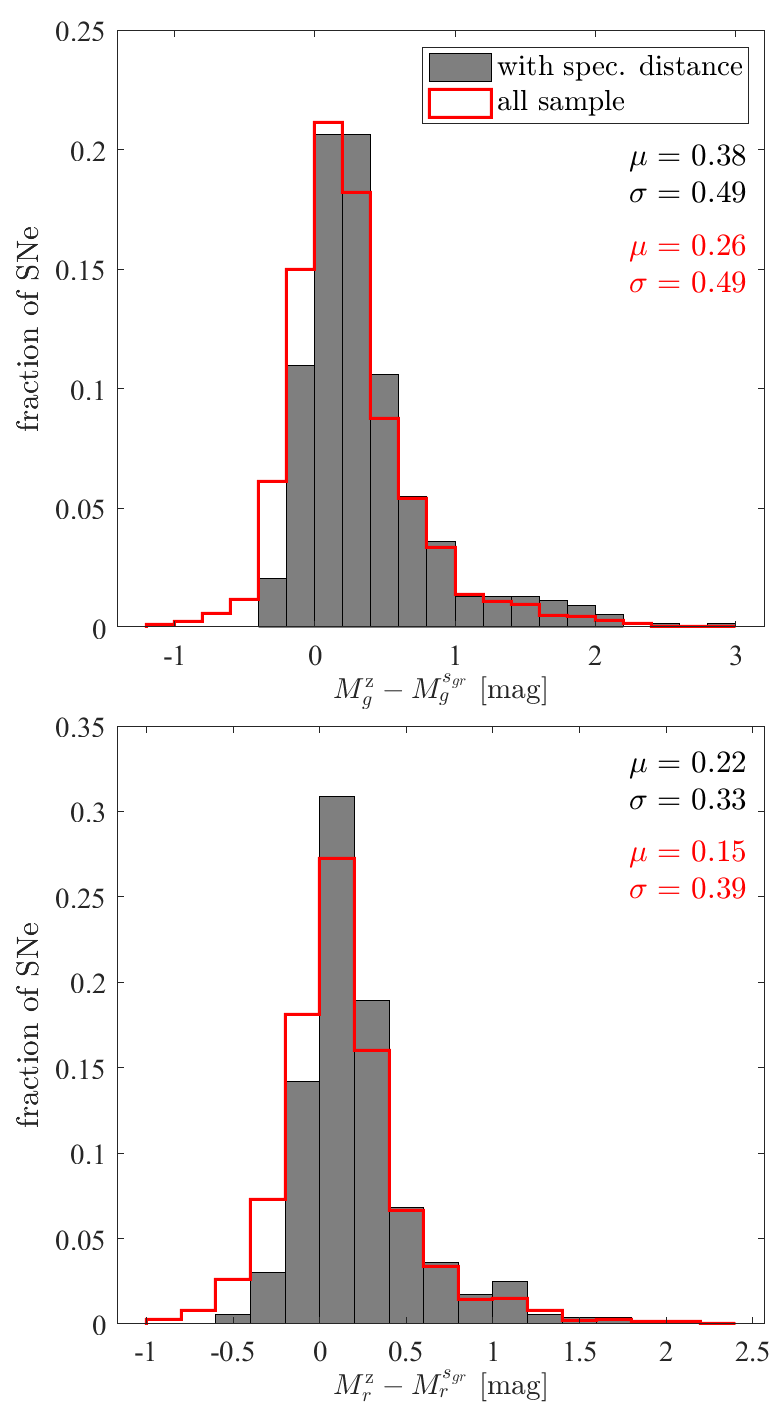}
	\caption{Distributions of the differences between the absolute magnitudes calculated with the redshift estimate, $ M^z $, and the absolute magnitudes calculated with $\msgr$, $ M^{\msgr}$. The distributions of the SNe with spectroscopic redshifts are indicated by grey-filled bars, and the whole sample by red lines. Top (bottom) panel: the $g$ ($r$) band distributions. The distributions have a bias of $ \approx0.3\text{--}0.4\,\text{mag} $ and $ \approx0.15\text{--}0.25 $ in the $ g $ and $ r $ bands, respectively.}
	\label{fig:peak mag hist}
\end{figure}

As can be seen in Figure \ref{fig:peak mag sgr}, the peak magnitudes are tightly correlated with \sgr\ for the CSP sample, with a sharp decrease in luminosity as \sgr\ decreases. To characterize this relation, we fit the peak magnitudes to a second-degree polynomial in $\ln(s_{gr})$ and obtained a scatter of $\approx0.06\,(0.08)$ mag in the $g$ ($r$) band. The fit parameters are given in Table~\ref{tab:fits sgr lum}. Since the scatter of the fits is dominated by an intrinsic scatter that is much larger than the typical error, the $\chi^2$ values of the fits are very large and are not shown. The scatter of the BTS peak magnitudes with similar \sgr\ values is significantly larger. Figure~\ref{fig:peak mag hist} shows the deviations between the BTS absolute magnitude, $ M^z $, calculated with the estimated redshifts and neglecting host extinction, and the absolute magnitude predicted by the CSP fit at the measured $\msgr$, $ M^{\msgr} $. SNe with spectroscopic redshifts are indicated by the grey bars while the whole sample is indicated by the red lines. To accurately compare the absolute magnitudes between the surveys, we applied corrections to the CSP results, taking into account the different filter transmission functions (see Appendix~\ref{app:k corr} for details). As can be seen in Figure~\ref{fig:peak mag hist}, the magnitude differences have a bias of $ \approx0.3\text{--}0.4\,\text{mag} $ in the $ g $ band and $ \approx0.15\text{--}0.25 $ in the $ r $ band, which is most likely the results of ignoring the host extinction. For the whole sample, a non-negligible fraction of SNe have negative values. Since the negative values cannot be attributed to host extinction, they are most likely the result of distance errors, supported by the less frequent negative values for SNe with spectroscopic redshifts.

\begin{table}
	\caption{Properties of the fits to the peak magnitudes with respect to $ \text{ln}(\msgr)$ for the SNe in the CSP sample.}	
	\label{tab:fits sgr lum}
	\begin{threeparttable}
	\renewcommand{\TPTminimum}{\linewidth}
	\makebox[\linewidth]{%
	\begin{tabular}{lcccccc}
	Parameter & Fit variable & $ p_0 $\tnote{a} & $ p_1 $\tnote{a} & $ p_2 $\tnote{a} & $ N $\tnote{b}  & Scatter \\ \midrule 
		 $ M_g $ & $ \text{ln}(\msgr) $ & -19.37 & -1.01 & 1.18 & 37 & 0.06  \\ 
		$ M_r $ & $ \text{ln}(\msgr) $ & -19.22 & -0.81 & 0.80 & 37 & 0.08  \\ 
		$ M_B $ & $ \text{ln}(\msgr) $ & -19.40 & -0.98 & 1.91 & 32 & 0.11  \\ 
		$ M_V $ & $ \text{ln}(\msgr) $ & -19.30 & -0.88 & 1.25 & 31 & 0.10  \\ 
	\end{tabular}}	
	\begin{tablenotes}
		\item [a] Polynomial coefficient to the equation $ p_0+p_1x+p_2x^2 $ 
		\item [b] Number of objects used to determine the fit
	\end{tablenotes}
	\end{threeparttable}
\end{table}

\section{Determination of the colour stretch parameter}
\label{sec:CSP}

We next sought to apply the CSP peak magnitude$ -\msgr$ relation to the BTS sample. However, we were able to directly determine \sgr\ values for only 204 out of the available 1519 SNe. The reason is that a determination of \sgr\ requires measurements of both bands for at least $ \sim30\,$d, which is the typical time between the $ g $-band-peak and the color maximum, and such late observations are not always available. Also, for many SNe, the $ g $-band-peak time cannot be determined, since there are no pre-peak observations. It is also possible that poor sampling does not allow the determination of neither of the two times. To overcome this issue, we measure several decline rate parameters, which are easier to determine, and attempt to find correlations between them and $ \msgr $, in order to recover \sgr\ from these decline rate parameters. We consider $\dmf(g) $, $\dmf(r) $, $\dm_8(g) $, and $\dm_{30}(r) $, which measure the magnitude difference at the specified number of days from the peak magnitude of the band. 

\begin{figure*}
	\includegraphics[width=0.95\textwidth]{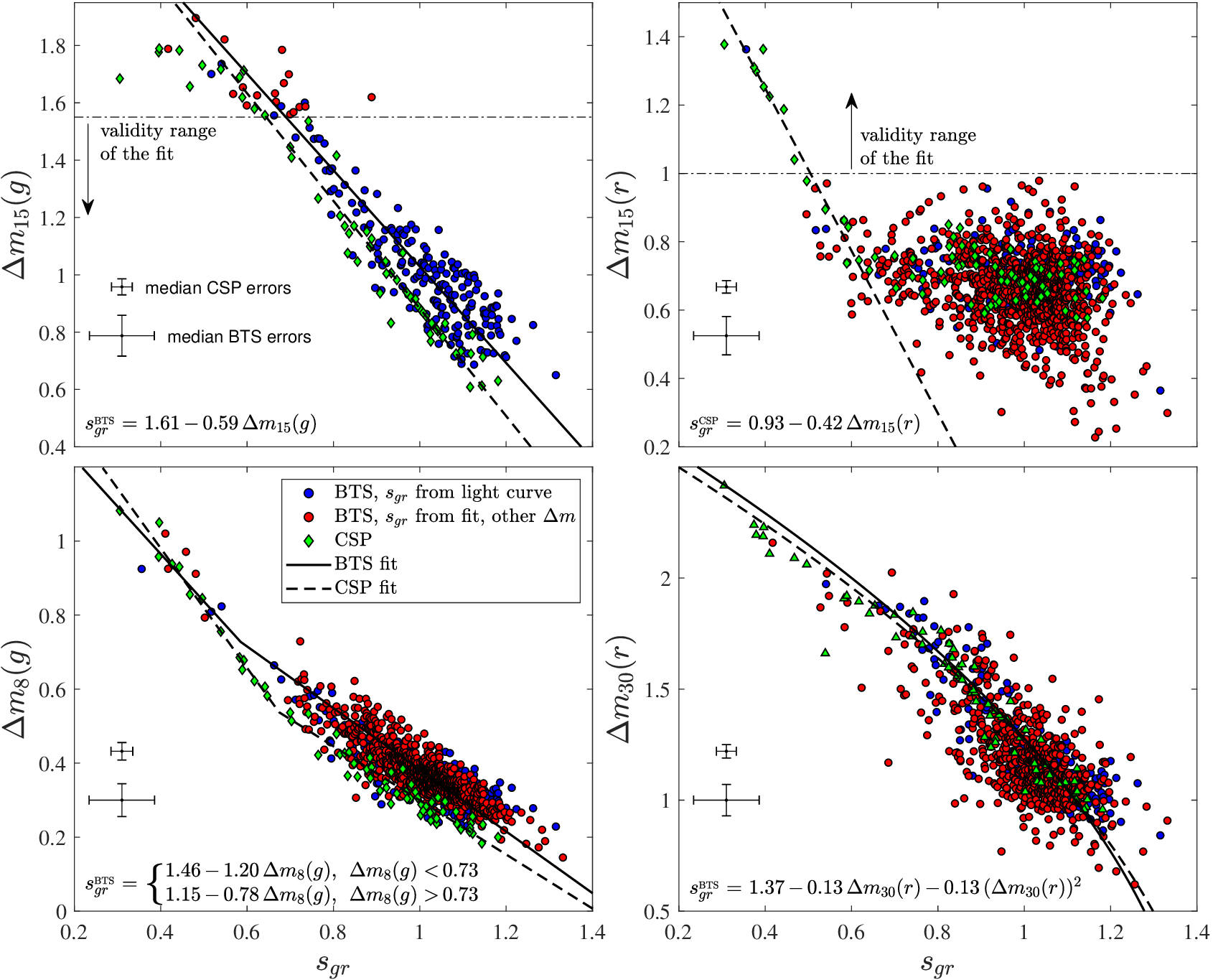}
	\caption{Decline rate parameters as a function of $\msgr$. BTS SNe with a direct \sgr\ measurement are shown in blue circles. BTS SNe whose \sgr\ is determined using a different decline rate parameter are shown in red circles. CSP SNe are marked by green diamonds. The median errors of the two samples are indicated in each plot. Fits of the BTS and CSP samples results are shown in black, solid lines and black, dashed lines, respectively. Also shown are the validity ranges of the $ \dmf(g) $ and $ \dmf(r) $ fits. SNe whose decline rate is outside the validity range are not fitted with a \sgr\ value from that decline rate.}
	\label{fig:BTS sgr vs dm}
\end{figure*}

Using SNe with both \sgr\ and some $\Delta m$ parameter measured, we fit \sgr\ as a first- or second-order polynomial in $\Delta m$:
\begin{equation}\label{eq:sgr dm}
\msgr = p_0+p_1\Delta m+p_2\left(\Delta m\right)^2,
\end{equation}
where $ p_i $ are the fit parameters ($ p_2 =0$ is zero for first-order polynomials). The fit to $ \dm_8(g) $ is divided into two linear fits. We make separate fits for the CSP and BTS samples, due to the differences between the redshift distribution and the filter response functions of the two surveys (see discussion below). Figure~\ref{fig:BTS sgr vs dm} shows the distribution of \sgr\ and the decline rate parameters of both samples. BTS SNe with both \sgr\ and decline rate measurements are shown as blue circles, with the fit to the results denoted by a black, solid line. Since the color stretch can often be determined from several decline rate parameters, and in order to check for consistency, for each panel, we also plot the \sgr\ values that were determined using a different decline rate parameter (red circles). CSP SNe are indicated by green diamonds, with the fit to the results in a black, dashed line. A similar plot that includes only the CSP results is given in Appendix~\ref{app:CSP}; there, the error bar of each measurement is indicated, and the SNe with both \sgr\ and decline rate measurements are presented separately from the SNe for which the \sgr\ values were determined using a different decline rate parameter. The parameters of the fits in each \sgr\ range are given in Table~\ref{tab:fits sgr dm}.

\begin{table*}
	\caption{The CSP and ZTF decline rates as a function of \sgr\ fit parameters, used in Equation~\eqref{eq:sgr dm}. For each fit, only SNe within the indicated \sgr\ range were part of the fit. The BTS $ \dm_{8}(g) $ fit for low \sgr\ values uses both BTS and CSP SNe.}
	\label{tab:fits sgr dm}
	\begin{threeparttable}
		\renewcommand{\TPTminimum}{\linewidth}
		\makebox[\linewidth]{%
	\begin{tabular}{llccccccc}
		& Decline rate & \sgr\ Range & $ p_0 $ & $ p_1 $ & $ p_2 $ & $ N $\tnote{a} & $ \chi_\nu^2 $\tnote{b} & Scatter\\\midrule 
	CSP & $ \Delta m_{15}(g)$ & $s_{gr}>0.5$ & 1.47 & -0.53 & -- & 29 & 3.37 & 0.035 \\
		& $ \Delta m_{15}(r)$ & $s_{gr}<0.7$ & 0.93 & -0.42 & -- & 10 & 2.70 & 0.021 \\
		& $ \Delta m_{8}(g)$ & $s_{gr}>0.65$ & 1.41 & -1.37 & -- & 27 & 2.70 & 0.037 \\
		& $ \Delta m_{8}(g)$ & $s_{gr}<0.65$ & 1.01 & -0.62 & -- & 10 & 1.34 & 0.016 \\
		& $ \Delta m_{30}(r)$ & --         & 1.41 & -0.16 & -0.13 & 35 & 4.37 & 0.036 \\\midrule 
	BTS & $ \Delta m_{15}(g)$ & $s_{gr}>0.5$ & 1.61 & -0.59 & -- & 192 & 1.83 & 0.050 \\
		& $ \Delta m_{8}(g)$ & $s_{gr}>0.65$ & 1.46 & -1.20 & -- & 198 & 2.44 & 0.056 \\
		& $ \Delta m_{8}(g)$ & $s_{gr}<0.65$ & 1.15 & -0.78 & -- & 9 & 0.95 & 0.025 \\
		& $ \Delta m_{30}(r)$ & -- & 1.37 & -0.13 & -0.13  & 202 & 1.09 & 0.051 \\
	\end{tabular}}	
	\begin{tablenotes}
		\item [a] Number of objects used to determine the fit
		\item [b] Reduced $ \chi^2 $
	\end{tablenotes}
\end{threeparttable}
\end{table*}

The differences between the BTS and the CSP relations are studied in detail in Appendix~\ref{app:k corr} with synthetic photometry of SNe Ia spectra. We find that the mean $ g $-band decline rates are larger when the BTS parameters are used, while they hardly change for the $ r $ band. This is in quantitative agreement with the differences seen in the $g$ band and with the similarity in the $r$ band between the two samples shown in Figure~\ref{fig:BTS sgr vs dm}.

Ideally, we would like to determine \sgr\ solely from the relations derived from the BTS data. However, the range of decline rate values with an observed \sgr\ is not entirely covered. Specifically, there is only one SNe with $ \dmf(r)>1 $, three SNe with $ \dm_{8}(g)>0.7 $ and no SNe with $ \dm_{30}(r)>2 $ that have an observed $ \msgr $, so it is not possible to determine a fit for these regimes from the BTS data alone. We are, therefore, forced to use the CSP fits in these regions\footnote{We use both BTS and CSP SNe for the $ \dm_{8}(g) $ fit at low, $<0.7$, \sgr\ values, and the CSP SNe with $ \dm_{30}(r)>2 $ for the $ \dm_{30}(r) $ fit.}. We expect this procedure to have a minimal effect, since $\Delta m_{8}(g)$ and the $r$-band decline rates are hardly affected by the redshift and the different filter transmission functions (see Figure~\ref{fig:kcorr time}). 

The validity ranges of the fits are determined by requiring the decline rate to provide a good estimate for \sgr\ (these ranges sometimes do not span the entire \sgr\ range over which the fit is determined). The \sgr\ scatter of the CSP fits is lower than $ \sigma_{\msgr}<0.04 $, and the reduced $ \chi^2 $($ \chi_\nu^2 $) values are higher than 2.5 for all fits except one, indicating that the scatter is most probably caused by the intrinsic scatter of the decline rate-color stretch relation. Due to the larger uncertainty in the shape parameters of the BTS sample, the scatter of the BTS fits is higher, and the $ \chi_\nu^2 $ values are lower, $ \sim1\text{--}2 $. This indicates that for the BTS fits the scatter could be either due to the uncertainty of the shape parameters (for $ \chi_\nu^2\approx1) $ or due to an intrinsic scatter (for $ \chi_\nu^2>1 $). 

Using the four decline rate parameters, we were able to determine the \sgr\ for another 1167 SNe from the BTS sample. This was done in the following order. First, 8 SNe with $ \dm_{15}(r)>1 $ were fitted using $ \dm_{15}(r)$. Then, due to its low scatter compared with the other decline rates, $ \dm_{15}(g) $ was used to fit another 853 SNe. 208 SNe with no $ \dmf(g) $ measurement or with high $ \dmf(g) $ values were fitted from $ \dm_8(g) $, and lastly, 98 SNe were fitted from $ \dm_{30}(r) $. 16 SNe have less than five observed epochs in both bands and are not considered in our analysis. An additional 89 SNe have $ \Delta m_{15}(r)<1 $ and no other decline rate measurement, so their \sgr\ value cannot be accurately estimated, but they are ruled out as being very dim, and therefore would not change the estimated LF significantly. For 30 SNe, we were unable to identify a peak magnitude in either band. This leaves 13 SN with an observed peak but no measurable decline rates\footnote{SN 2019fch had a peak $ g $ value from a single photometric measurement that was much higher than the peak $ r $ value and resulted in a very large $ \dm_8(g) $. These values were inconsistent with its $ E(g-r)_\text{h} $ evaluation or its typical $ \dmf(r) $ value, so we discarded this SN.}. These statistics are summarised in Table~\ref{tab:BTS numbers}.

\begin{table}
	
	\caption{BTS color stretch parameter statistics.}
	\hspace*{-0.75cm}\label{tab:BTS numbers}
	\begin{tabular}{lrl}
	Criteria & Number of SNe & Comments \\\midrule
	With \sgr\ & 1371 & \\
	\MyIndent \sgr\ directly & 204 & \\
	\MyIndent \sgr\ from $\Delta m$ & 1167 & \\
	\MyIndent \MyIndent \sgr\ from $\Delta m_{15}(r)$ & 8 & Only for $\Delta m_{15}(r)>1$, \\
	& & With CSP fit\\
	\MyIndent \MyIndent \sgr\ from $\Delta m_{15}(g)$ & 853 & Only for $\Delta m_{15}(g)<1.55$ \\
	\MyIndent \MyIndent \sgr\ from $\Delta m_{8}(g)$ & 208 & SNe with $\Delta m_{8}(g)>0.54$ \\
	& & are fit with combined fit\\
	\MyIndent \MyIndent \sgr\ from $\Delta m_{30}(r)$ & 98 & \\
	No \sgr, $\Delta m_{15}(r)<1$ & 89 & \\
	No recognized peaks & 30 & \\
	Less than 5 epochs in both bands & 16 & \\
	Undetermined & 13 & \\
	Total number of SNe & 1519 & \\
	\end{tabular}	
\end{table}

In the next section, we estimate the distances and extinction of the sample, which requires the observed peak magnitudes of both bands. Of the 1371 SNe with an estimated $ \msgr $, 1289 have observed peaks in both the $ g $ and $ r $ bands or a spectroscopic redshift. Out of the 82 that do not, one is very dim, with $ \msgr<0.4 $, which accounts for 25 per cent of the SNe in this \sgr\ regime. The observed peak statistics of the BTS sample are summarised in Table \ref{tab:BTS bands}.

\begin{table}
	\caption{BTS observed peak statistics.}
	\label{tab:BTS bands}
	\begin{tabular}{lr}
		Criteria & Number of SNe \\\midrule
		All SNe with \sgr\ & 1371  \\
		With peak $ r $ & 1319  \\
		With peak $ g $ & 1299  \\
		With peak $ g $ and $ r $ & 1247  \\
		With peak $ g $ and $ r $ or spec. distance & 1289  \\
	\end{tabular}	
\end{table}

A table with all the data used to construct the LF from the BTS SNe is given in Appendix~\ref{app:CSP}. The table includes the \sgr\ values, method used for their estimation (directly or the relevant decline rate parameter), decline rate parameters, estimated peak magnitudes, and $\mni$ (see Section \ref{sec:Ni}).

\section{Completeness and volume-limited sample}
\label{sec:completeness}

In this section, we investigate the completeness of the BTS sample. For the LF to represent the true distribution, the sample in each luminosity bin should be either complete or volume corrected. In order to analyse the sample's completeness, we consider a narrow range of peak luminosities, where the survey can be considered as volume limited, with the observed volume determined by the chosen luminosity. We choose to use the $ r $ band, because of the higher number of SNe with a peak magnitude in this band, and because of the more uniform $ K $-correction in this band, as compared to the $ g $ band. The cumulative comoving distance distribution of each luminosity bin should behave with distance like $ \propto V\sim D^3 $, assuming the SNe are uniformly distributed in the considered volume. This behaviour is expected to hold until $ D=D_\text{lim}(s_{gr}) $, the distance where the observed magnitude reaches the limiting magnitude of the survey. In reality, the behaviour is more complicated because of, e.g., extinction and incomplete classification. We analyse the completeness of the BTS survey using the spectroscopic redshifts, when available, and our method for determining the distances for the SNe that lack spectroscopic redshifts, as described below. We first use the CSP peak magnitude$ -\msgr$ relation (applying a $ K $-correction, which is somewhat different from the $ K $-correction of P20, see Appendix~\ref{app:k corr} for details). Then, we can write the differences between absolute and observed magnitudes in each band (for the $1247$ SNe with both $ g $ and $ r $ observed peak magnitudes) as:
\begin{equation}\label{eq:dist find}
\begin{split}
M_g = m_g-\mu-A^\text{gal}_g-A^\text{h}_g\\
M_r = m_r-\mu-A^\text{gal}_r-A^\text{h}_r,
\end{split}
\end{equation}
with the three unknown parameters: $ \mu$, the distance modulus, $A^\text{h}_g$, the host extinction in the $g$ band, and $A^\text{h}_r $, the host extinction in the $r$ band. In order to solve for the three unknowns, we assume a uniform ratio of the total-to-selective extinction $ R_r=A^\text{h}_r/E(g-r)_\text{h}\approx1.4 $, where $ E(g-r)_\text{h}=A^\text{h}_g-A^\text{h}_r $. This value, which corresponds to $ R_V\approx1.5 $, minimizes the mean error between the estimated distances and the distances to the SNe with spectroscopic distances, to $ \sim7$ per cent. This value of $ R_r $ is also close to the mean value of $ R_r\approx1.6 $ that was obtained in our analysis of the SNe with spectroscopic distances (see Section \ref{sec:extinction}), and is typical for SNe Ia hosts \citep{Burns2018}. Using a higher value of $  R_r=2.8 $, which corresponds to the Milky Way extinction law $ R_V=3.1 $, increases the total rate by $ \sim15$ per cent, but does not have a significant effect on the shape of the LF. SNe with negative values of $ E(g-r)_\text{h} $ were not corrected for host extinction.

\begin{figure}
	\includegraphics[width=\columnwidth]{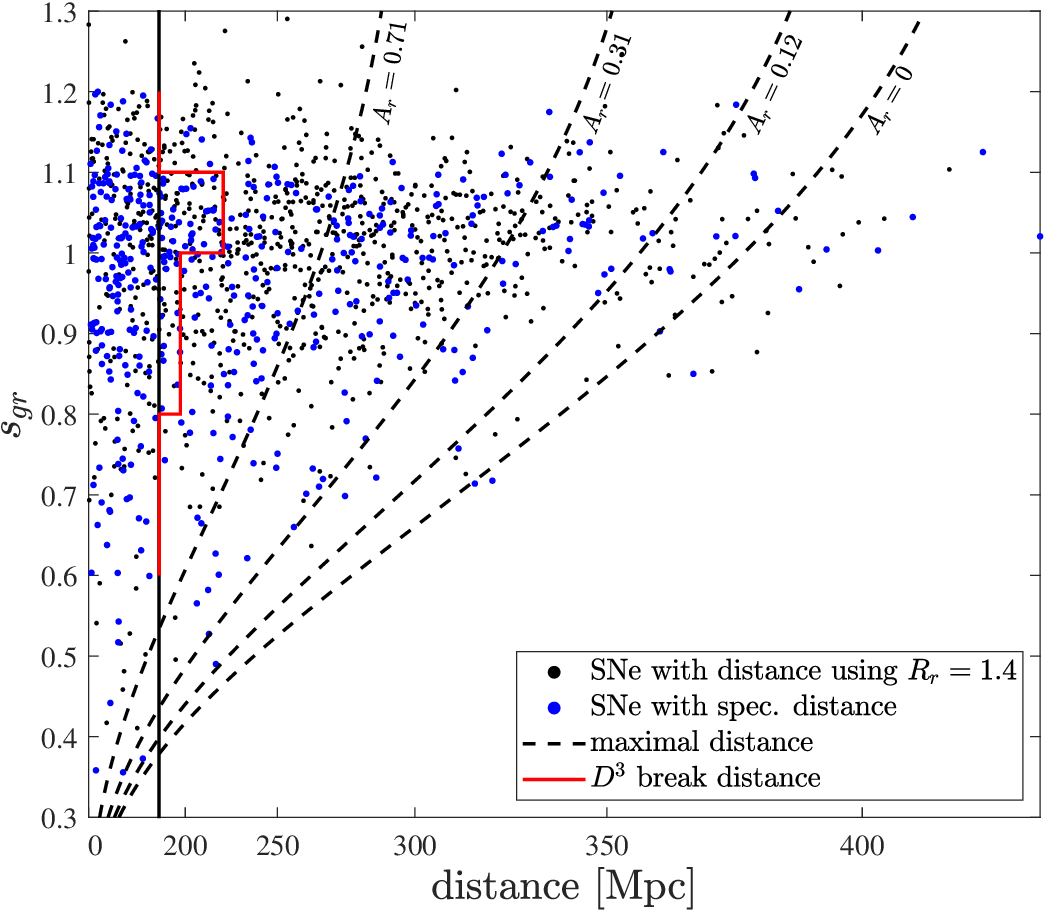}
	\caption{The $ \msgr$-distance distribution of the SNe in the sample, where the $x$-axis is scaled as $ D^3 $. Black circles denote the SNe with distances calculated with the methods described in the text, while SNe with spectroscopic distances are marked by blue circles. Dashed curves show the maximal visible distance for several extinction values and a $ m=18.5\,\text{mag} $ limiting magnitude. The extinction values represent the typical extinction of several \sgr\ bins (see Section~\ref{sec:extinction} and Appendix~\ref{app:volume corr}). The red line indicates, for each \sgr\ bin, the distance from where the $ D^3 $ law breaks, according to Figure~\ref{fig:completeness2}. The black vertical line at $ 180\,\textrm{Mpc} $ marks the distance of our volume-limited LF.}
	\label{fig:sgr dist}
\end{figure}

Figure \ref{fig:sgr dist} shows the $ \msgr$-distance distribution of the SNe in our sample, where the $x$-axis is scaled as $ D^3 $. SNe with distances calculated using the methods described above are shown as black circles, and SNe with spectroscopic distances are shown as blue circles. Dashed curves show the maximal visible distance for several values of extinction and a $ m=18.5\,\text{mag} $ limiting magnitude. These values are chosen to represent three \sgr\ regimes (see Section~\ref{sec:extinction} and Appendix~\ref{app:volume corr}). The red line shows $ D_\text{break} $(see below) for each \sgr\ bin. The black vertical line at $ 180\,\text{Mpc} $ marks the distance of our volume-limited LF, where most of the sample is approximately complete. As can be seen in the figure, the SNe are more concentrated at low distances, and their density becomes lower before the distance reaches the zero host-extinction limiting distance.

\begin{figure}
	\includegraphics[width=\columnwidth]{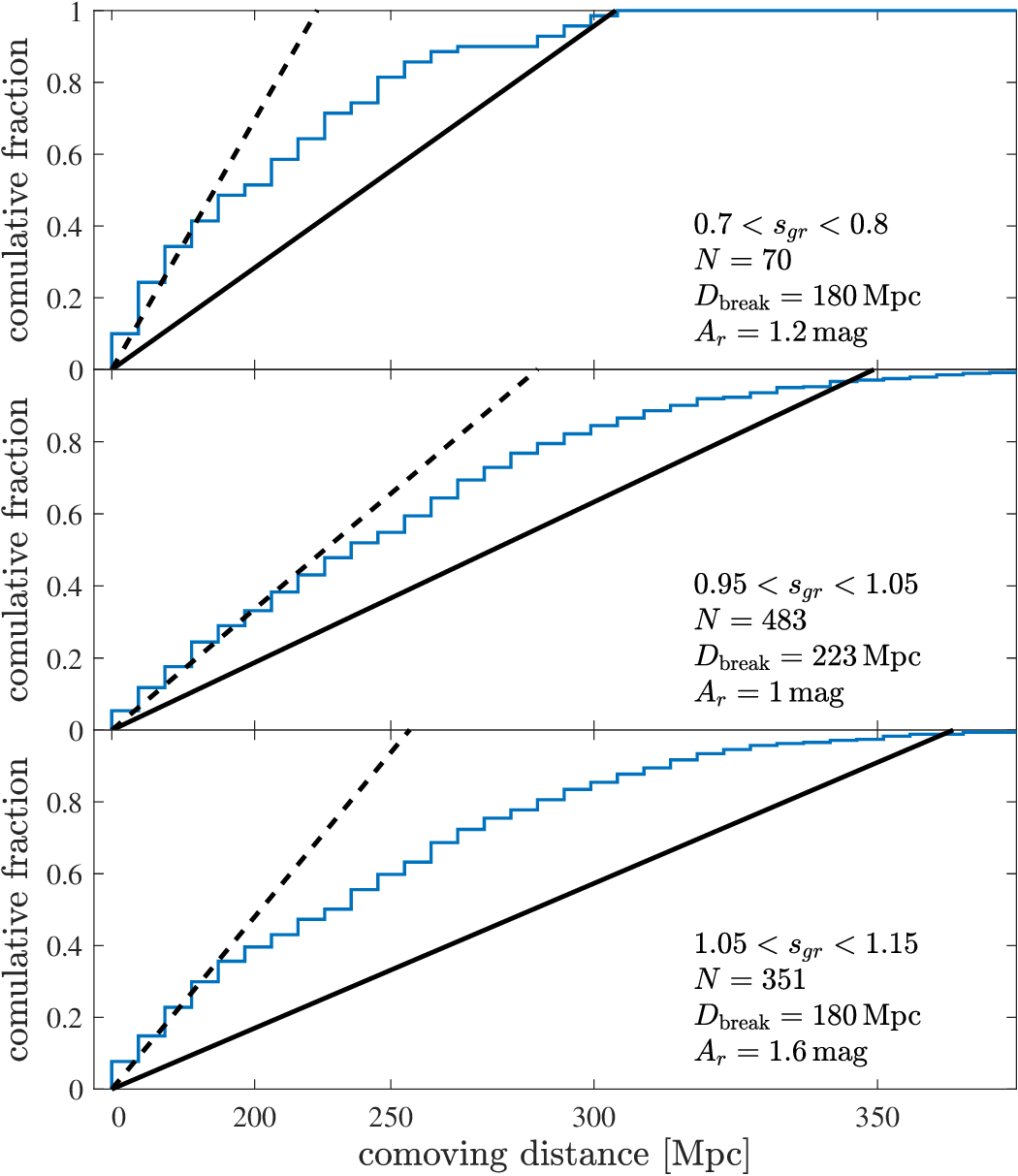}
	\caption{Cumulative fraction of SNe as a function of the comoving distance for different representative \sgr\ bins, where the $x$-axis is scaled as $ D^3 $, so that the observed volume increases linearly. For each \sgr\ bin, we present the plot of the $ (D/D_\text{lim})^3 $ curve (black, solid line) and $ (D/D_\text{break})^3 $ (black, dashed line). The number of SNe in the bin, $N$, the break distance, $ D_\text{break} $, and the corresponding extinction according to Equation \eqref{eq:Abreak} $, A_r^\text{break} $, are indicated in each panel.}
	\label{fig:completeness2}
\end{figure}

We next plot the comoving distance distributions for several \sgr\ bins in Figure~\ref{fig:completeness2}, where the $x$-axis is scaled as $ D^3 $. For each bin, a curve of $ (D/D_\text{lim})^3 $, assuming zero extinction, is also plotted in a black solid line. As can be seen in the figure, all distributions deviate from the $ D^3 $ law at some distance $ D_\text{break}<D_\text{lim} $. This is due to Milky Way and host galaxy extinction that increase the magnitude of events that are within the observed volume to above the limiting magnitude, and possibly due to incomplete classification of dim events. In order to estimate $ D_\text{break} $, we find the highest distance where the maximal deviation of a linear fit to the cumulative fraction as a function of the volume is larger than $0.03$. These fits are shown by black, dashed lines, and the break distance, $ D_\text{break} $, is displayed in each panel. Also shown is the extinction required to decrease the limiting distance to the break distance, given by:
\begin{equation}\label{eq:Abreak}
A_r^\text{break}=10^{\frac{D_\text{lim}-D_\text{break}}{5}}.
\end{equation}
This quantity roughly marks the extinction that affects a non-negligible fraction of the SNe for each \sgr\ bin. As can be seen in Figure \ref{fig:completeness2}, SNe with large \sgr\ values suffer from large extinction values of $ A_r^\text{break}\approx1.5\,\text{mag} $ (the implication of this result is further discussed in Section~\ref{sec:discussion}), while the distribution of SNe with low and medium values of \sgr\ are affected only at lower values, $ A_r\lesssim 1.1\,\text{mag} $. This is in agreement with the host extinction distributions obtained in Section~\ref{sec:extinction} and presented in Figure~\ref{fig:BTS host ext sgr}. However, the break distance for low-luminosity SNe, with $\msgr\lesssim0.8$, is lower than expected from the typical extinction values of these SNe, and possibly related to incomplete classification.

In Appendix~\ref{app:volume corr}, we study in detail different choices to limit the sample (either in volume or in magnitude) in order to reach a balance between statistic and systematic errors. We find an optimum with a $ 180\,\text{Mpc} $ volume-limited subsample (indicated in Figure~\ref{fig:sgr dist}) with 298 SNe in the considered time range (which is not the entire survey duration, see Appendix \ref{app:volume corr} for details regarding the LF time range), which require a non-negligible volume correction only for the dimmest bin. We verify in Appendix~\ref{app:volume corr} that different choices to limit the sample provide consistent results, although with larger errors (statistical or systematic). In the next section, we construct the LF from the $180\,\text{Mpc} $ volume-limited subsample.


\section{Luminosity function constructed from the colour stretch parameter}
\label{sec:LF}

After obtaining the color stretch parameter \sgr\ for the vast majority of the BTS sample and choosing a subsample that is nearly complete, we can construct the LF by differentiating Equation~\eqref{eq:rate} with respect to the chosen parameter (e.g., $ \msgr $, and $M_r$). The calculation and comparison of LFs from different subsamples are described in Appendix~\ref{app:volume corr}.

The \sgr\ distribution is shown in Figure~\ref{fig:LF Ni}. As can be seen in the figure, the distribution peaks at $ \msgr\approx1 $ and drops monotonically on both sides, up to $ \msgr\approx1.3 $ at the high end and down to $ \msgr\approx0.35 $ at the low end. However, this is probably not the lower limit, as SN 2006mr from the CSP sample has a color stretch value of $ \msgr\approx0.3 $. The upper limit marks the highest rate to detect no events at 95 per cent confidence for an \sgr\ bin size of 0.1. We show in Section~\ref{sec:unimodal} that the \sgr\ distribution is consistent with a unimodal, featureless distribution.

The intrinsic LF of the peak $ r $ magnitude is shown in Figure~\ref{fig:LF peak r}. The volume-corrected (-uncorrected) rates are indicated by blue (red) symbols. Only the dimmest bin requires a non-negligible volume correction. The pseudo-LF of P20, where the BTS absolute magnitudes are calculated using the estimated redshift, is presented as green symbols. The LFs are scaled so that the total rate matches the total rate obtained in this work, which is calculated by integrating the LF. Also plotted is the pseudo-LF of the LOSS survey \citep{Li2011}, scaled to match the total rate and shown with respect to the $ R $ filter used in the LOSS survey. The $x$-axis is shifted so that the LOSS peak's location matches the other LFs. The upper limit is the same as in Figure~\ref{fig:LF Ni}. Despite the similar observation lengths (although the observation dates are different, see Appendix~\ref{app:volume corr}), the upper limits of the low-luminosity SNe in this work are higher than in P20 because we included a host extinction completeness factor (see Appendix~\ref{app:volume corr}) that reduces the observed volume, and the bin size used in our analysis is smaller by $ \approx35 $ per cent, increasing the chance that an SN in an empty bin will be missed. The $ g $-band peak magnitude LF is similar to the $ r $ band, and is not shown here.

As can be seen in Figure~\ref{fig:LF peak r}, the LF of this work peaks at $ M_r\approx-19.2 $, in agreement with P20. However, the rate of more luminous and dimmer events drops much more rapidly than in P20, where the rate of peak magnitudes dimmer than $ M_r\gtrsim-19\,\text{mag}$ is higher by almost an order of magnitude than in this work. In addition, the pseudo-LF of P20 reaches low-luminosity values of $ M_r\approx-16.7\,\text{mag} $, while our dimmest events have peak magnitudes of $ M_r\approx-17.5\,\text{mag} $. As explained in Section~\ref{sec:II}, this is the result of distance uncertainties and the host extinction being neglected in P20, which smears the pseudo-LF in comparison to our derived LF, where distance estimates are more accurate and host extinction is taken into account. We obtained a total SN Ia rate of $2.91^{+0.58}_{-0.45}\times10^4\,\text{Gpc}^{-3}\,\text{yr}^{-1} $, higher than the result of P20, $\approx 2.35\times10^4\,\text{Gpc}^{-3}\,\text{yr}^{-1} $. The error indicates the statistical uncertainty due to $ 1\text{--}\sigma $ Poisson noise. The total rate drops to $2.56^{+0.58}_{-0.46} \times10^4\,\text{Gpc}^{-3}\,\text{yr}^{-1} $ when the observation dates are the same as in P20, within $ \approx10$ per cent of their total rate. This is a good consistency check of our procedure, as the total rates derived from the intrinsic LF and the pseudo-LF should be similar [neglecting the host extinction in the pseudo-LF leads to larger volume corrections, which are compensated by the larger maximal observed volume in Equation~\eqref{eq:max vol}, calculated assuming zero extinction].

Though the range of magnitudes is not as wide, the LOSS pseudo-LF is quite similar to P20 (but with larger error-bars, as the LOSS sample size is smaller), since the host extinction is neglected also when calculating the LOSS pseudo-LF. We study the LOSS sample in more detail in Appendix~\ref{app:LOSS}.


\section{The \nickel mass distribution}
\label{sec:Ni}

In this section, we transform the distributions calculated in Section~\ref{sec:LF} to a $\mni$ distribution, which is more useful for constraining models. This is done by using the CSP sample to calibrate a relation between \sgr\ and $\mni$. We use the sample of 20 SNe from \cite{Sharon2020} with derived $\mni$, supplemented with an additional 5 SNe, where $\mni$ is calculated using the methods of \cite{Sharon2020}. The bolometric luminosity, photometry, and processed photometry for the additional SNe is given in the supplementary material.

The obtained relation between \sgr\ and $\mni$ is shown in the middle panel of Figure~\ref{fig:LF Ni}. As can be seen in the figure, \sgr\ and $\mni$ are tightly correlated, and the best fit of the form:
\begin{equation}\label{eq:Ni sgr}
\mni/\msol  = p_0+p_1\msgr+p_2\msgr^2,
\end{equation}
has a very small scatter of $\sim0.03\,M_{\odot}$. The fit is plotted with a solid line, and its parameters are given in Table~\ref{tab:fits sgr ni}.

\begin{table}
	\hskip-0.4cm
		\caption{Properties of the $\mni$ fits with respect to \sgr\ and \sbv for the CSP SNe [Equations~\eqref{eq:Ni sgr} and~\eqref{eq:Ni sbv}].}
	\label{tab:fits sgr ni}
	\begin{threeparttable}
		\renewcommand{\TPTminimum}{\linewidth}
		\makebox[\linewidth]{%
	\begin{tabular}{lccccccc}
		Parameter & Fit variable & $ p_0 $ & $ p_1 $ & $ p_2 $ & $ N $\tnote{a} & $ \chi_\nu^2\tnote{b} $ & Scatter \\ \midrule 
		$ \mni/M_\odot $ & \sgr\ & -0.11 & 0.37 & 0.34 & 16 & 0.31 & 0.03 \\ 
		$ \mni/M_\odot $ & \sbv & -0.19 & 0.62 & 0.18 & 20 & 0.19 & 0.03 \\  
	\end{tabular}	}	
	\begin{tablenotes}
	\item [a] Number of objects used to determine the fit
	\item [b] reduced $ \chi^2 $
	\end{tablenotes}
\end{threeparttable}

\end{table}

Using Equation~\eqref{eq:Ni sgr}, we evaluate $\mni$ for each SN in the BTS sample with an estimated \sgr\ value. Then, using the same procedure as in the previous section, we calculate the $\mni$ distribution of the BTS sample. The result is shown in blue symbols in the bottom panel of Figure~\ref{fig:LF Ni}. As can be seen in the figure, the rate peaks at $ \mni \approx0.6\,M_\odot $, and there are no other distinguishable peaks. The rate of dim ($ \mni\lesssim0.3\,M_\odot $) and luminous ($ \mni\gtrsim0.8\,M_\odot $) events is significantly lower than the rate of typical $ \mni $ events. We show in Section~\ref{sec:unimodal} that the $\mni$ distribution is consistent with a unimodal, featureless, distribution. 

\cite{Piro2014} constructed the $\mni$ distribution (red symbols in the figure), using the volume-limited LOSS survey \citep{Leaman2011,Li2011}. $\mni$ was estimated from the $ B $ band decline rate, $ \dmf(B) $, using the equation:
\begin{equation}\label{eq:piro ni-dm15}
\mni/\msol = 1.34-0.67\dmf(B).
\end{equation}
We perform our own analysis on the LOSS survey, and obtain the $\mni$ through the color stretch parameter \sbv (green symbols in the figure). A detailed description is provided in Appendix~\ref{app:LOSS}. To allow comparison between the two LOSS distributions, in the figure, the total rates of both LOSS distributions were normalized to match the BTS total rate. All the distributions peak at roughly the same mass, $ \mni\approx0.6\,M_\odot $, and behave quite similarly at $ 0.4\,M_\odot\lesssim\mni\lesssim0.7\,M_\odot $. Differences arise at the low and high $ \mni $ regimes. In the low-luminosity end, the rate of the BTS LF is smaller and drops more or less monotonically with $ \mni $, while the behaviours of the LOSS distributions are more erratic. Additionally, the dip in the \sbv distribution of the LOSS sample (see Appendix~\ref{app:LOSS}) remains after the transformation to $\mni$ ($0.2\,M_\odot\lesssim\mni\lesssim0.4\,M_\odot $), and a secondary peak appears at $ \approx0.2\,M_\odot $. This secondary peak also appears in the distribution of \cite{Piro2014}, which is explained there as a large fraction of 91bg-like events. However, the secondary peak is not seen in the BTS results, and it is therefore most likely a result of the low number statistics of the LOSS survey (consistent with the large error bars of this bin). The small sample size might also explain the erratic behavior of the LOSS distribution throughout the entire dim end and the differences at the luminous ( $ \mni\gtrsim0.7\, $) end. 

A note is in place regarding the sharp peak of the LF (Figure~\ref{fig:LF peak r}) compared with the more smooth distribution of the $\mni$ (the bottom panel of Figure~\ref{fig:LF Ni}). While the LF's sharp peak is somewhat driven by our procedure that uses the peak magnitude$ \,-\,\msgr $ relation, which ignores the small scatter around this relation (see Figure~\ref{fig:peak mag sgr}), the main reason for the difference between the distributions is the nonlinear relation between $\mni$ and the peak magnitude. The vast majority of SNe have $ \mni\sim0.3\text{--}0.8
\,M_\odot $ (see Figure~\ref{fig:LF Ni}), and since the peak flux is roughly proportional to $\mni$, this range of $\mni$ is $ \sim1\, $mag in magnitude space. Additionally, the logarithmic definition of magnitude forces a narrower spacing of high $\mni$ values compared to lower values, which makes the LF appear very narrow, as the $\mni$ distribution is already concentrated at relatively high $\mni$ values.


\section{The derived distributions are consistent with unimodal, featureless, distributions}
\label{sec:unimodal}

The LF and the distributions of this work are based on the shape parameter $\msgr$. Given the non-negligible errors in the evaluation of this parameter, we examine the effects of these errors on the derived distributions and check whether the small features seen in the results are consistent with a unimodal, featureless distribution. We do this by guessing an underlying distribution, and with MC simulations, we calculate the distribution that would be observed after applying random errors to the \sgr\ values. We can then compare the simulated distribution to the observed distribution in order to constrain the parameters of the underlying distribution and the errors that are applied to it. The underlying \sgr\ distribution is given by one rising and one decaying exponential:
\begin{equation}\label{eq:init LF}
R(\msgr)=\begin{cases}
C\times\left(\exp(\msgr/\tau_1)-1\right) & 0\le\msgr\le\mu\\
R(\msgr=\mu)\cdot\exp(-(\msgr-\mu)/\tau_2) & \msgr\ge\mu\\
\end{cases},      
\end{equation}
where $ \tau_1 $ and $ \tau_2 $ are the scale parameters and $ \mu $ is the turnover point. The rate of the rising exponential is modified so that it is zero for $ \msgr=0 $, and the decaying exponential is multiplied by $ R(\msgr=\mu) $ for continuity. Parameter $ C $ is the overall normalization. This model includes five free parameters: two scale parameters $ (\tau_1,\tau_2) $, one turnover points $ (\mu) $, the overall scale and the \sgr\ (Gaussian) error.

\begin{figure}
	\includegraphics[width=\columnwidth]{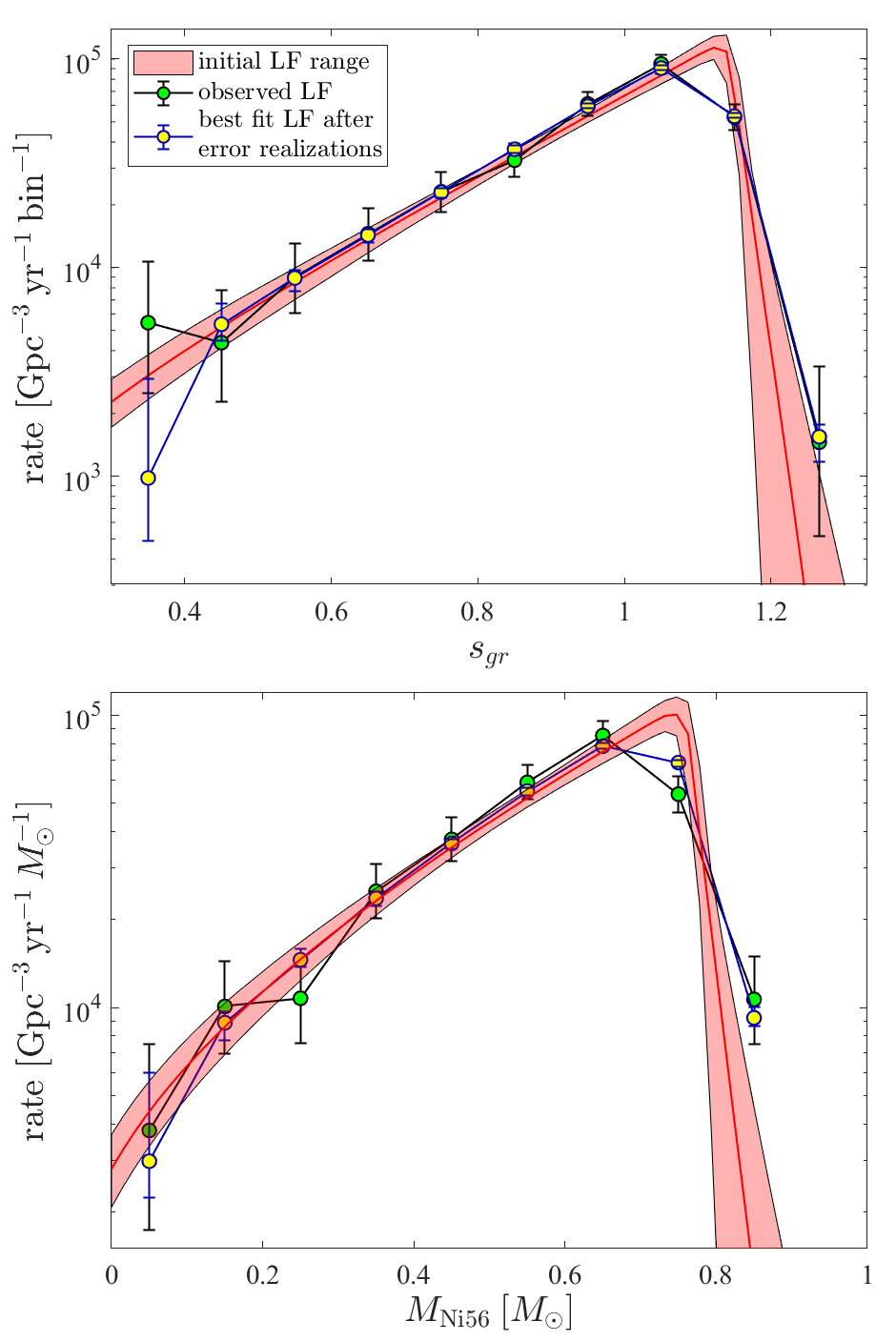}
	\caption{Upper panel: Model \sgr\ distribution and best-fitting after error realizations. Shown are the $68$ per cent confidence levels of the underlying distribution (red shaded area) and its median (red line), together with the observed distribution (green symbols) and the best-fitting distribution after error realizations (yellow symbols). Lower panel: same as the upper panel for the $\mni$ distribution. The parameters of the underlying and best-fitting distributions are the same as those of the $\msgr$ distribution; the distribution was transformed using Equation \eqref{eq:Ni sgr}.}
	\label{fig:LF_realizations}
\end{figure}

A Markov Chain Monte Carlo (MCMC), using the \textsc{mcmcstat} \textsc{matlab} package\footnote{https://mjlaine.github.io/mcmcstat/}, is performed in order to find the distribution of the parameters and their best fit to the observed distribution. The results are shown in the upper panel of Figure~\ref{fig:LF_realizations}. The red shaded area shows the $68$ per cent confidence levels of the underlying distribution, obtained from the MCMC realizations over the set of model parameters, where the red line within marks the median. The best-fitting model distribution lies within the shaded region and is similar to the median distribution, although it is somewhat narrower. The observed distribution is presented by green symbols, and the best-fitting model distribution after applying error realizations by yellow symbols. The error for the best-fitting model distribution is the $68$ per cent confidence bounds of the error realizations using the same underlying best-fitting distribution parameters. As expected, the resulting underlying distribution is somewhat narrower than the observed one, since applying errors widens the distribution. The best-fitting distribution is within the error margin of the observed distribution, and reproduces it quite well. The obtained error in \sgr\ is $ \sim0.028 $, which is lower than the BTS mean estimated \sgr\ uncertainty of $ \sim0.11 $. However, the \sgr\ uncertainties, given by SNooPy through MC simulations, might be overestimated, and the low $ \chi^2_{\nu} $ values of the BTS fits in Table \ref{tab:fits sgr dm} support this claim. The parameters of the underlying distributions are $\mu = 1.14^{+0.02}_{-0.02}$, $ \tau_1=0.22^{+0.03}_{-0.02}$, $ \tau_2 = 0.017^{+0.013}_{-0.014}$, and $C=8.5^{+5.0}_{-3.2}\times10^2\,\text{Gpc}^{-3}\,\text{yr}^{-1} $. 

In the lower panel of Figure~\ref{fig:LF_realizations}, we show the same results for the $\mni$ distribution. We use the same underlying distribution with the parameters that were found for the \sgr\ distribution, transformed using Equation \eqref{eq:Ni sgr}. The best-fitting results also agree with the observed $\mni$ distribution.

The total rate of the underline LF is $ R\approx2.81^{+0.72}_{-0.62}\times10^4\,\text{Gpc}^{-3}\,\text{yr}^{-1} $, consistent with the total rate of the observed LF. We conservatively add to the Poisson noise the error due to the uncertainty of the fitted model parameters. We obtain for the iron production rate $ 1.61^{+0.36}_{-0.32}\times10^4\,M_\odot\,\text{Gpc}^{-3}\,\text{yr}^{-1}  $. 

The results in this section show that the features on top of the unimodal-derived distributions are all consistent with statistical noise on top of featureless distributions. The derived distributions, therefore, are consistent with a single progenitor channel for the explosions.


\section{Galaxy type comparison}
\label{sec:host}

In this section, we calculate the $\mni$ distribution in different types of galaxies for the $ 180\,\text{Mpc} $ subsample. We follow the procedure in P20, where galaxy types are determined according to the absolute $ i $ magnitude and the $ g-i $ color of the host galaxy. The hosts were required to be in the SDSS photometric field, and have redshift values between $ 0.015<z<0.1 $. P20 found that the color of the SNe Ia hosts forms a bimodal distribution, with a "red sequence", dominated by early type galaxies, and a "blue cloud" that consists mostly of spirals and irregulars, separated by a "green valley". The distinction between the different types of galaxies is given in table 2 of P20.

\begin{figure}
	\includegraphics[width=\columnwidth]{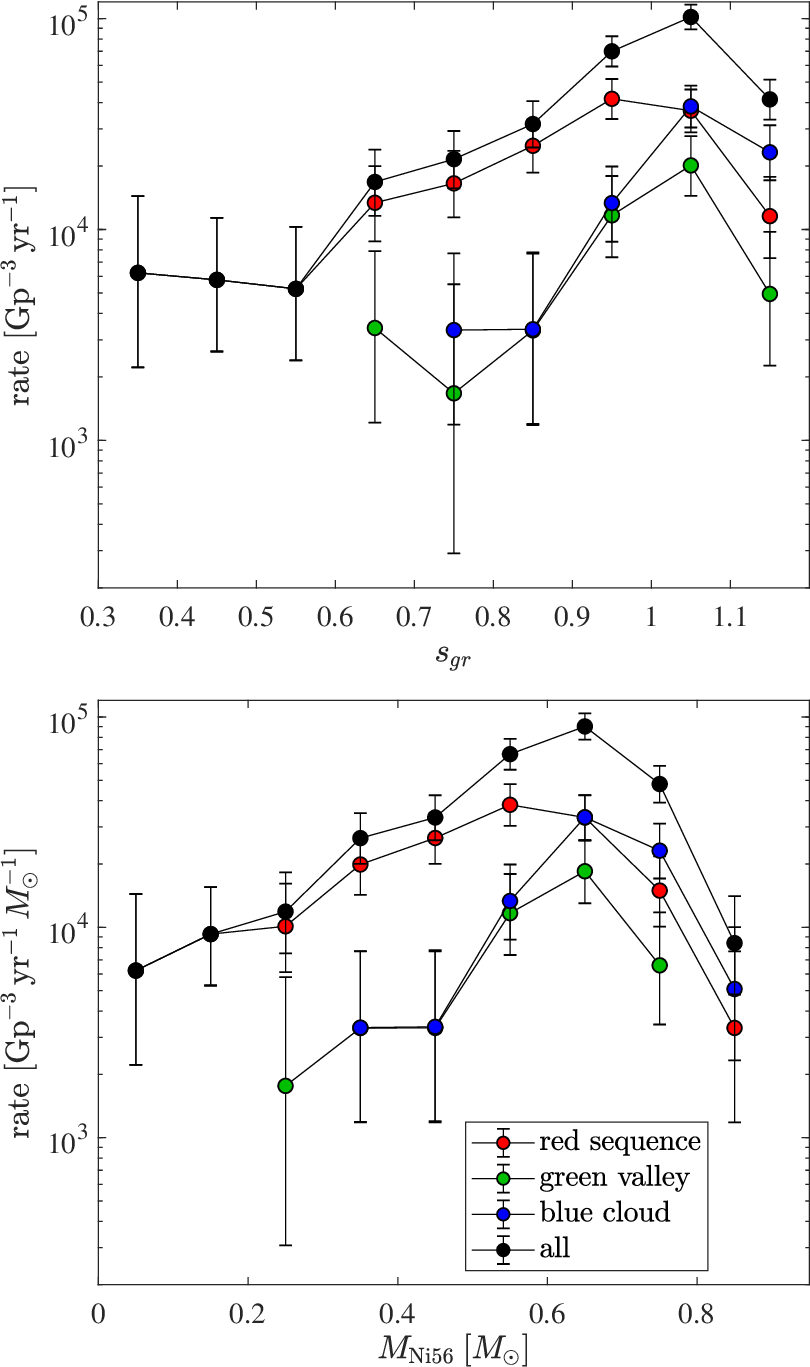}
	\caption{BTS volume-corrected distributions for different galaxy types. Top panel: $ \msgr $. Red sequence, green valley, and blue cloud galaxies are shown as red, green, and blue symbols, respectively. See table 2 of P20 for details about the classification procedure. The total distribution is shown as black symbols. Only SNe in the SDSS photometric field are considered, so the distribution of all the galaxies does not exactly match the distribution in Figure~\ref{fig:LF Ni}. Bottom panel: Same as the top panel for $\mni$. The red sequence, which contains early-type galaxies, is much more skewed towards dim events, while normal and luminous events occur mostly in the blue cloud, which contains spirals and irregulars.}
	\label{fig:host gal types}
\end{figure}

We use the same host types as in P20 (note that the $ 180\,\text{Mpc} $ subsample is well within the redshift upper limit), and we recalculate the LF and the $\mni$ distribution for each type. We apply an additional global correction of $ \approx1.71 $ that takes into account the fraction of SNe that are in the SDSS photometric field. The results are shown in Figure~\ref{fig:host gal types}. Note that the SDSS field requirement introduces a selection bias (possibly distance-dependent), which introduces systematic biases to our results in this section (e.g. the black lines in the figure do not exactly match the distributions in Figure~\ref{fig:LF Ni}). As can be seen in the figure, the relative rates strongly depend on the galaxy type. The red-sequence galaxies are the main hosts of low-luminosity events, with $ \approx82$ per cent of the SNe with $ \mni\lesssim0.4 $ found in these galaxies. Additionally, the fraction of low-luminosity SNe is $ \approx23$ per cent of the whole sample, and $ \approx36 $ per cent of the red sequence. Luminous events of $ \mni\gtrsim0.7 $ take place mostly in the blue cloud spiral and irregular galaxies, which host $ \approx53 $ per cent of these SNe. For the green valley distributions, the fraction of dim (luminous) events is lower (higher) than the red sequence and higher (lower) than the blue cloud. The results are in qualitative agreement with the results of \cite{Ashall2016} and the results of LOSS in \cite{Li2011}. It is also consistent with our analysis of the LOSS sample, although we do not provide the LOSS LF for different types of galaxies, as it contains a small number of SNe in each galaxy type.

\section{Host extinction analysis}
\label{sec:extinction}

\begin{figure}
	\includegraphics[width=\columnwidth]{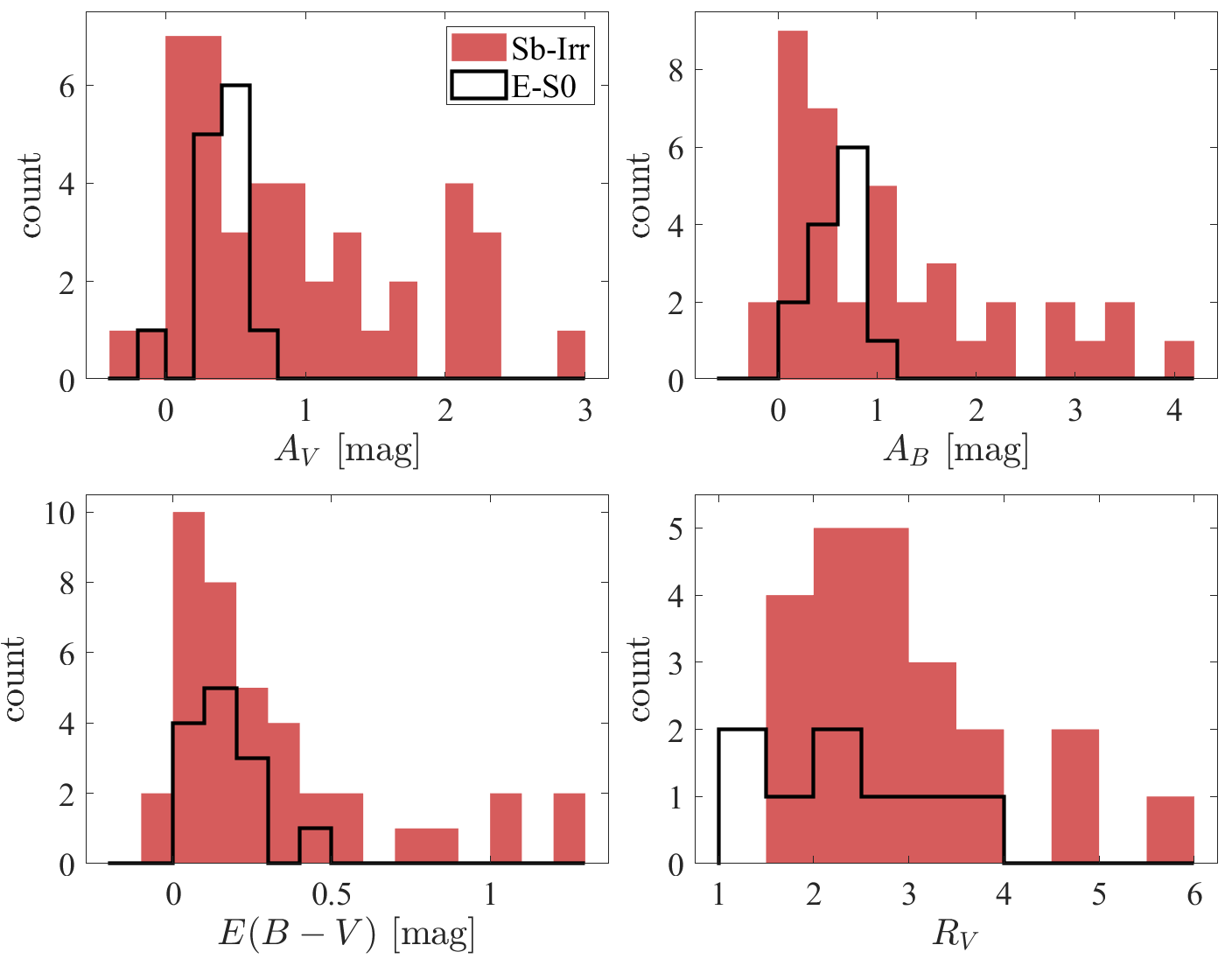}
	\caption{Distributions of the estimated host extinction in the $ V $ and $ B $ bands, the selective host extinction $ E(B-V)_\text{h} $, and the ratio of the total-to-selective extinction $ R_V^\text{h} $ of the LOSS sample. Sb-Irr galaxies are shown in red-filled bars, and E-S0 galaxies in black, empty bars.}
	\label{fig:LOSS host ext}
\end{figure}

In this section, we estimate the distribution of the host galaxy extinction for SNe Ia, using the BTS and the LOSS samples. The LOSS sample includes accurate distances to the SNe, but lacks host galaxy extinction evaluation, as it is not required for the calculation of the pseudo-LF. In Appendix~\ref{app:LOSS}, we present a reanalysis of the LOSS sample, in which we calculate absolute magnitudes from the light curve shape parameters. We can therefore use the known distances and galactic extinction $A^\text{gal}_x$, the observed magnitudes and the estimated absolute magnitudes to estimate the host extinction $ A^\text{h}_x $ in some band $ x $:
\begin{equation}\label{eq:host ext}
A^\text{h}_x=m_x-M_x-\mu-A^\text{gal}_x.
\end{equation}
The results are shown in the top panels of Figure~\ref{fig:LOSS host ext}, where the distributions of the estimated host extinction in the $ V $ and $ B $ bands are shown for two groupings of host galaxy types: spiral and irregular galaxies (red-filled bars) and elliptical and lenticular galaxies (empty black bars). Note that the galaxy types are grouped differently than in \cite{Li2011}. Around $ 10 $ per cent of $ A^\text{h}_V $ and $ \sim6 $ per cent of $ A^\text{h}_B $ have negative extinction values, which is the result of the error in the observed peak value and of the intrinsic scatter of the absolute magnitude fit.

We can further calculate the selective extinction, $ E(B-V)_\text{h}=A^\text{h}_B-A^\text{h}_V $, and the ratio of the total-to-selective extinction, $ R^\text{h}_V = A^\text{h}_V/E(B-V)_\text{h}$. The results are shown in the bottom panels of Figure~\ref{fig:LOSS host ext}. For the $ R^\text{h}_V $ results, we only consider SNe with $ E(B-V)_\text{h}>0.1 $ and positive $ A^\text{h}_V $ and $ A^\text{h}_B $. Smaller values of $ E(B-V)_\text{h} $ result in large, unrealistic values of $R^\text{h}_V $ for several SNe, since the extinction at these values is comparable to its uncertainty. However, there are still a few SNe with $R^\text{h}_V>4 $, which is probably due to these errors.

The figures show that most SNe exhibit some amount of host extinction, and a non-negligible fraction are highly reddened by their host. All the SNe with large extinction values are known to be highly reddened, such as 2003cg \citep{Elias-Rosa2006}, 1999cl \citep{Krisciunas2017} and 2006X \citep{Wang2008}. As expected, extinction is stronger in spiral and irregular galaxies, and its distribution in these galaxies shows a tail extending up to $ \approx 3 $ mag in the $ V $ band. SNe in early type galaxies also suffer from reddening values of a few tenths of a magnitude, but no more than $ 1\,\text{mag} $. The distribution of $ R^\text{h}_V $ is mostly between $ 1 $ and $ 4 $, (with a few outliers that have higher values), but the sample size is rather low and biased towards SNe with large extinction values.

Determining the distribution of the BTS host galaxy extinction is more difficult, as most SNe have no accurate distances, and the ones that do might have some selection bias. None the less, we have calculated the host extinction in the $ g $ and $ r $ bands for 455 SNe with spectroscopic redshifts and measured \sgr\ with the same method as above, including $ K $-corrections. Because of the low number of objects in this analysis, we did not apply a volume limit as in the previous sections; rather, we include all the objects that passed our criteria (with a proper volume correction, see below). The selective extinction, $ E(g-r)_\text{h}=A^\text{h}_g-A^\text{h}_r $, can be calculated from the observed magnitudes and the absolute magnitudes for all SNe (if they have an observed peak magnitude measurement in both bands), since it is independent of distance:
\begin{equation}\label{eq:BTS ext}
E(g-r)_\text{h} = m_g-M_g -(m_r-M_r)-(A^\text{gal}_g-A^\text{gal}_r).
\end{equation}

\begin{figure}
	\includegraphics[width=\columnwidth]{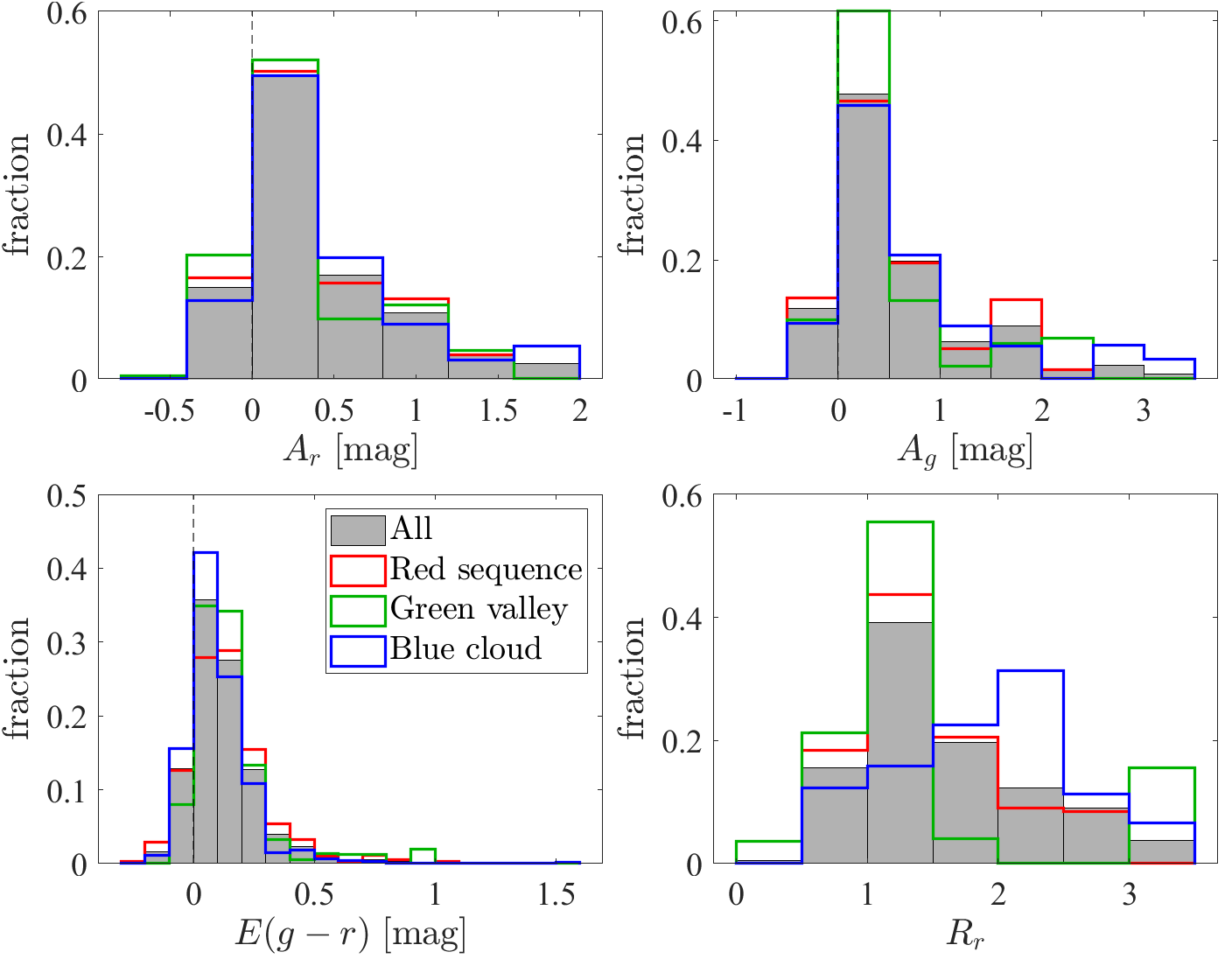}
	\caption{Distributions of the host galaxy extinction of the BTS sample. Upper-left-hand (right-hand) panel: $ A_r^\text{h} $ ($ A_g^\text{h} $) distribution of the BTS SNe with spectroscopic redshifts for different galaxy types. All the distributions are volume-corrected and normalized by the total count of the whole sample. Lower left-hand panel: The selective host extinction $ E(g-r)_\text{h} $ distribution for the whole sample. Lower right-hand panel: The ratio of the total-to-selective extinction $ R^\text{h}_r$ for SNe with $ E(g-r)_\text{h}>0.2\,\text{mag} $.}
	\label{fig:BTS host ext}
\end{figure}

\begin{figure}
	\includegraphics[width=\columnwidth]{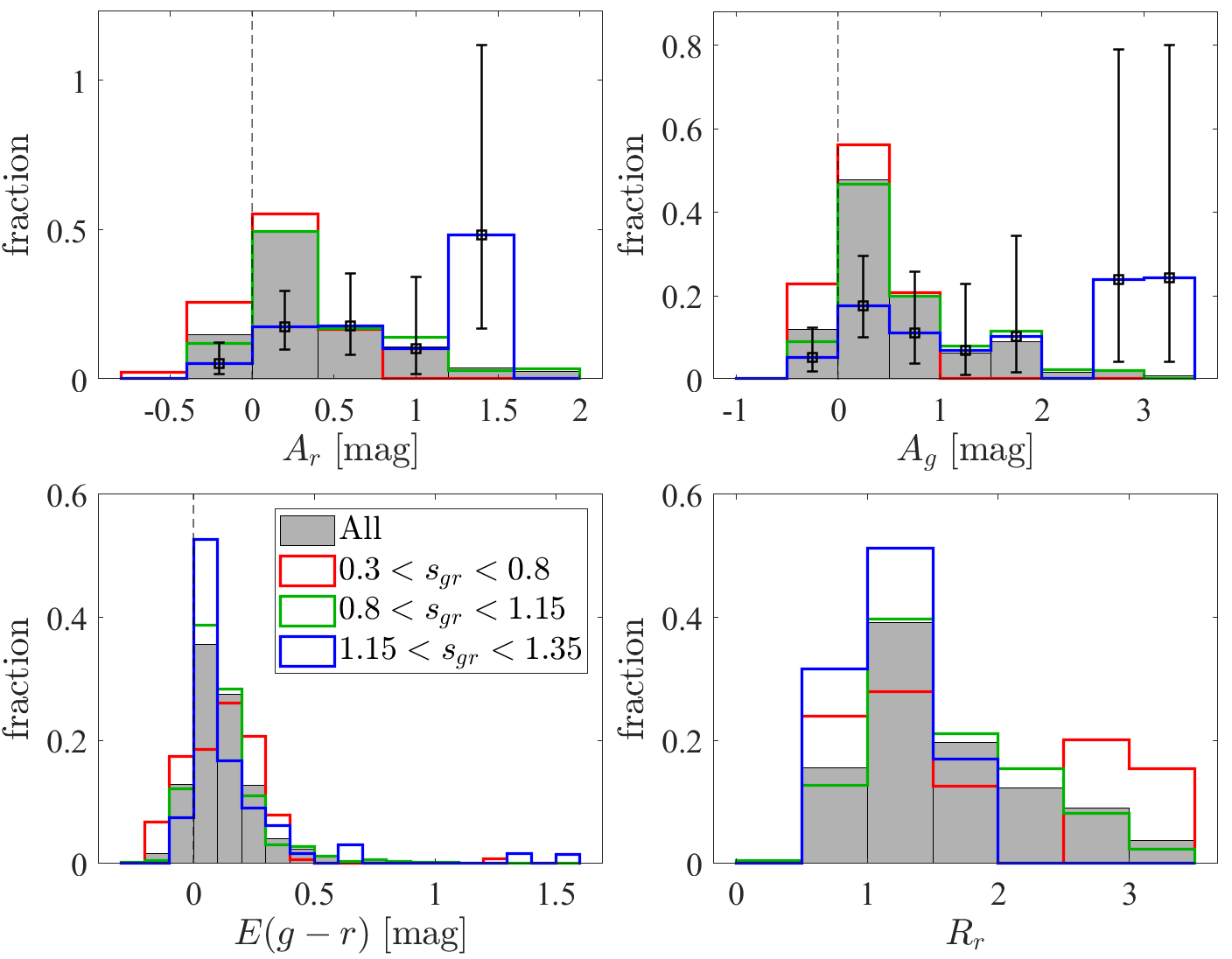}
	\caption{Distributions of the host galaxy extinction of the BTS sample for different \sgr\ ranges. Upper left-hand (right-hand) panel: The $ A_r^\text{h} $ ($ A_g^\text{h} $) distribution of the BTS SNe with spectroscopic redshifts for different \sgr\ bins. All the distributions are volume-corrected, and error bars are shown for the $ \msgr>1.15 $ bin distribution. Lower left-hand panel: The selective host extinction $ E(g-r)_\text{h} $ distribution for the whole sample. Lower right-hand panel: The ratio of the total-to-selective extinction $ R^\text{h}_r$ for SNe with $ E(g-r)_\text{h}>0.2\,\text{mag} $ and with positive extinction values in both bands.}
	\label{fig:BTS host ext sgr}
\end{figure}

Figures \ref{fig:BTS host ext} and \ref{fig:BTS host ext sgr} show the distributions of the host galaxy extinction $ A_r^\text{h} $ (upper left-hand panel), $ A_g^\text{h} $ (upper right-hand panel), the selective extinction of all SNe, $ E(g-r)_\text{h} $ (lower left-hand panel), and the ratio of the total-to-selective extinction, $ R^\text{h}_r = A^\text{h}_r/E(g-r)_\text{h}$. The relations to the Johnson B and V bands are given by $ E(g-r)/E(B-V)\approx0.8-1 $ and  $ R_V/R_r\approx1.13\text{--}1.17 $ for $ R_V $ values in the range of $1\text{--}4$. In Figure \ref{fig:BTS host ext}, the distributions are divided into galaxy types, classified as in Section \ref{sec:host}, while in Figure \ref{fig:BTS host ext sgr}, the distributions are divided into \sgr\ bins. The distributions are volume-corrected according to their luminosities and their host extinction value, modifying Equation \eqref{eq:max vol} to
\begin{equation}
\label{eq:voll_ext}
V_\text{max,i} = \frac{4\pi}{3}\left(10^{\frac{m_\text{lim}-M_i-A^\text{h}_{x,i}}{5}-8}\right)^3\,\text{Gpc}^3.
\end{equation}
This correction substantially increases the weights of high-extinction SNe, and may bias the distribution when the sample size is small. This is the case for the $ A_r^\text{h} $ and $ A_g^\text{h} $ distributions of the $ \msgr>1.15 $ bin, where the sample size is 13 SNe and the extinction values are relatively high. Therefore, these distributions have been supplemented with error bars that illustrated the high uncertainty. The uncertainties of the other distributions uncertainties are much smaller, and are not shown for visualization purposes. The $ R_r $ distribution only contains SNe with $ E(g-r)_\text{h}>0.2\,\text{mag} $ and $ A_g^\text{h},A_r^\text{h}>0.1 $.

The BTS $ A_r^\text{h} $ and $ A_g^\text{h} $ distributions resemble the LOSS host extinction distribution, with an extinction value of a few $ 0.1\,\text{mag} $ for most SNe and a tail of high extinction values.
The typical uncertainties in the calculation of the extinction are larger for BTS than for LOSS because of larger errors in the observed peak magnitudes and color stretch values. The mean error in the observed $ r $ peak magnitude is $ \approx0.05\,\text{mag}$, and the mean error in the absolute peak magnitude due to the uncertainty in the color stretch is $ \approx0.13\,\text{mag} $, so their root sum squared is $ \approx0.14\,\text{mag} $. Variations in the $ K $-corrections and the filter transmission function might also contribute to the error. Therefore, the fraction of SNe with a negative extinction is higher, reaching $ \approx13$ per cent and $ \approx10$ per cent in the $ r $ and $ g $ bands, respectively, while for $ E(g-r)_\text{h} $ the fraction is $ \approx17$ per cent.

In order to quantify the host extinction distribution and the observational errors, we fitted a three-parameter model of a gamma distribution that represents the intrinsic extinction distribution, convolved with a Gaussian that represents the extinction uncertainty:
\begin{equation}\label{eq:ext model}
f(A_r^\text{h}) = \Gamma(k,\theta)*\mathcal{N}(0,\sigma^2),
\end{equation}
where $ \Gamma(k,\theta) $ is the gamma distribution with a shape parameter $ k $ and a scale parameter $ \theta $, and $\mathcal{N}(0,\sigma^2)$ is a Gaussian with a zero mean and a standard deviation $ \sigma $. 
The results of $ k $ and $ \theta $ are highly affected by the binning method, but the results of the Gaussian standard deviation and the distribution mean are much more stable. The Gaussian standard deviation is $ \sigma \approx 0.11\text{--}0.14\,\text{mag}$, and is consistent with the uncertainties in the observed and absolute peak magnitudes. The fitted distribution expectation value is $ E[A_r^\text{h}]\approx0.49\text{--}0.54\,\text{mag} $, and is consistent with the mean of the observed distribution of $ \approx0.46\,\text{mag} $. Fitting the same model for $E(g-r)_\text{h}$ results in a standard deviation of $ \sigma \approx 0.065$ and an expectation value at $E[E(g-r)_\text{h}]\approx0.11\,\text{mag} $, while the mean of the observed distribution is also $ 0.11\,\text{mag} $.

The $ A_r^\text{h} $ and $ A_g^\text{h} $ distributions show that highly reddened SNe are more common in the blue cloud than in the red sequence. However, we do not find considerable differences between the galaxy types in the $ E(g-r)_\text{h}$ plot. This might be the result of the large errors in the extinction, or the insufficiency of the galaxy type classification, which is based on the host color and absolute magnitude, for this analysis. From the $ R^\text{h}_r $ distribution, it can be gleaned that blue-cloud galaxies favour large $ R^\text{h}_r $ values, in comparison to the red sequence. The differences in Figure \ref{fig:BTS host ext sgr} are more pronounced, and there are no low \sgr\ SNe with an $ A_r^\text{h} $ or $ A_g^\text{h} $ larger than 1 mag. However, this may be the result of an observational bias, since these SNe are intrinsically dim and would be hard to detect if they are highly reddened as well.

The large \sgr\ SNe are characterized by exceptionally high extinction values. These results are consistent with the relatively small distances at which these SNe are complete (see Section~\ref{sec:completeness}), which is smaller than the distances of dimmer SNe. The small sample size of large \sgr\ SNe does not allow us to accurately measure the shape of the $ A_r^\text{h} $ and $ A_g^\text{h} $ distributions at high extinctions. This situation could be improved with future observations (see discussion in Section~\ref{sec:discussion}).


\section{Summary and Discussion}
\label{sec:discussion}

In this work, we used the large ZTF BTS data set to construct the intrinsic LF of SNe Ia. In general, constructing a LF requires knowledge of the redshift and host extinction for each SN, which is a difficult task to perform in a survey with a very high discovery rate such as the BTS. This challenge can be partly bypassed for SNe Ia, by using the tight correlations between the light curve color stretch, $\msgr$, and the intrinsic luminosity, calibrated from the CSP Ia sample (see Section~\ref{sec:II} and Figure~\ref{fig:peak mag sgr}). We determined the \sgr\ values of the vast majority of the SNe Ia in the BTS survey (Section~\ref{sec:CSP}) and we chose a $180\,\text{Mpc} $ volume-limited subsample that is nearly complete (with 298 SNe, Section~\ref{sec:completeness}). We then constructed, for the first time, the intrinsic SNe Ia LF (Section~\ref{sec:LF} and Figure~\ref{fig:LF peak r}). In Section~\ref{sec:Ni}, we used the CSP data to calibrate a novel tight relation between the \sgr\ and $\mni$ (the middle panel of Figure~\ref{fig:LF Ni}), which served to construct the $\mni$ distribution of SNe Ia (the bottom panel of Figure~\ref{fig:LF Ni}). We also constructed the SN Ia distribution for different galaxy types (Section~\ref{sec:host} and Figure \ref{fig:host gal types}), and, for the first time, the host galaxy extinction distributions (Section~\ref{sec:extinction}).

We find that the $\mni$ distribution peak at $ \approx0.6\,M_\odot $, and that the LF peaks are at an $ r $ magnitude of $ M_r\approx-19.2 $, in agreement with previous results. The rate of dim and luminous events, with $ \mni\lesssim0.4\,M_\odot $ or $ \mni\gtrsim0.8\,M_\odot  $, is lower by at least a factor of 5 than the rate at the peak. We find a total rate of $ R\approx2.91^{+0.58}_{-0.45}\times10^4,\text{Gpc}^{-3}\,\text{yr}^{-1} $ (per comoving element in the redshift range $ z\approx [0.01,0.04] $), consistent with previous studies \citep{Dilday2010,Graur2011,Li2011b,Frohmaier2019,Perley2020}. This result has an additional non-negligible systematic error due to uncertainties in the correction factors (e.g., host extinction and the $ K $-correction), and is probably a few tens of a percent. A detailed estimate of the survey's efficiency \citep[see e.g.][]{Frohmaier2019} is required in order to establish the systematic uncertainty of the total rate. 

The dimmest end of the LF in this work is at $ M_r\approx-17.5\,\text{mag} $. While dimmer SNe Ia exist, such as SN 2007mr \citep{Burns2018}, the non-detection of such events within our sample is an indication of their rarity. The ZTF Census of the Local Universe \citep[CLU;][]{De2020} reported additional low-luminosity SNe Ia, SN 2019gau and SN 2019ofm. These objects have peak magnitudes of $ -16.75 $ and $ -16.84 \,$mag, respectively. However, the quality of their light curves does not allow one to determine their intrinsic luminosity from the light curve shape and it is unclear whether they are intrinsically dim or suffer from a large host extinction.

We performed MC simulations to show that the features on top of the unimodal-derived distributions are all consistent with statistical noise (Section~\ref{sec:unimodal}). The derived distributions, therefore, are consistent with a single progenitor channel for the explosions.

We show that the red sequence galaxies are the main hosts of low-luminosity events, with $ \approx82$ per cent  of the SNe with $ \mni\lesssim0.4 $  found to explode in these galaxies. Luminous events of $ \mni\gtrsim0.7 $ take place mostly in the blue-cloud spiral and irregular galaxies, which host $ \approx53 $ per cent of these SNe. These results are in a qualitative agreement with the results of \cite{Ashall2016} and the results of LOSS  \citep{Li2011}. From the host galaxy extinction distributions, we find mean values of $ A_r^\text{h}\sim0.5\,\text{mag} $ and $ E(g-r) \sim0.11\,\text{mag} $, and that a non-negligible fraction of SNe, especially in star-forming galaxies, exhibit large extinction values, $\gtrsim1\,\text{mag} $, in the optical bands. We further find that a host galaxy's extinction highly depends on the luminosity of the SNe. A non-negligible fraction of the luminous SNe with $\msgr\gtrsim1.15$ suffer from $ A_r^\text{h} \gtrsim1\,\rm{mag}$, with no low \sgr\ SNe exhibiting an $ A_r^\text{h} $ or $ A_g^\text{h} $ larger than 1 mag. These results are in line with the idea that luminous SNe Ia originate from young populations, while low-luminosity SNe Ia require old progenitors. 

\begin{figure}
	\includegraphics[width=\columnwidth]{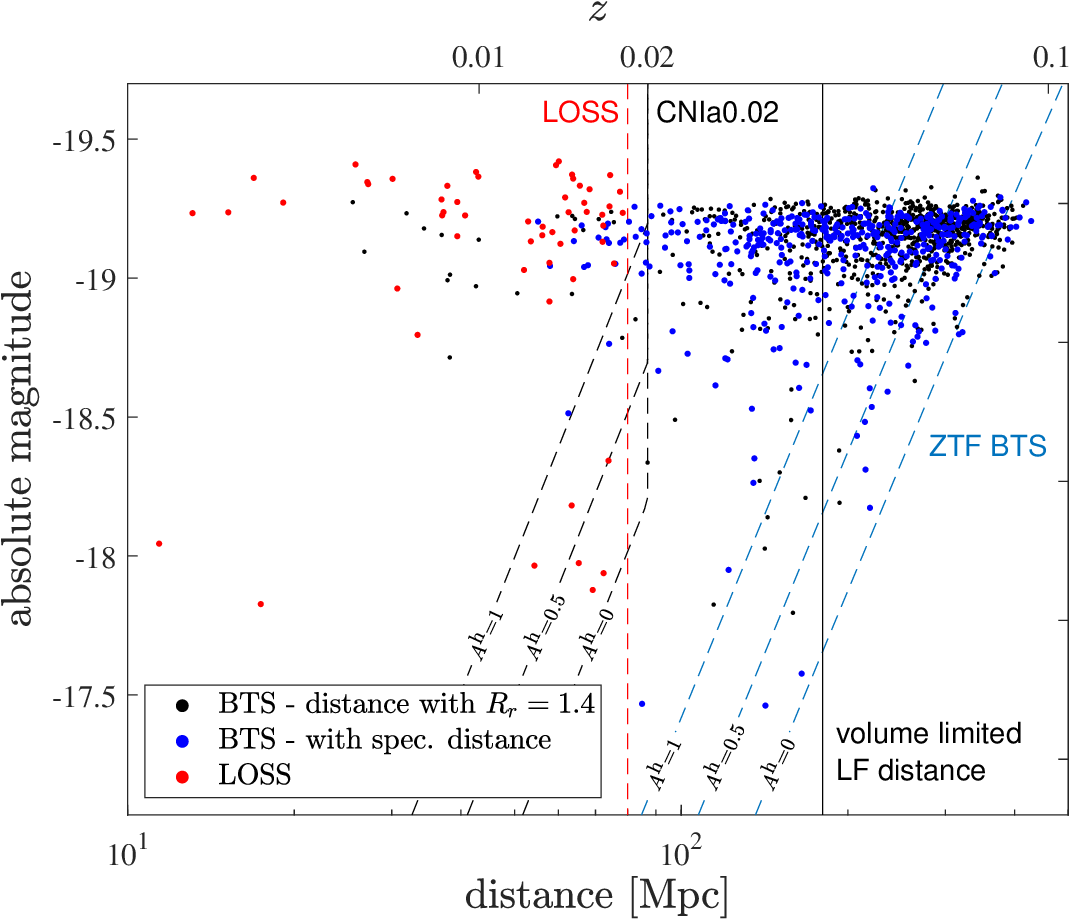}
	\caption{The distance--magnitude observation limits and SNe Ia distributions of several SNe surveys. The BTS observation limit is shown as blue, dashed lines for several host extinction values. The BTS SNe Ia with distances calculated in this work are shown as black symbols, and the SNe with spectroscopic redshifts as blue symbols. The peak magnitudes of these SNe are the ones determined in this work. The LOSS distance limit and SNe, with the peak $ V $ band magnitudes calculated in this work, are presented as a red dashed line and red symbols, respectively. The CNIa0.02 observation limits for several extinction values are shown as black, dashed lines. The black solid line at $ 180\,\textrm{Mpc} $ marks the distance of our volume-limited LF.}
	\label{fig:dist mag}
\end{figure}

We conclude the discussion of SNe Ia LF with a visual summary of the considered surveys. Figure~\ref{fig:dist mag} shows the distance and absolute magnitude detection limits, for several extinction values, of the ZTF BTS (blue), LOSS (red), and CNIa0.02 (black) surveys. The absolute $ r $ magnitudes, as calculated in this work, and the distances of the BTS SNe Ia calculated in this work (with spectroscopic redshifts), are marked by black (blue) circles. The LOSS sample distances and absolute peak $ V $ magnitudes, as calculated in this work, are shown as red circles. Despite the difficulties that arise in a large survey, the high cadence and the large distances at which the BTS classifies transients are unmatched by the other surveys. However, as we have shown, accurate distances are essential in order to calculate the LF. Acquiring the accurate redshift for all BTS hosts is possible, and would significantly improve the scientific output of the survey. For example, there are 100 SNe with $ E(g-r)_{\rm{h}}>0.2 $ and spectroscopic redshifts in our sample that comprise the $ R_r $ distribution (except for 8 with negative $ A_g^\text{h}<0.1 $ or $ A_r^\text{h}<0.1 $); obtaining the redshifts of all the SNe would increase this number to 241. This information is available for the CNIa0.02 survey, which would allow the cross-checking of the results presented in this paper.

An accurate distance estimation is even more important for core-collapse SNe. As the photometric evolution of these SNe is more diverse than that of SNe Ia, it is more difficult to estimate their intrinsic luminosity from the light curve shapes, as we have done for SNe Ia. In addition, due to the younger environments of these SNe, the host extinction is probably higher, on average, than the ones we found for SNe Ia, leading to even higher uncertainties in their intrinsic luminosity. The transformation from the published pseudo-LF of these SNe \citep[][,P20]{Li2011} to the intrinsic LF is, therefore, highly uncertain.

\section*{Data Availability}
The data underlying this article are available in the article and in its
online supplementary material.

\section*{Acknowledgements}

We thank Boaz Katz and Eran Ofek for useful discussions. DK is supported by the Israel Atomic Energy Commission --The Council for Higher Education --Pazi Foundation -- by a research grant from The Abramson Family Center for Young Scientists and by ISF grant.

\bibliographystyle{mnras}
\bibliography{bibliography} 

\appendix

\section{A reanalysis of the LOSS sample}
\label{app:LOSS}
\begin{figure*}
	\includegraphics[width=\textwidth]{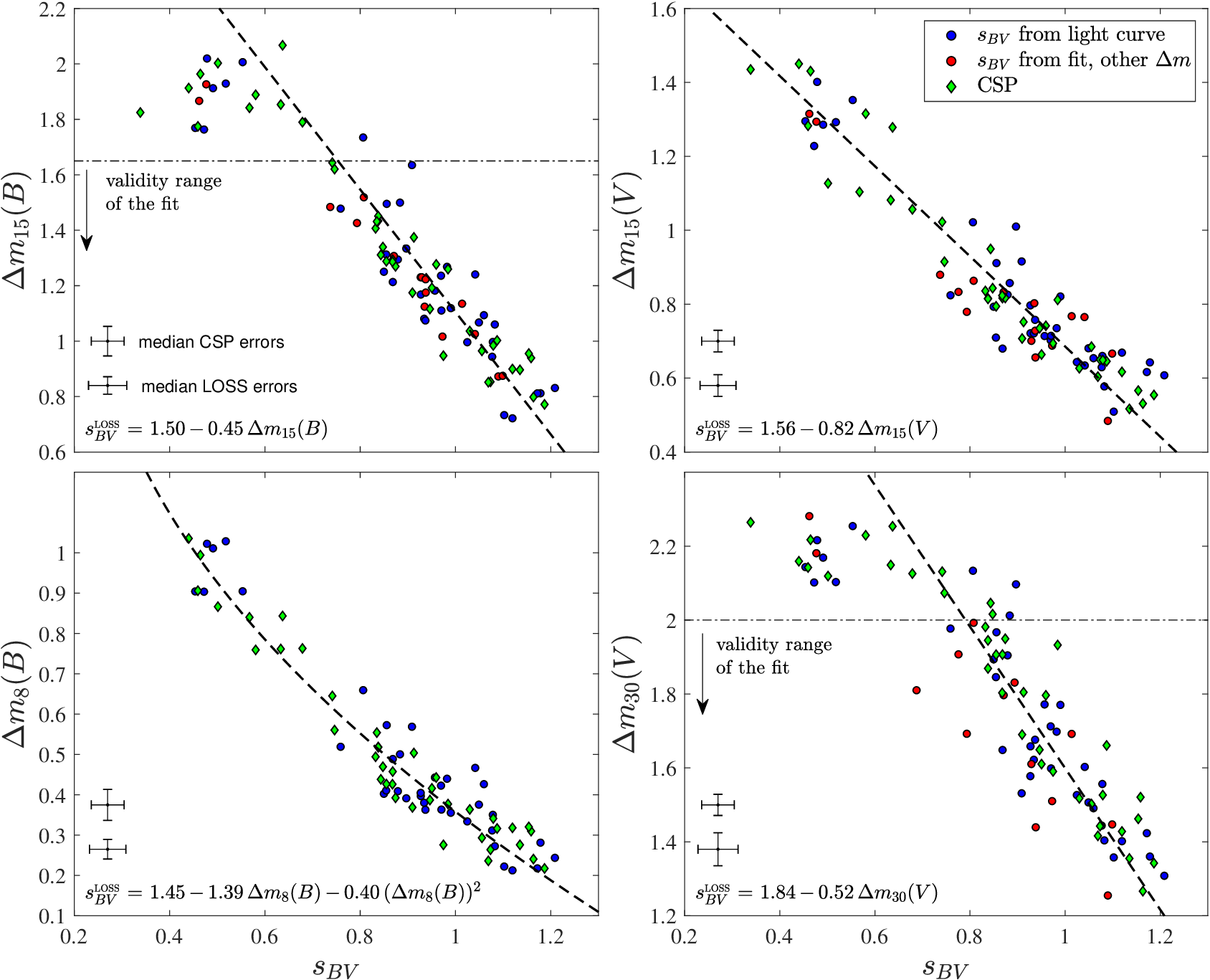}
	\caption{LOSS decline rate parameters as a function of \sbv. LOSS SNe with a direct \sbv measurement are shown as blue circles, while red circles correspond to SNe whose \sbv was determined using a different decline rate parameter. CSP SNe are shown as green diamonds. The median errors of the two samples are shown for each panel. The fits of the decline rate to the \sbv of the BTS and CSP samples are shown as a black, solid line and black dashed line, respectively. Also shown are the validity ranges of the $ \dmf(B) $ and $ \dm_{30}(V) $ fits.}
	\label{fig:LOSS sbv-dms}
\end{figure*}

The SNe of the LOSS sample \citep{Leaman2011,Li2011,Li2011b,Maoz2011} were mostly observed in the Johnson filters, specifically, with the $ B $ and $ V $ bands, and we assume that there are no significant differences between the LOSS and the CSP filters. We, therefore, use the color stretch parameter \sbv to estimate peak magnitude, which allows one to distinguish between the low luminosity events properly \citep{Burns2018}. We calculate the color stretch parameter and decline rates using SNooPy \citep{Burns2011}. When comparing our results of $ \dmf(B) $ to \cite{Piro2014}, we find a small average deviation of 5 per cent. We obtain the light curves of 66 out of the 76 SNe in \cite{Li2011}\footnote{The other 10 LOSS SNe have only unfiltered clear band data (B. Stahl, private communication).}. We are able to directly estimate the \sbv values of 37 SNe, while for another $20$ SNe, we estimate \sbv using the decline rate parameters $ \dm_{8}(B) $, $ \dmf(B) $, and $ \dm_{15}(V) $. The \sbv values and the decline rates are given in Table \ref{tab:LOSS sample}, and the $s_{BV}-\Delta m $ distributions are shown in Figure~\ref{fig:LOSS sbv-dms}.

While the LOSS SNe span the entire $ \dm_{15}(B) $ range, the  region $ 0.55\lesssim s_{BV}\lesssim0.75 $ contains only one SN. This is in contrast to the ZTF BTS sample, where the SNe span the entire \sgr\ range. We attribute this to the low number of SNe in the LOSS sample, an issue further discussed in Section~\ref{sec:Ni}.

\begin{figure}
	\includegraphics[width=\columnwidth]{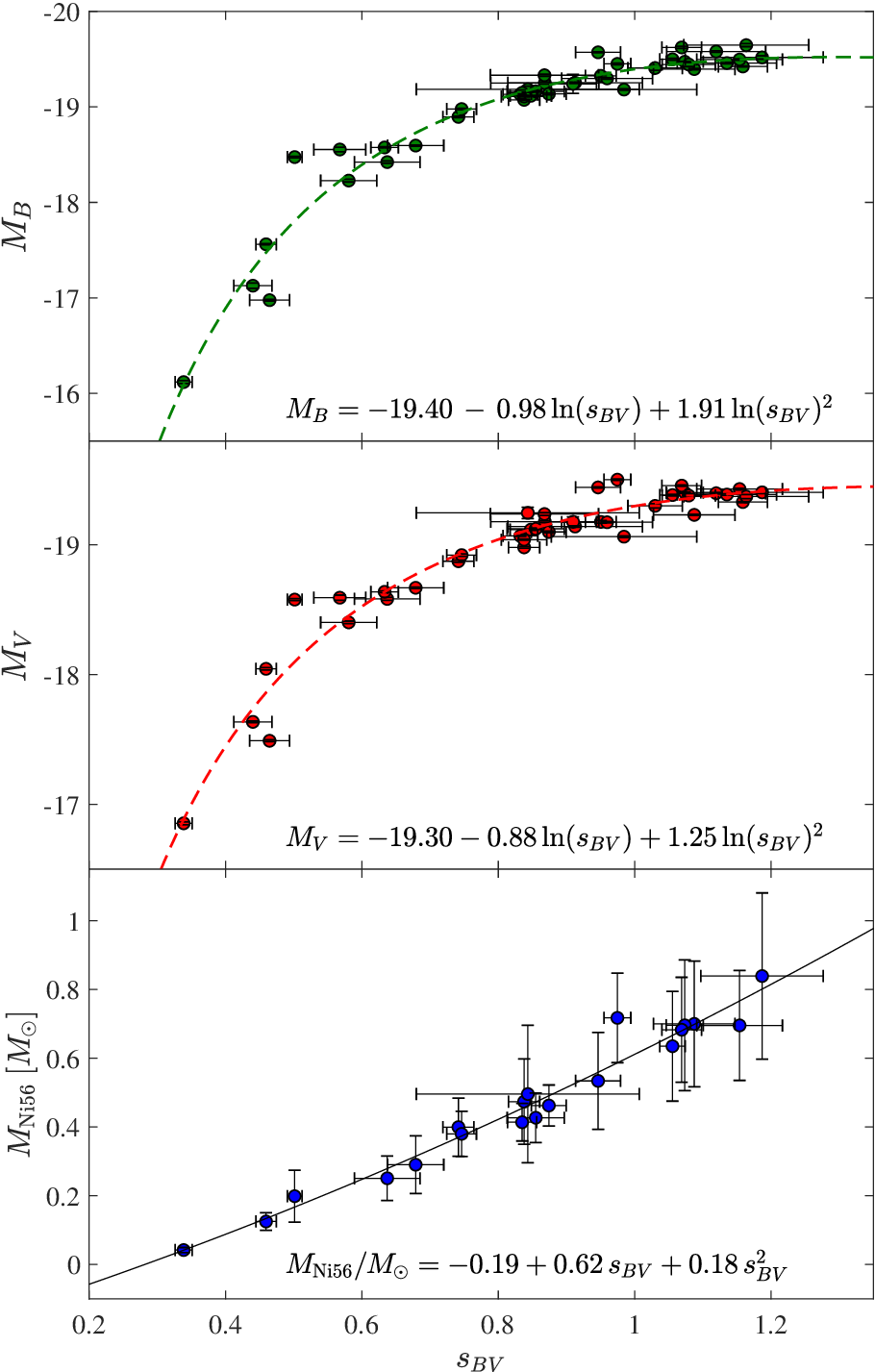}
	\caption{Peak $ M_B $ (upper panel), peak $ M_V $ (middle panel), and $\mni$ (lower panel) as a function of \sbv for the CSP sample. The fits are shown in lines, with the fit coefficients displayed at the bottom of each panel. The error bars for the peak magnitudes are smaller than the symbol size.}
	\label{fig:CSP sbv lum}
\end{figure}

We use observations in the $ B $ and $ V $ bands of the CSP sample to relate the color stretch \sbv to $\mni$ and to the peak magnitudes. We use a second-order polynomial for $\mni$:
\begin{equation}\label{eq:Ni sbv}
M_\text{Ni56}/\msol = p_0+p_1 s_{BV}+p_2 s_{BV}^2,
\end{equation}
and similar expressions for the peak magnitudes with $ \text{ln}(s_{BV}) $ as the argument. The coefficients of the peak magnitudes fits and of the $\mni$ fit are given in Tables~\ref{tab:fits sgr lum} and~\ref{tab:fits sgr ni}, respectively. Figure~\ref{fig:CSP sbv lum} shows the peak $ M_B $ (upper panel), peak $ M_V $ (middle panel), and $\mni$ (lower panel) values of the CSP sample as a function of $ s_{BV} $. For each parameter, the fit to \sbv (or to $ \text{ln}(s_{BV}) $) and the fit coefficients are displayed. The $\mni$-\sbv and $\mni$-\sgr\ distribution are very similar (Figure~\ref{fig:LF Ni}), as expected from the close relation between \sgr\ and \sbv \citep{Ashall2020}. 

The LOSS SNe Ia survey has only minor completeness corrections, and we ignore them in this work. The obtained LF are shown and discussed in Section~\ref{sec:Ni}.

\section{$ K $-corrections}
\label{app:k corr}

In this section, we investigate the effects of the supernova redshift and of the filter transmission function on the observed decline rates and peak magnitudes. Since our analysis compares multiple surveys, the decline rates distribution could be affected by the specific parameters of each survey. This is apparent in Figure \ref{fig:BTS sgr vs dm}, where the $ \dmf(g) $ and $ \dm_{8}(g) $ of the BTS sample have higher values than the CSP sample for the same $\msgr$. This could be potentially related to the filter transmission functions, where the Las Campanas Observatory filters \citep{Stritzinger2011} are used by the CSP, and the Palomar observatory filters \citep{Bellm2019} are used by the ZTF, and to the redshift distribution of the samples, as the BTS survey extends to higher redshifts. The mean redshift of the BTS sample is $ z\approx0.06 $ (and $ z\approx0.04 $ for the subsample with a direct \sgr\ measurement), while the mean redshift of the CSP sample is $ z\approx0.025 $. We show below that the decline rate differences are mostly the result of the different characteristic redshifts of the two surveys.

\begin{figure}
	\includegraphics[width=\columnwidth]{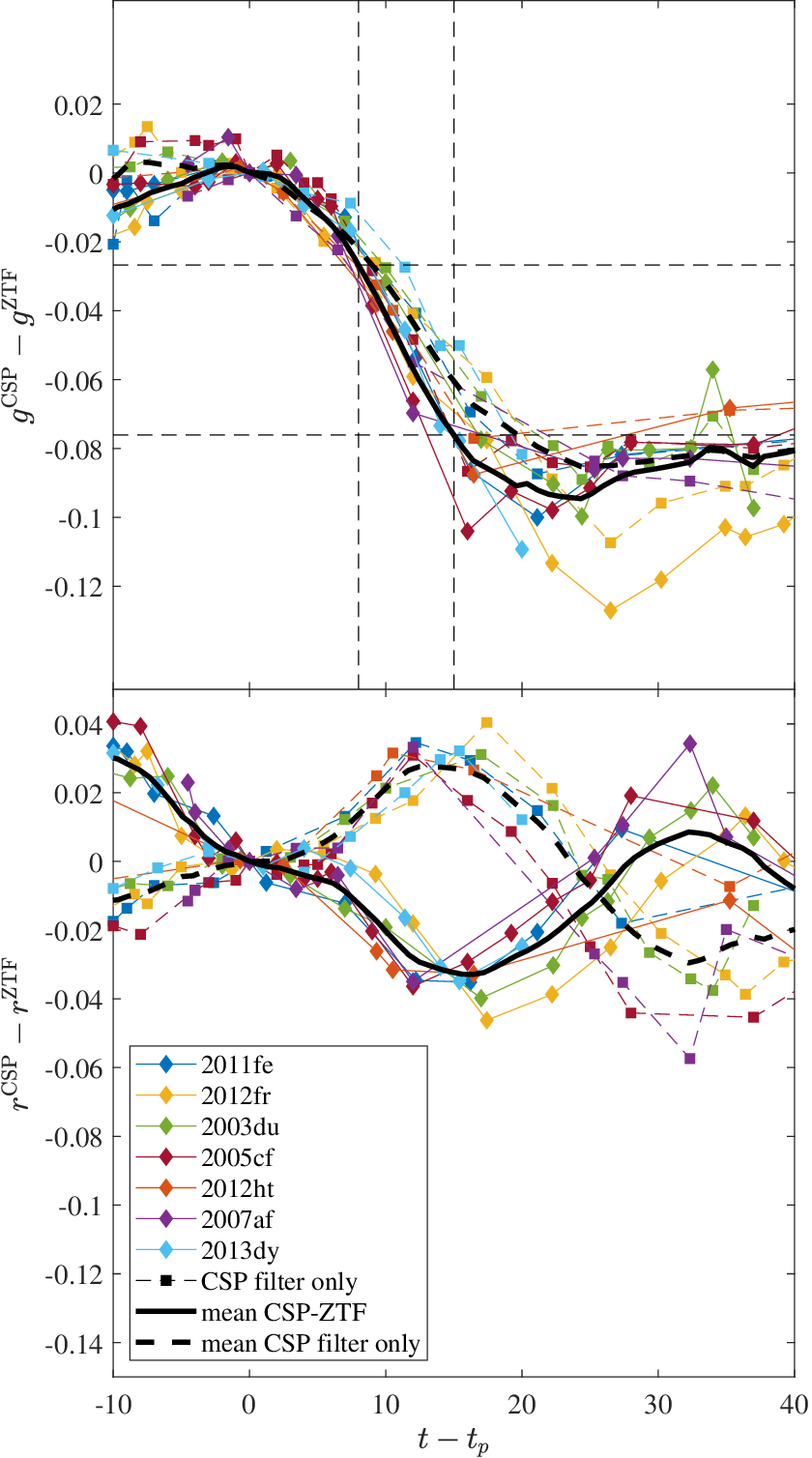}
	\caption{CSP and BTS synthetic magnitude difference as a function of the time since peak (top panel: $g$ band; bottom panel: $r$ band). The values for each SN are shifted so that they are zero at peak time. The differences between the CSP filter at the CSP typical redshift of $z=0.025 $ to the ZTF filter at the BTS typical redshift of $ z=0.04 $ are shown in diamonds and solid lines. Squares and dashed lines show the results for the CSP filters where the only difference is the redshift. Black curves show the mean difference in both cases. The dashed vertical lines in the top panel mark 8 and 15 d after peak, and the dashed horizontal lines mark their intersection with the mean difference. The intersection values indicate the mean difference in the decline rate that would have been calculated.}
	\label{fig:kcorr time}
\end{figure}

Using the spectra of several well-observed SNe Ia (see list in Figure~\ref{fig:kcorr time}), we measured the synthetic photometry of the $ g $ and $ r $ bands at multiple epochs for each SN. The spectra were taken from WiSeREP\footnote{https://wiserep.weizmann.ac.il}\citep{Yaron2012} and the filters transmission functions were taken from the SVO Filter Profile Service\footnote{http://svo2.cab.inta-csic.es/theory/fps/} \citep{Rodrigo2020}. Since the transmission functions of the ZTF filters does not include atmospheric extinction, we use the atmospheric transmission measured at the Kitt Peak National Observatory with an airmass of 1.1. For each epoch and SN, we measured the difference between the magnitude obtained with the CSP filter at the CSP typical redshift of $z=0.025 $, to the magnitude obtained with the ZTF filter at the BTS typical redshift of $ z=0.04 $. The results are shown in Figure~\ref{fig:kcorr time}, where the values, shown in diamonds and solid lines, are shifted so that they are zero at peak time. Also plotted is the mean difference of all the SNe (black solid lines). As can be seen in the figure, the differences between the $ g $ magnitudes increase until $ \sim20$ d since peak. The dashed vertical lines mark $8$ and $15$ d after peak, and the dashed horizontal lines mark their intersection with the mean difference. The intersection values are $ \approx0.03\,\text{mag} $ and $ \approx0.08\,\text{mag} $ at $8$ and $15$ d after peak, respectively. This means that the $ g $-band decline rate values would increase for the filter and mean redshift of the BTS, compared to the CSP sample. The differences in the $ r $ band are less pronounced. To check whether the effect arises from the different filters or from the redshift difference, we also measured the differences using only the CSP filter for both redshift values. The results are shown in squares and dashed lines, and the mean difference in black dashed lines. This decreases the differences by $ \sim30$ per cent in the $ g $ band, so we conclude that they are mainly caused by the redshift differences. In the $ r $ band, the differences continue to be relatively small.

The difference between the $ g $ band decline rates obtained using the spectra agrees with the observed decline rate difference using photometry, where the fit of $ \dmf(g) $ to the \sgr\ is higher by $ \approx0.08\,\text{mag} $ for the BTS sample, and the fit to $ \dm_{30}(r) $ is almost identical (Figure \ref{fig:BTS sgr vs dm}). We confirmed that the color stretch values are unaffected in this exercise. This might also explain the scatter of the BTS decline rates seen in Figure \ref{fig:BTS sgr vs dm}, since the redshift of the BTS sample is distributed up to $ z\sim0.1 $, although the typical uncertainties in the decline rate measurements dominate the scatter.

\begin{figure}
	\includegraphics[width=\columnwidth]{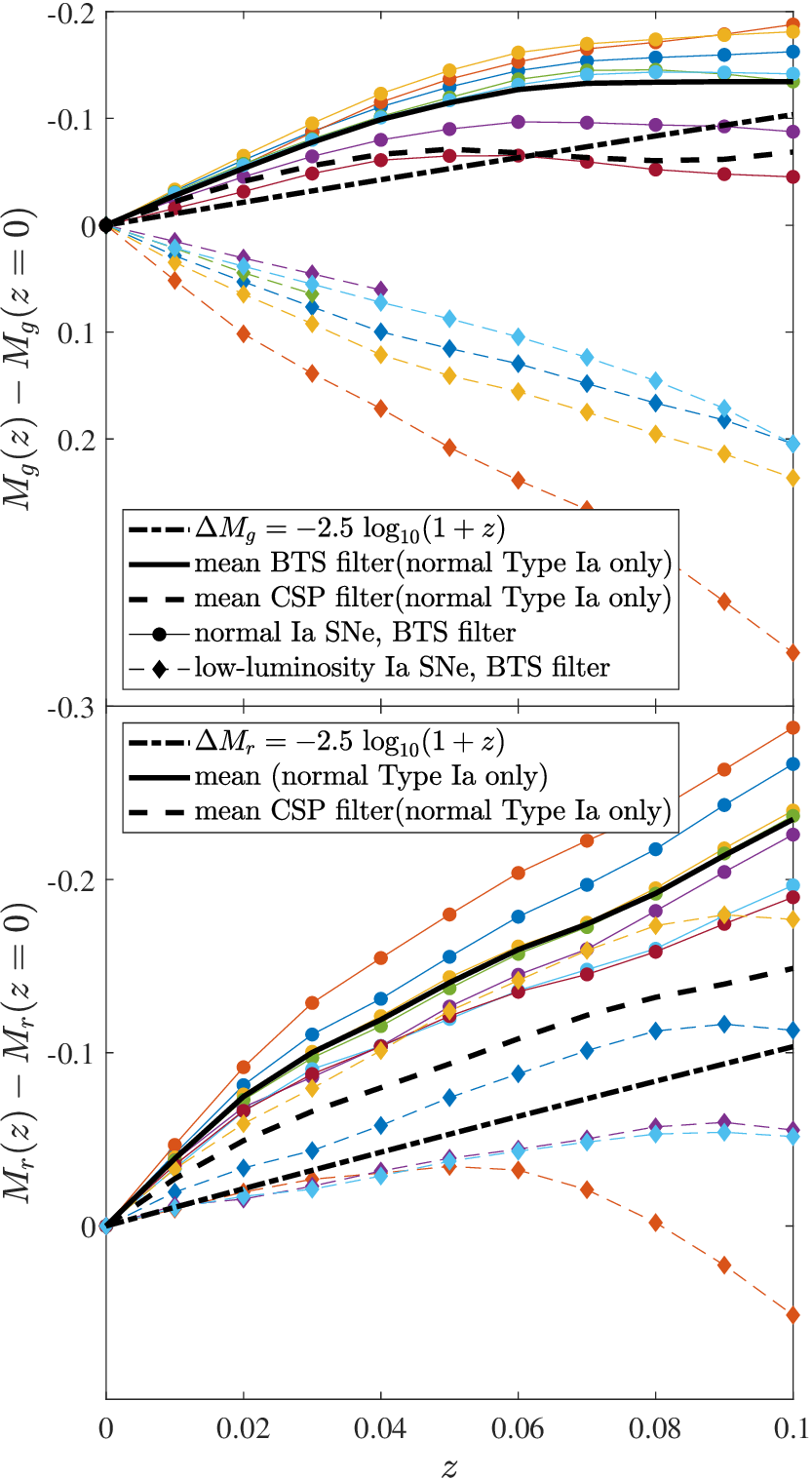}
	\caption{Top panel: The change in the synthetic $ g $ band peak magnitude as a function of redshift, calculated with the BTS filter transmission function for several SNe Ia. The solid (dashed) black line indicates the mean BTS (CSP) curve, while the dashed--dotted black line indicates the correction given by P20. Bottom panel: Same as the top panel for the $ r $ band. The normal SNe Ia are 2011fe, 2012fr, 2003du, 2005cf, 2012ht, 2007af, and 2013dy. The low-luminosity SNe are: 2005ke, 2005bl, 1999by, 1998de, 2006mr, and 2007ax.}
	\label{fig:kcorr z}
\end{figure}

We have also studied the relations between the redshift and the peak magnitude of the ZTF and CSP filters. For this purpose, we measured the change of the peak magnitude as a function of the redshift that is applied to the spectra. For each SN, the spectra closest to the peak time was chosen, and redshifted by several values, up to $ z\sim0.1 $. We repeated this procedure for known low-luminosity SNe Ia. The results using the BTS filters are shown in Figure \ref{fig:kcorr z}. As can be seen in the figure, the peak magnitude of normal SNe Ia becomes more luminous with the redshift in both the $ g $ and $ r $ bands, and its evolution is rather uniform among the different SNe. Low-luminosity SNe Ia have quite a different behaviour, especially in the $ g $ band. Also plotted is the $ K$-correction used by P20 (dashed-dotted black line), which is linear to a good approximation for the observed range of redshift values, and a fit that consists of the mean difference of the normal SNe Ia of the BTS filters (solid black line) and the CSP filters (dashed black line). The mean $ r $ band correction of the fit is higher by a factor of $ \approx2.7 $ on average, and we use this fit for volume corrections throughout our analysis, and, together with the $ g $ band $ K $-correction, for calculating the host extinction (see Section \ref{sec:extinction}). The values of the fits for the BTS and CSP filters for several redshifts are given in Table \ref{tab:kcorr}. The dim SNe in Figure \ref{fig:kcorr z} have very low \nickel mass ($ \sim0.1\,M_\odot $) and are a small part of the sample, therefore we ignore them in our fit and apply a single correction function for all SNe based on normal SNe Ia.

\begin{table}	
	\caption{The $ g $ and $ r $ peak magnitude $ K $-correction of the BTS and CSP filters. The function values are the mean peak magnitude difference of the considered SNe.}
	\centering
	\label{tab:kcorr}
	\begin{tabular}{lrrrr}
		$z$ & BTS $\Delta M_g $ & BTS $\Delta M_r $ & CSP $\Delta M_g $ & CSP $\Delta M_r $\\\midrule
		0.00 &  0.000 &  0.000 &  0.000 &  0.000 \\ 
		0.01 & -0.028 & -0.039 & -0.022 & -0.027 \\ 
		0.02 & -0.053 & -0.075 & -0.041 & -0.049 \\ 
		0.03 & -0.078 & -0.100 & -0.055 & -0.066 \\ 
		0.04 & -0.099 & -0.119 & -0.065 & -0.080 \\ 
		0.05 & -0.115 & -0.140 & -0.069 & -0.094 \\ 
		0.06 & -0.127 & -0.160 & -0.065 & -0.108 \\ 
		0.07 & -0.133 & -0.174 & -0.060 & -0.122 \\ 
		0.08 & -0.134 & -0.192 & -0.057 & -0.132 \\ 
		0.09 & -0.134 & -0.214 & -0.058 & -0.140 \\ 
		0.10 & -0.134 & -0.235 & -0.064 & -0.149 \\  
\end{tabular}
\end{table}

\begin{figure}
	\includegraphics[width=\columnwidth]{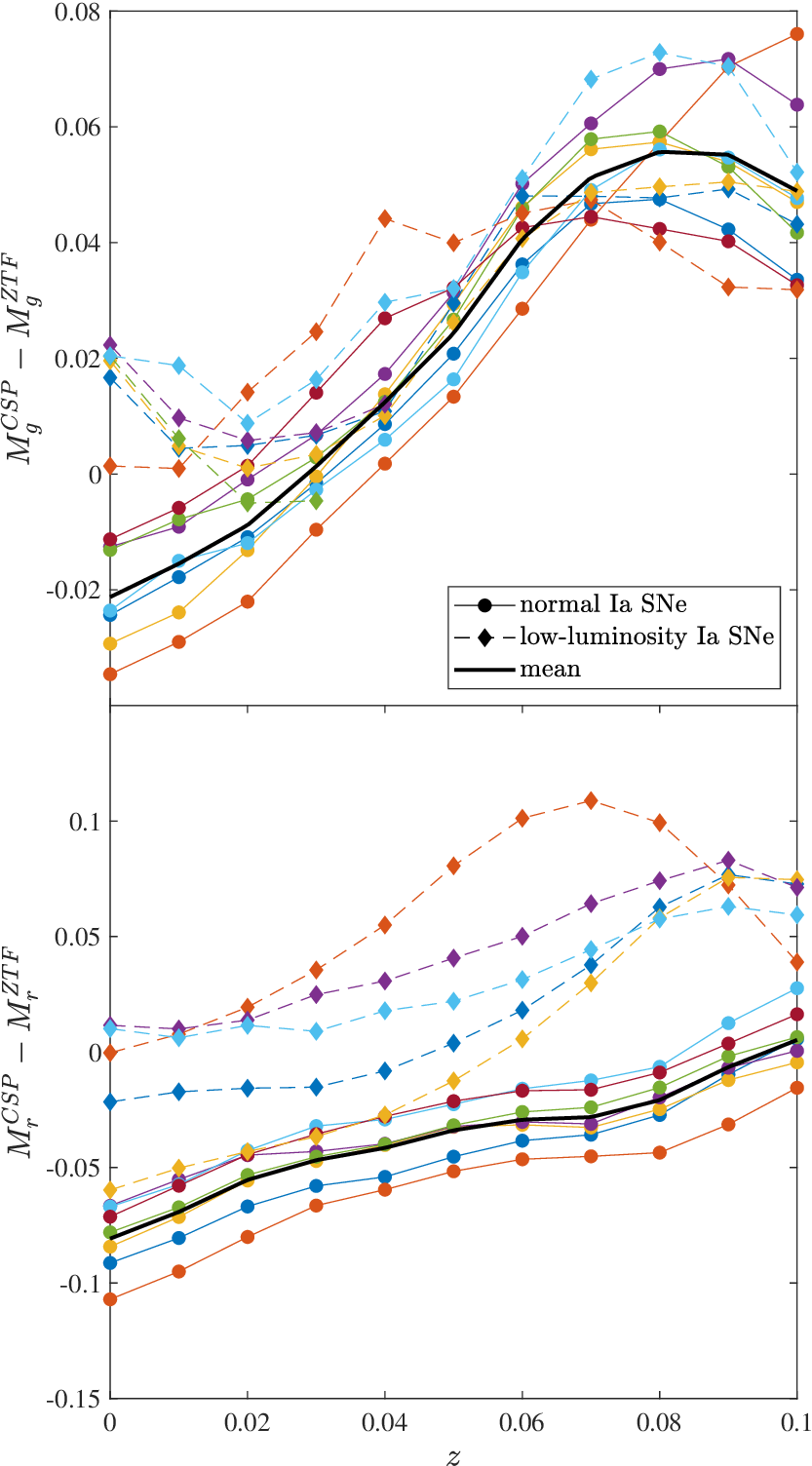}
	\caption{Top panel: Difference between the synthetic $ g $-band peak magnitude calculated with the CSP filter and with the ZTF filter, as a function of redshift. The solid black line indicates the mean curve. Bottom panel: Same as the top panel for the $ r $ band. The normal SNe Ia are 2011fe, 2012fr, 2003du, 2005cf, 2012ht, 2007af, and 2013dy. The low-luminosity SNe are 2005ke, 2005bl, 1999by, 1998de, 2006mr, and 2007ax.}
	\label{fig:mpeak diff z}
\end{figure}

Lastly, we used the SNe Ia spectra to calculate the peak magnitude difference between the CSP and the ZTF filters. Since we calibrate the absolute peak magnitude with the CSP sample and filters, this correction is required when calculating the distances, the volume corrections and the host extinction, where the observed magnitude is measured in the ZTF filters. To determine this correction, we once again chose the spectra closest to the peak time for each SN, and measured the difference between the synthetic magnitudes as a function of redshift. We performed the same procedure for normal and for low-luminosity SNe Ia (although we only use the correction for the normal SNe Ia). The results are presented in Figure \ref{fig:mpeak diff z}. As can be seen in the figures, the difference behaviour is quite uniform for all non-low-luminosity SNe, with a scatter of $ \lesssim0.02\;\text{mag} $. A curve showing the mean difference is shown in a black solid line, and its values for several redshifts are given in Table \ref{tab:peakmag diff}. The peak magnitudes are corrected using the mean difference of normal SNe Ia for each filter:
\begin{equation}\label{eq:peakmag diff}
M^\text{ZTF}=M^\text{CSP}(\msgr)-\Delta M(z),
\end{equation}
where $ M^\text{CSP}(\msgr) $ is the peak magnitude given by the CSP fit and $ \Delta M(z) $ is the filter correction function. The correction function is given by a linear interpolation of the mean difference function using the redshift, as provided by the BTS explorer. Although we argue throughout this work that these redshift values are not accurate, the peak magnitude error resulting from this uncertainty is small. This correction has a small effect on the LF, but a substantial effect on the selective extinction $ E(g-r)_{\rm{h}} $, reducing the fraction of SNe with negative $ E(g-r)_{\rm{h}} $ from $ \approx41$ to $ \approx17$ per cent.

\begin{table}	
	\caption{The $ g $- and $ r $-peak magnitude correction function used in Equation \eqref{eq:peakmag diff}. The function values are the mean peak magnitude difference of the considered SNe.}
	\centering
	\label{tab:peakmag diff}
		\begin{tabular}{lrr}
		$z$ & $\Delta M_g $ &  $\Delta M_r $\\\midrule
		0.00 & -0.021 & -0.081 \\ 
		0.01 & -0.016 & -0.069 \\ 
		0.02 & -0.009 & -0.055 \\ 
		0.03 &  0.001 & -0.047 \\ 
		0.04 &  0.012 & -0.042 \\ 
		0.05 &  0.024 & -0.034 \\ 
		0.06 &  0.041 & -0.029 \\ 
		0.07 &  0.051 & -0.028 \\ 
		0.08 &  0.056 & -0.021 \\ 
		0.09 &  0.055 & -0.007 \\ 
		0.10 &  0.049 &  0.005 \\ 
		\end{tabular}
\end{table}

\section{Construction of the LF and volume corrections}
\label{app:volume corr}
In this section, we describe a few different methods for calculating the LF and we compare between volume-limited LF and magnitude-limited LF. Specifically, we show that a $ 180\,\text{Mpc} $ volume-limited subsample is an optimal choice for balancing between statistical and systematic errors. We verify that different limits on the sample provide consistent results, although with larger errors (statistical or systematic).

For the calculation of the LF, we only consider SNe from 2019 March 15, as the detection and classification rate of the BTS had substantially increased compared to earlier epochs (see fig 13 in P20), narrowing the sample size to 1371 SNe. In all the rest of the analysis, the whole sample of 1519 SNe is used. There is also a gap in March 2020 caused by bad weather that we take into account, so the total observation time of the LF is 2.27 yr. the classification efficiency $ f_{\text{cl},i} $ remains the same as in P20. We measure the recovery fraction of the BTS sample at the considered times to be $ \approx0.68 $, slightly higher than the value in P20, measured at different observation dates. Since we only use SNe with an estimated $\msgr$ and observed peak $ g $ and $ r $ magnitudes (note that one observed peak is sufficient for the magnitude-limited LF), we apply additional global corrections, i.e., we divide by the fraction of SNe with an estimate for \sgr\ and by the fraction of SNe with estimated distances ($ \approx0.9 $ for each factor; see Section \ref{sec:completeness} for details regarding the distance estimation).  

The calculation of the magnitude-limited LF is similar to the procedure described in P20, but the volume corrections in Equation \eqref{eq:max vol} are based on the estimated peak magnitudes of the $ r $ band. Additionally, the LF has another global correction factor that takes into account the reduction of the effective volume due to host extinction (note that this factor is not required for the construction of the pseudo-LF). For a given extinction $ A_r^\text{h} $, the fractional volume reduction is
\begin{equation}\label{eq:vol red}
\frac{V_{A_r^\text{h}}}{V_0}=10^{-\frac{3}{5}A_r^\text{h}},
\end{equation}
where $ V_{A_r^\text{h}} $ is the observed volume for an extinction value $ A_r^\text{h} $ and some limiting magnitude. To calculate the mean fractional volume reduction, we use the host extinction distribution from Section~\ref{sec:extinction}, $ f_{A_r} $:
\begin{equation}\label{eq:vol red mean}
f_\text{ext,h} = \left\langle\frac{V_{A_r^\text{h}}}{V_0}\right\rangle = \int dA_r^\text{h} f_{A_r} 10^{-\frac{3}{5}A_r^\text{h}}\approx0.65,
\end{equation}
which increases the SN Ia rate by $ \approx50$ per cent.
 
The volume-limited LF is constructed from the SNe in the chosen volume, $ V_\text{LF}=4\pi D^3_\text{LF}/3 $, with the weights in Equation~\eqref{eq:rate} differing only because of the small variation in the classification efficiency, $w_i=(f_{\text{cl},i}V_\text{LF})^{-1}  $.
In contrast to the magnitude-limited LF, a global correction factor for the Milky Way and host extinction is not applied (the other factors remain). However, as can be seen in Figures~\ref{fig:completeness2} and~\ref{fig:sgr dist}, volume corrections are still required for dim-enough SNe, depending on the chosen distance, $ D_\text{LF} $. The simplest form of correction is to correct for SNe with $D_\text{lim}<D_\text{LF} $, where  $D_\text{lim}$ is the limiting distance, neglecting galactic and host extinction:
\begin{equation}\label{eq:dlim}
D_\text{lim} = 10^{\frac{m_\text{lim}-M_i}{5}-5}\,\text{Mpc}^3.	
\end{equation}
This correction increases the weights to $w_i=(f_{\text{cl},i}V_\text{lim})^{-1} $, where $ V_\text{lim} $ is the volume corresponding to $ D_\text{lim} $. As can be seen in Section~\ref{sec:completeness}, the sample is not uniformly distributed in $D^3$ up to $D_\text{lim} $, so this correction is an underestimate. Since the host extinction distribution depends strongly on \sgr\ (see Figure~\ref{fig:BTS host ext sgr}), each \sgr\ value has a different correction. In order to take into account the host extinction when applying the volume correction, we use again the host extinction distributions of Section~\ref{sec:extinction}. In this case, we divide the distribution into the three regimes of Figure~\ref{fig:BTS host ext sgr}: $ \msgr<0.8,\,0.8<\msgr<1.15 $ and $ \msgr>1.15 $. For each distribution $ f_{A_r} $, we calculate the mean volume reduction using Equation~\eqref{eq:vol red mean}. We find that the corresponding extinction values for each bin are as follows:
\begin{equation}\label{eq:host ext mag}
\left\langle A_r^\text{h}\right\rangle=-\frac{5}{3}\log\left(\left\langle\frac{V_{A_r^\text{h}}}{V_0}\right\rangle\right) = \begin{cases}
0.16\,\text{mag} & \msgr<0.8 \\
0.34\,\text{mag} & 0.8<\msgr<1.15 \\
0.71\,\text{mag} & \msgr>1.15. \\
\end{cases}
\end{equation}
Note that $ \left\langle A_r^\text{h}\right\rangle $ is not the mean extinction and is always smaller than it. A similar value for the Milky Way extinction can be calculated from the volume reduction in P20, $ f_\text{ext,MW} $:
\begin{equation}\label{eq:gal ext volume}
\left\langle A^\textrm{MW}_r\right\rangle=-\frac{5}{3}\log\left(f_\text{ext,MW}\right)=0.144\,\text{mag}.
\end{equation}
Together, the maximal distance of each SN is modified by 
\begin{equation}\label{eq:modified distance}
D_\text{lim}\rightarrow D_\text{lim}\times10^{-\frac{\langle A^\text{h}_r\rangle+\langle A^\text{MW}_r\rangle}{5}}.
\end{equation}
The curves of the modified distance for the three values of Equation \eqref{eq:host ext mag} are shown in Figure \ref{fig:sgr dist} in black, dashed lines. The volume corrections are then applied as above, by using the modified distance.

A different method that can be utilized to correct for the extinction is to use the entire host distribution in each \sgr\ bin and to sum the corrections for the extinction values that reduce the limiting distance below $ D_\text{LF} $. When using this method, the weight of each SN is:
\begin{equation}\label{eq:VC dist}
w_i=\frac{1}{f_{\text{cl},i}}\times \left(\int dA^\text{h}_r f_{A_r}\min\left\{V_\text{LF},V_\text{lim}\times 10^{-\frac{3}{5}(A^\text{h}_r+\langle A^\text{MW}_r\rangle)} \right\}  \right)^{-1}.
\end{equation}
The corrections provided by this method are equal or larger than the corrections obtained with the previous method. However, these modifications have a small effect on the LF for the chosen distance of $ 180\,\text{Mpc} $.

\begin{figure}
	\includegraphics[width=\columnwidth]{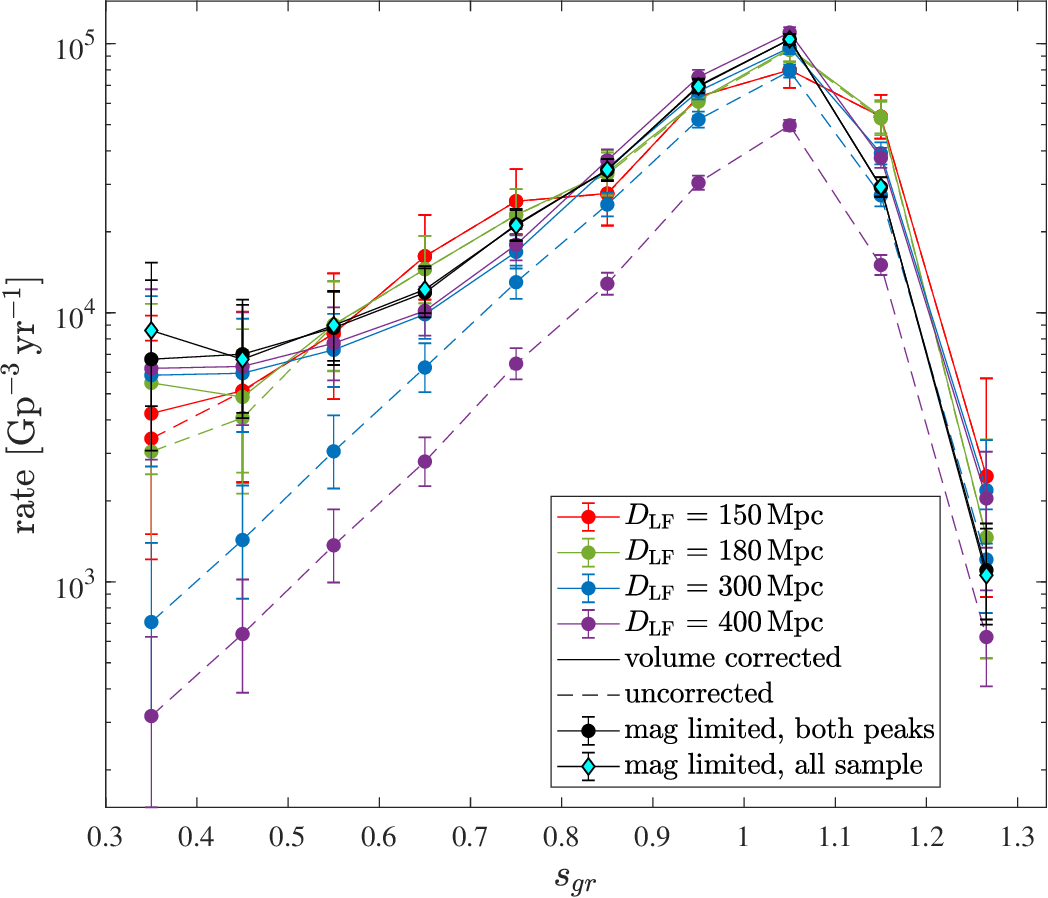}
	\caption{Comparison between the volume-limited \sgr\ distributions at various distances and the magnitude-limited distributions. Volume-corrected (uncorrected) distributions are shown as solid (dashed) lines, and the magnitude-limited distribution of SNe with both $ g $ and $ r $ observed peaks (all SNe) is shown as black (cyan) symbols. The volumetric rates for the uncorrected distributions at $ D_\text{LF}\gtrsim250\,\text{Mpc} $ are significantly lower in all bins, as the sample is not complete at these distances for all luminosity bins. At $ D_\text{LF}\lesssim200\,\text{Mpc} $, the differences are mainly in the very dim and very luminous bins. The corrected distributions are in better agreement with each other. The magnitude-limited distributions are mostly in agreement with the volume-corrected distributions, with differences arising in high luminosity bins, where the impact of the host extinction is the largest.}
	\label{fig:LF comp}
\end{figure}

The \sgr\ distributions are shown in Figure~\ref{fig:LF comp}. The figure shows the volume-corrected (uncorrected) distribution in solid (dashed) lines, where the volume correction was applied using Equation \eqref{eq:modified distance}. As expected from Figure~\ref{fig:completeness2}, the rates of the uncorrected LFs are considerably lower for $ D_\text{LF}\gtrsim250\,\text{Mpc} $, since the survey is not complete at these distances for all luminosity bins. At $ D_\text{LF}=400\,\text{Mpc} $, the total rate is less than half the rate at $ 180\,\text{Mpc} $. 

The volume-corrected distributions are in better agreement with each other, and their total rate typically deviates by $ \sim5$ to $\sim10 $ per cent from each other. The figure also shows the magnitude-limited distribution of the SNe with both $ g $ and $ r $ observed peaks (black symbols), and the magnitude-limited distribution of all SNe (cyan symbols). The differences between the two magnitude-limited distributions are minor, except for the dimmest bin, $ \msgr<0.4 $, that has a $ \approx25$ per cent difference due to one SN with no observed peaks in both the $ g $ and $ r $ bands. The magnitude-limited distributions are in agreement with the volume-corrected distributions, with rates higher in low-luminosity bins and lower in high-luminosity bins. This is because the host extinction correction factor of the magnitude-limited distributions affects all bins equally, while the impact of host extinction, as seen in Sections~\ref{sec:completeness} and~\ref{sec:extinction}, is much more prominent in high-luminosity bins. The magnitude-limited total rate is within $ \sim5$ per cent of the volume-limited rates, and within $ \sim1$ per cent of the rate of the $ 180\,\text{Mpc} $ subsample that is used for the main results. We chose this distance because it is the largest distance where the vast majority of the luminosity bins are complete, and therefore, requires minimal volume corrections (which are inherently uncertain, and therefore, increase the systematic error). 

\section{CSP, BTS, and LOSS sample properties}
\label{app:CSP}

\begin{figure*}
	\includegraphics[width=\textwidth]{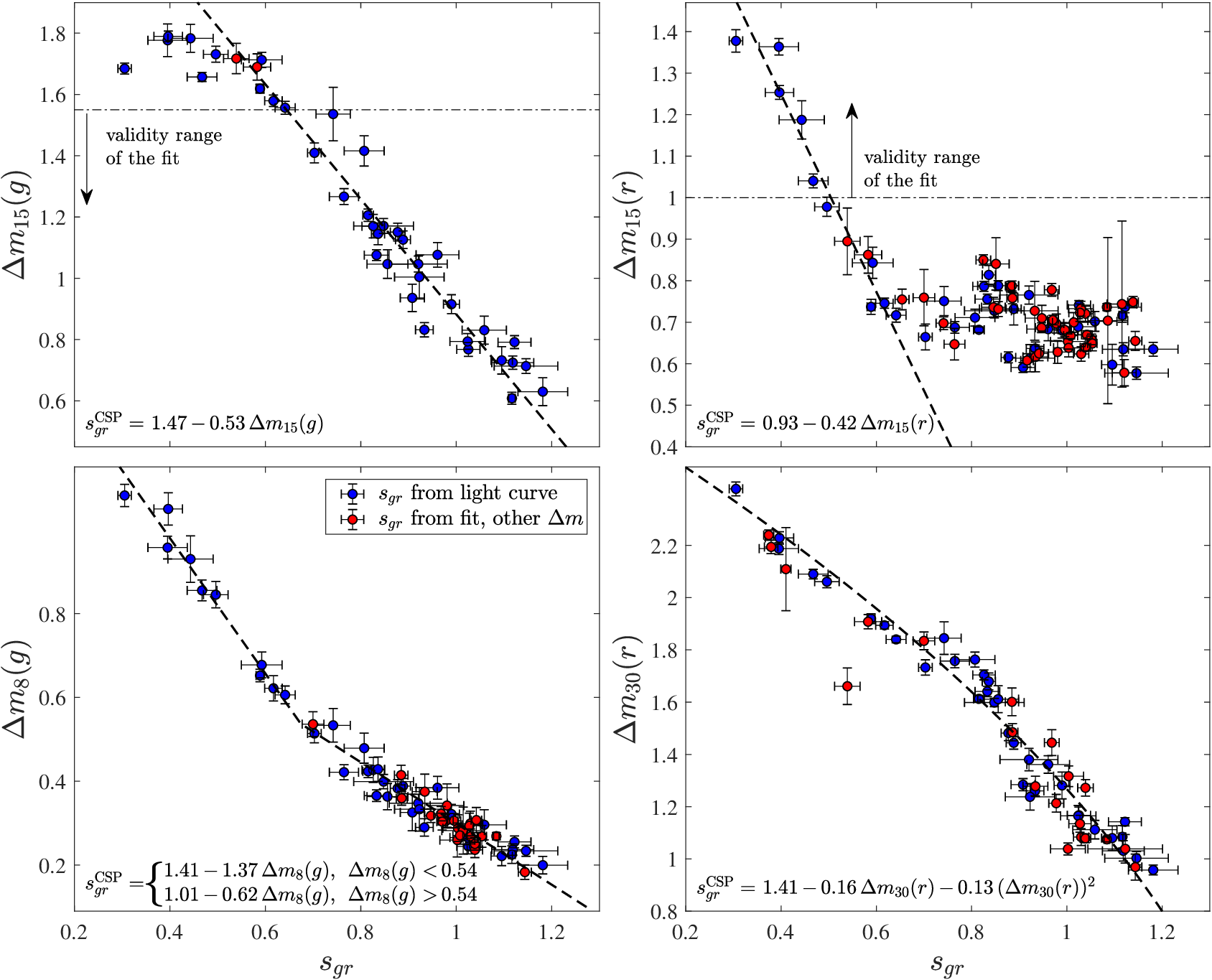}
	\caption{The CSP decline rate parameters as a function of $\msgr$. SNe with a direct \sgr\ measurement are indicated by blue symbols, and SNe whose \sgr\ that was determined using a different decline rate parameter are marked by red symbols. Also shown are the fits of the decline rate (black dashed lines) and the validity range of the $ \dmf(g) $ and $ \dmf(r) $ fits.}
	\label{fig:sgr vs dm}
\end{figure*}

Figure~\ref{fig:sgr vs dm} shows the CSP decline rate parameters as a function of \sgr\ for the CSP SNe. We also provide in Tables~\ref{tab:CSP sample} and~\ref{tab:CSP sampleBV} the CSP sample data used to calibrate the relations between the light-curve shape parameter and the intrinsic luminosity of the SNe. The tables contain the color stretch parameter and the source used for its determination (directly or from which decline rate), decline rates, peak magnitudes, and $\mni$. The peak magnitudes and $\mni$ were calculated using the independent distance and extinction estimates, from the constructed bolometric luminosity.

We similarly provide in Tables~\ref{tab:BTS sample} and~\ref{tab:LOSS sample} the data for the BTS and LOSS samples, respectively.

\begin{table*}
	\caption{CSP sample $ g $- and $ r $-band color stretches, decline rates, and luminosity parameters. The complete table is available in the electronic supplementary material.}
	\label{tab:CSP sample}
	\begin{threeparttable}
	\renewcommand{\TPTminimum}{\linewidth}
	\makebox[\linewidth]{%
	\begin{tabular}{lccccccccc}
		Name & \sgr\ & \sgr\ src\tnote{a} &    $ \dmf(g)  $& $ \dmf(r) $  & $ \dm_8(g) $   &     $ \dm_{30}(r) $   & $ M_g $ & $ M_r $  & $ \mni (M_\odot) $ \\	\midrule 
	2004ef   & $  0.82\,\pm\,0.01  $  & dir& $  1.21\,\pm\,0.02  $ & $  0.68\,\pm\,0.01  $ & $  0.42\,\pm\,0.02  $ & $  1.61\,\pm\,0.01  $ & $-19.17\,\pm\,0.00  $ & $-19.10\,\pm\,0.00  $ & $   -   $ \\
	2004eo   & $  0.88\,\pm\,0.02  $  & dir& $  1.15\,\pm\,0.03  $ & $  0.62\,\pm\,0.01  $ & $  0.38\,\pm\,0.02  $ & $  1.48\,\pm\,0.03  $ & $-19.12\,\pm\,0.01  $ & $-19.00\,\pm\,0.00  $ & $  0.47\,\pm\,0.12  $ \\
	2004ey   & $  1.12\,\pm\,0.03  $  & dir& $  0.79\,\pm\,0.02  $ & $  0.74\,\pm\,0.01  $ & $  0.26\,\pm\,0.01  $ & $  1.14\,\pm\,0.02  $ & $-19.44\,\pm\,0.00  $ & $-19.27\,\pm\,0.00  $ & $   -   $ \\
	2004gc   & $  0.92\,\pm\,0.01  $  & $ \dm_{30}(r) $& $   -   $ & $  0.61\,\pm\,0.02  $ & $   -   $ & $  1.43\,\pm\,0.02  $ & $   -   $ & $-19.01\,\pm\,0.00  $ & $   -   $ \\
	2004gs   & $  0.70\,\pm\,0.01  $  & dir& $  1.41\,\pm\,0.03  $ & $  0.66\,\pm\,0.03  $ & $  0.51\,\pm\,0.02  $ & $  1.73\,\pm\,0.03  $ & $-18.96\,\pm\,0.00  $ & $-18.87\,\pm\,0.01  $ & $  0.40\,\pm\,0.08  $ \\
	2004gu   & $  1.05\,\pm\,0.06  $  & $ \dm_{8}(g) $& $  0.88\,\pm\,0.07  $ & $  0.65\,\pm\,0.02  $ & $  0.26\,\pm\,0.04  $ & $  1.04\,\pm\,0.02  $ & $-19.45\,\pm\,0.01  $ & $-19.48\,\pm\,0.01  $ & $   -   $ \\
	2005ag   & $  1.02\,\pm\,0.04  $  & $ \dm_{8}(g) $& $  0.88\,\pm\,0.06  $ & $  0.64\,\pm\,0.02  $ & $  0.29\,\pm\,0.03  $ & $  1.32\,\pm\,0.04  $ & $-19.53\,\pm\,0.01  $ & $-19.46\,\pm\,0.01  $ & $   -   $ \\
	2005al   & $  0.82\,\pm\,0.02  $  & $ \dm_{30}(r) $& $   -   $ & $  0.85\,\pm\,0.01  $ & $   -   $ & $  1.60\,\pm\,0.03  $ & $   -   $ & $-19.12\,\pm\,0.00  $ & $   -   $ \\
	2005am   & $  0.74\,\pm\,0.02  $  & $ \dm_{30}(r) $& $   -   $ & $  0.70\,\pm\,0.02  $ & $   -   $ & $  1.74\,\pm\,0.03  $ & $   -   $ & $-18.90\,\pm\,0.01  $ & $   -   $ \\
	2005A    & $  0.91\,\pm\,0.03  $  & dir& $  0.94\,\pm\,0.04  $ & $  0.59\,\pm\,0.01  $ & $  0.33\,\pm\,0.04  $ & $  1.28\,\pm\,0.02  $ & $-19.27\,\pm\,0.01  $ & $-19.36\,\pm\,0.01  $ & $   -   $ \\
	2005bg   & $   -   $  & none& $   -   $ & $  0.79\,\pm\,0.12  $ & $   -   $ & $   -   $ & $   -   $ & $-18.51\,\pm\,0.04  $ & $   -   $ \\
	2005bl   & $  0.43\,\pm\,0.02  $  & $ \dm_{8}(g) $& $   -   $ & $   -   $ & $  0.94\,\pm\,0.03  $ & $   -   $ & $-17.69\,\pm\,0.01  $ & $-17.99\,\pm\,0.02  $ & $   -   $ \\
	2005bo   & $  0.85\,\pm\,0.03  $  & $ \dm_{30}(r) $& $   -   $ & $  0.84\,\pm\,0.06  $ & $   -   $ & $  1.55\,\pm\,0.05  $ & $   -   $ & $-19.11\,\pm\,0.01  $ & $   -   $ \\
	2005eq   & $  1.18\,\pm\,0.05  $  & dir& $  0.63\,\pm\,0.05  $ & $  0.63\,\pm\,0.02  $ & $  0.20\,\pm\,0.02  $ & $  0.96\,\pm\,0.02  $ & $-19.54\,\pm\,0.01  $ & $-19.37\,\pm\,0.01  $ & $  0.84\,\pm\,0.24  $ \\
	2005hc   & $  1.04\,\pm\,0.02  $  & $ \dm_{8}(g) $& $  0.83\,\pm\,0.03  $ & $  0.62\,\pm\,0.02  $ & $  0.27\,\pm\,0.02  $ & $  1.08\,\pm\,0.03  $ & $-19.57\,\pm\,0.01  $ & $-19.42\,\pm\,0.00  $ & $  0.70\,\pm\,0.16  $ \\
	2005hj   & $  1.10\,\pm\,0.02  $  & dir& $  0.73\,\pm\,0.04  $ & $  0.60\,\pm\,0.05  $ & $  0.22\,\pm\,0.02  $ & $  1.08\,\pm\,0.05  $ & $-19.51\,\pm\,0.01  $ & $-19.44\,\pm\,0.01  $ & $   -   $ \\
	2005iq   & $  0.92\,\pm\,0.02  $  & $ \dm_{8}(g) $& $  1.10\,\pm\,0.02  $ & $  0.76\,\pm\,0.01  $ & $  0.36\,\pm\,0.02  $ & $  1.49\,\pm\,0.03  $ & $-19.33\,\pm\,0.01  $ & $-19.13\,\pm\,0.01  $ & $   -   $ \\
	2005ir   & $  1.02\,\pm\,0.08  $  & $ \dm_{8}(g) $& $  0.86\,\pm\,0.08  $ & $  0.70\,\pm\,0.02  $ & $  0.28\,\pm\,0.06  $ & $   -   $ & $-19.52\,\pm\,0.01  $ & $-19.47\,\pm\,0.01  $ & $   -   $ \\
	2005kc   & $  0.97\,\pm\,0.02  $  & $ \dm_{8}(g) $& $  0.98\,\pm\,0.02  $ & $  0.69\,\pm\,0.01  $ & $  0.32\,\pm\,0.02  $ & $   -   $ & $-19.28\,\pm\,0.00  $ & $-19.11\,\pm\,0.00  $ & $   -   $ \\
	2005ke   & $  0.47\,\pm\,0.03  $  & dir& $  1.66\,\pm\,0.01  $ & $  1.04\,\pm\,0.02  $ & $  0.86\,\pm\,0.02  $ & $  2.09\,\pm\,0.02  $ & $-17.85\,\pm\,0.01  $ & $-18.15\,\pm\,0.01  $ & $  0.13\,\pm\,0.03  $ \\		
	\end{tabular}}
	\begin{tablenotes}
	\item [a] The source for the \sgr\ value. The options are either `dir' (directly from the light curve) or one of the decline rates. 
	\end{tablenotes}
\end{threeparttable}
\end{table*}

\begin{table*}
	\caption{CSP sample $ B $- and $ V $-band color stretches, decline rates and luminosity parameters. The complete table is available in the electronic supplementary material.}
	\label{tab:CSP sampleBV}
	\begin{threeparttable}
		\renewcommand{\TPTminimum}{\linewidth}
		\makebox[\linewidth]{%
			\begin{tabular}{lcccccccc}
					Name & \sbv  &    $ \dmf(B)  $& $ \dmf(V) $  & $ \dm_8(B) $   &     $ \dm_{30}(V) $   & $ M_B $ & $ M_V $  & $ \mni (M_\odot) $ \\	\midrule 
				2004ef   & $  0.83\,\pm\,0.03  $ & $  1.41\,\pm\,0.03  $ & $  0.84\,\pm\,0.02  $ & $  0.49\,\pm\,0.02  $ & $  1.98\,\pm\,0.02  $ & $-19.13\,\pm\,0.00  $ & $-19.07\,\pm\,0.00  $ & $   -   $ \\
				2004eo   & $  0.84\,\pm\,0.02  $ & $  1.43\,\pm\,0.03  $ & $  0.82\,\pm\,0.04  $ & $  0.52\,\pm\,0.02  $ & $  1.87\,\pm\,0.03  $ & $-19.07\,\pm\,0.00  $ & $-18.98\,\pm\,0.01  $ & $  0.47\,\pm\,0.12  $ \\
				2004ey   & $  1.16\,\pm\,0.04  $ & $  0.94\,\pm\,0.10  $ & $   -   $ & $  0.31\,\pm\,0.02  $ & $  1.52\,\pm\,0.02  $ & $-19.42\,\pm\,0.00  $ & $-19.33\,\pm\,0.00  $ & $   -   $ \\
				2004gs   & $  0.74\,\pm\,0.02  $ & $  1.64\,\pm\,0.04  $ & $  1.02\,\pm\,0.03  $ & $  0.65\,\pm\,0.02  $ & $  2.13\,\pm\,0.03  $ & $-18.90\,\pm\,0.01  $ & $-18.87\,\pm\,0.01  $ & $  0.40\,\pm\,0.08  $ \\
				2004gu   & $   -   $ & $   -   $ & $  0.55\,\pm\,0.07  $ & $   -   $ & $  1.36\,\pm\,0.06  $ & $   -   $ & $-19.41\,\pm\,0.01  $ & $   -   $ \\
				2005ag   & $   -   $ & $   -   $ & $  0.58\,\pm\,0.07  $ & $   -   $ & $  1.72\,\pm\,0.06  $ & $   -   $ & $-19.47\,\pm\,0.01  $ & $   -   $ \\
				2005al   & $   -   $ & $   -   $ & $  0.79\,\pm\,0.02  $ & $   -   $ & $  1.87\,\pm\,0.03  $ & $   -   $ & $-19.16\,\pm\,0.00  $ & $   -   $ \\
				2005am   & $   -   $ & $   -   $ & $  0.85\,\pm\,0.04  $ & $   -   $ & $  2.00\,\pm\,0.02  $ & $   -   $ & $-18.90\,\pm\,0.00  $ & $   -   $ \\
				2005A    & $   -   $ & $  1.15\,\pm\,0.05  $ & $  0.73\,\pm\,0.02  $ & $  0.35\,\pm\,0.04  $ & $  1.65\,\pm\,0.03  $ & $-19.37\,\pm\,0.01  $ & $-19.27\,\pm\,0.01  $ & $   -   $ \\
				2005bl   & $   -   $ & $   -   $ & $   -   $ & $  1.03\,\pm\,0.04  $ & $   -   $ & $-17.43\,\pm\,0.01  $ & $-17.87\,\pm\,0.01  $ & $   -   $ \\
				2005bo   & $   -   $ & $   -   $ & $  0.81\,\pm\,0.06  $ & $   -   $ & $  1.85\,\pm\,0.07  $ & $   -   $ & $-19.11\,\pm\,0.01  $ & $   -   $ \\
				2005el   & $  0.84\,\pm\,0.16  $ & $  1.31\,\pm\,0.14  $ & $  0.95\,\pm\,0.06  $ & $  0.44\,\pm\,0.11  $ & $  2.05\,\pm\,0.05  $ & $-19.19\,\pm\,0.03  $ & $-19.25\,\pm\,0.04  $ & $  0.50\,\pm\,0.20  $ \\
				2005eq   & $  1.19\,\pm\,0.09  $ & $  0.77\,\pm\,0.08  $ & $  0.55\,\pm\,0.05  $ & $  0.22\,\pm\,0.04  $ & $  1.34\,\pm\,0.04  $ & $-19.52\,\pm\,0.02  $ & $-19.41\,\pm\,0.01  $ & $  0.84\,\pm\,0.24  $ \\
				2005hc   & $  1.15\,\pm\,0.06  $ & $  0.96\,\pm\,0.05  $ & $  0.57\,\pm\,0.02  $ & $  0.32\,\pm\,0.03  $ & $  1.46\,\pm\,0.03  $ & $-19.50\,\pm\,0.01  $ & $-19.43\,\pm\,0.01  $ & $  0.70\,\pm\,0.16  $ \\
				2005hj   & - & $  0.76\,\pm\,0.06  $ & $  0.50\,\pm\,0.04  $ & $  0.31\,\pm\,0.03  $ & $  1.36\,\pm\,0.03  $ & $-19.38\,\pm\,0.01  $ & $-19.41\,\pm\,0.01  $ & $   -   $ \\
				2005iq   & $  0.87\,\pm\,0.08  $ & $  1.29\,\pm\,0.04  $ & $  0.82\,\pm\,0.02  $ & $  0.46\,\pm\,0.03  $ & $  1.91\,\pm\,0.03  $ & $-19.25\,\pm\,0.01  $ & $-19.18\,\pm\,0.01  $ & $   -   $ \\
				2005ir   & $   -   $ & $  0.95\,\pm\,0.09  $ & $  0.78\,\pm\,0.04  $ & $  0.31\,\pm\,0.06  $ & $  1.91\,\pm\,0.06  $ & $-19.49\,\pm\,0.01  $ & $-19.55\,\pm\,0.02  $ & $   -   $ \\
				2005kc   & $   -   $ & $  1.24\,\pm\,0.04  $ & $  0.74\,\pm\,0.02  $ & $  0.39\,\pm\,0.03  $ & $   -   $ & $-19.33\,\pm\,0.00  $ & $-19.13\,\pm\,0.01  $ & $   -   $ \\
				2005ke   & $  0.46\,\pm\,0.01  $ & $  1.77\,\pm\,0.02  $ & $  1.28\,\pm\,0.02  $ & $  0.91\,\pm\,0.03  $ & $  2.14\,\pm\,0.02  $ & $-17.56\,\pm\,0.01  $ & $-18.04\,\pm\,0.01  $ & $  0.13\,\pm\,0.03  $ \\
				2005ki   & $  0.87\,\pm\,0.03  $ & $  1.27\,\pm\,0.04  $ & $  0.82\,\pm\,0.02  $ & $  0.39\,\pm\,0.03  $ & $  1.95\,\pm\,0.02  $ & $-19.14\,\pm\,0.01  $ & $-19.10\,\pm\,0.01  $ & $  0.46\,\pm\,0.06  $ \\
			\end{tabular}}
	\end{threeparttable}
\end{table*}

\begin{table*}
		\caption{BTS sample  $ g $- and $ r $-band color stretches and decline rates. The complete table is available in the electronic supplementary material. }
	\label{tab:BTS sample}
	\begin{threeparttable}
		\renewcommand{\TPTminimum}{\linewidth}
		\makebox[\linewidth]{%
		\begin{tabular}{llcccccrrrr}
		Name & \sgr\ & \sgr\ src\tnote{a} &    $ \dmf(g)  $& $ \dmf(r) $  & $ \dm_8(g) $   &     $ \dm_{30}(r) $   & dist\tnote{b} & LF flag\tnote{c}  & hostabs\tnote{d} & hostcol\tnote{e}  \\	\midrule 
		2018bpd    & $  1.06\,\pm\,0.06  $ & dm30r   & $   -   $ & $  0.80\,\pm\,0.04  $& $   -   $ & $  1.13\,\pm\,0.06  $ & 160.84 & 0 &$-$18.60 & 0.49   \\
		2018bxo    & $  1.04\,\pm\,0.06  $ & dm15g   & $  0.97\,\pm\,0.06  $ & $  0.65\,\pm\,0.09  $& $  0.35\,\pm\,0.04  $ & $  1.13\,\pm\,0.13  $ & 278.14 & 0 &$-$20.86 & 0.58   \\
		2018ccj    & $  0.72\,\pm\,0.07  $ & dm15g   & $  1.50\,\pm\,0.09  $ & $   -   $& $  0.62\,\pm\,0.09  $ & $   -   $ & --    & 0 &$-$22.20 & 1.11   \\
		2018ccl    & $  0.90\,\pm\,0.07  $ & dir     & $  1.28\,\pm\,0.07  $ & $   -   $& $  0.50\,\pm\,0.05  $ & $  1.56\,\pm\,0.04  $ & 117.39 & 0 &$-$22.33 & 1.19   \\
		2018cdt    & $  0.79\,\pm\,0.08  $ & dm15g   & $  1.39\,\pm\,0.10  $ & $  0.78\,\pm\,0.04  $& $  0.50\,\pm\,0.09  $ & $   -   $ & 161.80 & 0 &$-$22.12 & 1.15   \\
		2018cdu    & $  0.88\,\pm\,0.09  $ & dm15g   & $  1.22\,\pm\,0.13  $ & $  0.78\,\pm\,0.07  $& $  0.44\,\pm\,0.07  $ & $   -   $ & 356.93 & 0 &$-$22.64 & 1.13   \\
		2018cfa    & $  0.88\,\pm\,0.07  $ & dm15g   & $  1.23\,\pm\,0.08  $ & $  0.77\,\pm\,0.04  $& $  0.41\,\pm\,0.08  $ & $  1.52\,\pm\,0.08  $ & 147.88 & 0 &$-$18.48 & 0.89   \\
		2018cng    & $  1.09\,\pm\,0.09  $ & dm30r   & $   -   $ & $  0.70\,\pm\,0.07  $& $   -   $ & $  1.06\,\pm\,0.16  $ & --    & 0 &$-$17.83 & 0.96   \\
		2018cnw    & $  1.09\,\pm\,0.12  $ & dir     & $  0.76\,\pm\,0.05  $ & $  0.70\,\pm\,0.10  $& $  0.26\,\pm\,0.03  $ & $  0.99\,\pm\,0.09  $ & 97.27  & 0 &$-$22.61 & 1.21   \\
		2018cny    & $  0.99\,\pm\,0.17  $ & dm15g   & $  1.05\,\pm\,0.28  $ & $  0.64\,\pm\,0.08  $& $  0.35\,\pm\,0.16  $ & $  1.18\,\pm\,0.08  $ & 212.98 & 0 &$-$21.65 & 1.26   \\
		2018cod    & $  1.24\,\pm\,0.09  $ & dm15g   & $  0.63\,\pm\,0.12  $ & $  0.64\,\pm\,0.14  $& $  0.27\,\pm\,0.04  $ & $  0.99\,\pm\,0.05  $ & 209.20 & 0 &$-$13.49 & 0.17   \\
		2018coi    & $  0.89\,\pm\,0.12  $ & dm15g   & $  1.21\,\pm\,0.18  $ & $  0.67\,\pm\,0.09  $& $  0.39\,\pm\,0.20  $ & $  1.59\,\pm\,0.20  $ & 266.01 & 0 &$-$22.48 & 0.88   \\
		2018cri    & $  0.93\,\pm\,0.10  $ & dm15g   & $  1.15\,\pm\,0.14  $ & $  0.75\,\pm\,0.08  $& $  0.38\,\pm\,0.07  $ & $  1.61\,\pm\,0.14  $ & 313.35 & 0 &$-$17.16 & 2.49   \\
		2018crk    & $  1.11\,\pm\,0.07  $ & dm15g   & $  0.84\,\pm\,0.09  $ & $  0.30\,\pm\,0.12  $& $  0.30\,\pm\,0.06  $ & $  0.87\,\pm\,0.09  $ & 277.45 & 0 &$-$21.36 & 1.11   \\
		2018crn    & $  0.96\,\pm\,0.15  $ & dm15g   & $  1.10\,\pm\,0.25  $ & $  0.68\,\pm\,0.10  $& $  0.32\,\pm\,0.15  $ & $  1.37\,\pm\,0.12  $ & 266.12 & 0 &$-$19.38 & 1.23   \\
		2018ctm    & $  1.07\,\pm\,0.08  $ & dm15g   & $  0.91\,\pm\,0.11  $ & $  0.67\,\pm\,0.11  $& $  0.39\,\pm\,0.05  $ & $  1.28\,\pm\,0.20  $ & 307.82 & 0 &$-$21.23 & 0.67   \\
		2018cto    & $  0.89\,\pm\,0.07  $ & dm30r   & $  1.62\,\pm\,0.22  $ & $  0.70\,\pm\,0.12  $& $   -   $ & $  1.51\,\pm\,0.08  $ & 221.98 & 0 &$-$22.07 & 1.17   \\
		2018cuw    & $  1.07\,\pm\,0.06  $ & dir     & $  0.91\,\pm\,0.07  $ & $  0.75\,\pm\,0.07  $& $  0.36\,\pm\,0.05  $ & $  1.14\,\pm\,0.11  $ & 129.67 & 0 &$-$19.40 & 0.71   \\
		2018cvd    & $  0.88\,\pm\,0.10  $ & dm15g   & $  1.24\,\pm\,0.15  $ & $   -   $& $  0.53\,\pm\,0.09  $ & $   -   $ & 385.81 & 0 &$-$15.17 & 1.05   \\
		2018cvq    & $  0.93\,\pm\,0.14  $ & dm15g   & $  1.14\,\pm\,0.23  $ & $  0.74\,\pm\,0.12  $& $  0.42\,\pm\,0.19  $ & $  1.21\,\pm\,0.11  $ & 300.37 & 0 &$-$22.12 & 0.88   \\
	\end{tabular}}
		\begin{tablenotes}
		\item [a] The source for the \sgr\ value. The options are either `dir' (directly from the light curve) or one of the decline rates. 
		\item [b] The distance to the SNe in Mpc, calculated from the spectroscopic redshift, if available. If not, it is estimated from the peak magnitudes absolute and observed magnitudes. SNe without \sgr\ or without observed peak magnitudes in both bands do not have an estimated distance. 
		\item [c] A flag indicating if the SN is within the time range of the LF.
		\item [d] Host galaxy absolute $ i $-band magnitude.
		\item [e] Host galaxy $ g-i $ color.
	\end{tablenotes}
	\end{threeparttable}
\end{table*}

\begin{table*}
	\caption{LOSS sample $ B $ and $ V $ band color stretch, decline rates, and distances. The complete is available in the electronic supplementary material.}
	\label{tab:LOSS sample}
	\begin{threeparttable}
		\renewcommand{\TPTminimum}{\linewidth}
		\makebox[\linewidth]{%
	\begin{tabular}{llcccccr}
		Name & \sbv & \sbv src\tnote{a} &    $ \dmf(B)  $& $ \dmf(V) $  & $ \dm_8(B) $   &     $ \dm_{30}(V) $ & dist [Mpc]\tnote{b}  \\	\midrule 
		1998de     & $  0.46\,\pm\,0.04  $ & $\dm_{8}(B)$    & $  1.87\,\pm\,0.08  $ & $  1.31\,\pm\,0.03  $& $  0.99\,\pm\,0.06  $ & $  2.28\,\pm\,0.12  $ & $  69.20 $\\
		1998dh     & $  0.93\,\pm\,0.06  $ & dir     & $  1.17\,\pm\,0.02  $ & $  0.72\,\pm\,0.04  $& $  0.40\,\pm\,0.02  $ & $  1.66\,\pm\,0.07  $ & $  37.00 $\\
		1998dk     & $   -   $ & none    & $   -   $ & $   -   $& $   -   $ & $   -   $ & $  53.80 $\\
		1998dm     & $  1.18\,\pm\,0.09  $ & dir     & $  0.81\,\pm\,0.05  $ & $  0.64\,\pm\,0.05  $& $  0.28\,\pm\,0.03  $ & $  1.36\,\pm\,0.06  $ & $  25.80 $\\
		1998ef     & $  0.96\,\pm\,0.05  $ & dir     & $  1.18\,\pm\,0.02  $ & $  0.71\,\pm\,0.02  $& $  0.44\,\pm\,0.03  $ & $  1.77\,\pm\,0.06  $ & $  74.20 $\\
		1998es     & $  1.09\,\pm\,0.04  $ & $\dm_{8}(B)$    & $  0.87\,\pm\,0.07  $ & $  0.48\,\pm\,0.04  $& $  0.28\,\pm\,0.04  $ & $  1.25\,\pm\,0.07  $ & $  43.00 $\\
		1999aa     & $  1.21\,\pm\,0.06  $ & dir     & $  0.83\,\pm\,0.09  $ & $  0.61\,\pm\,0.06  $& $  0.24\,\pm\,0.05  $ & $  1.31\,\pm\,0.07  $ & $  60.10 $\\
		1999ac     & $  0.93\,\pm\,0.07  $ & dir     & $  1.23\,\pm\,0.04  $ & $  0.80\,\pm\,0.04  $& $  0.40\,\pm\,0.04  $ & $  1.58\,\pm\,0.03  $ & $  40.70 $\\
		1999bh     & $   -   $ & none    & $   -   $ & $   -   $& $   -   $ & $   -   $ & $  74.00 $\\
		1999by     & $  0.49\,\pm\,0.02  $ & dir     & $  1.91\,\pm\,0.02  $ & $  1.29\,\pm\,0.02  $& $  1.01\,\pm\,0.02  $ & $  2.17\,\pm\,0.02  $ & $  11.40 $\\
		1999cl     & $  0.93\,\pm\,0.10  $ & dir     & $  1.08\,\pm\,0.03  $ & $  0.72\,\pm\,0.03  $& $  0.38\,\pm\,0.02  $ & $  1.62\,\pm\,0.05  $ & $  13.10 $\\
		1999cp     & $  0.97\,\pm\,0.02  $ & $\dm_{8}(B)$    & $  1.02\,\pm\,0.02  $ & $  0.69\,\pm\,0.02  $& $  0.38\,\pm\,0.02  $ & $  1.51\,\pm\,0.02  $ & $  39.40 $\\
		1999da     & $  0.48\,\pm\,0.03  $ & $\dm_{8}(B)$    & $  1.93\,\pm\,0.06  $ & $  1.29\,\pm\,0.03  $& $  0.96\,\pm\,0.05  $ & $  2.18\,\pm\,0.06  $ & $  54.30 $\\
		1999dk     & $  0.99\,\pm\,0.09  $ & dir     & $  1.12\,\pm\,0.04  $ & $  0.82\,\pm\,0.03  $& $  0.36\,\pm\,0.04  $ & $  1.77\,\pm\,0.07  $ & $  61.70 $\\
		1999dq     & $  1.17\,\pm\,0.13  $ & dir     & $  0.81\,\pm\,0.05  $ & $  0.62\,\pm\,0.08  $& $  0.22\,\pm\,0.04  $ & $  1.42\,\pm\,0.06  $ & $  59.40 $\\
		1999ej     & $  0.81\,\pm\,0.02  $ & $\dm_{8}(B)$    & $  1.52\,\pm\,0.03  $ & $  0.86\,\pm\,0.03  $& $  0.54\,\pm\,0.02  $ & $  1.99\,\pm\,0.03  $ & $  57.80 $\\
		1999ek     & $  0.93\,\pm\,0.02  $ & $\dm_{8}(B)$    & $  1.23\,\pm\,0.03  $ & $  0.70\,\pm\,0.03  $& $  0.42\,\pm\,0.02  $ & $  1.61\,\pm\,0.06  $ & $  72.00 $\\
		1999gd     & $   -   $ & none    & $   -   $ & $   -   $& $   -   $ & $   -   $ & $  77.00 $\\
		2000dm     & $  0.88\,\pm\,0.08  $ & dir     & $  1.50\,\pm\,0.05  $ & $  0.86\,\pm\,0.03  $& $  0.50\,\pm\,0.03  $ & $  2.01\,\pm\,0.05  $ & $  64.20 $\\
		2000dr     & $  0.81\,\pm\,0.08  $ & dir     & $  1.73\,\pm\,0.10  $ & $  1.02\,\pm\,0.03  $& $  0.66\,\pm\,0.06  $ & $  2.13\,\pm\,0.06  $ & $  75.60 $\\	
	\end{tabular}}
	\begin{tablenotes}
	\item [a] The source for the \sbv value. The options are either 'dir' (directly from the light curve) or one of the decline rates. 
	\item[b] Distance to the host galaxy, as given in \cite{Li2011}.
	\end{tablenotes}
\end{threeparttable}
\end{table*}

\bsp	
\label{lastpage}
\end{document}